\documentclass[twocolumn]{aastex631}

\usepackage{amssymb,amsmath,graphicx,natbib,hyperref}
\usepackage{booktabs}
\usepackage{longtable}
\usepackage{multirow}
\usepackage{url}
\usepackage{graphicx}
\usepackage{color}
\usepackage{enumitem}

\newcommand{\Ha}{$\rm{H}\alpha$}

\DeclareRobustCommand{\ION}[2]{%
\relax\ifmmode
\ifx\testbx\f@series
{\mathbf{#1\,\mathsc{#2}}}\else
{\mathrm{#1\,\mathsc{#2}}}\fi
\else\textup{#1\,{\mdseries\textsc{#2}}}%
\fi}
\newcommand{\kms}{\mathrm{\,km\,s^{-1}}}

\newcommand{\hii}{\ION{H}{ii}}

\newcommand{\nii}{[\ION{N}{ii}]}

\newcommand{\oiii}{[\ION{O}{iii}]}
\newcommand{\sii}{[\ION{S}{ii}]}

\newcommand{\sn}{\mathrm{SN}}

\newcommand{\ha}{H$\alpha$} 
\newcommand{\hb}{H$\beta$}

\newcommand{\XS}{$\mathtt{XS}$}

\shorttitle{Physical properties of gas exhibiting noncircular motions}
\shortauthors{L\'opez-Cob\'a et al.}

\begin{document}

\title{Physical properties of gas departing from circular rotation at 50 pc scales using the PHANGS-MUSE galaxies} 

\correspondingauthor{C. L\'opez-Cob\'a}
\email{calopez@asiaa.sinica.edu.tw}

\author[0000-0003-1045-0702]{Carlos~L\'opez-Cob\'a}
\affiliation{Institute of Astronomy and Astrophysics, Academia Sinica, No. 1, Section 4, Roosevelt Road, Taipei 10617, Taiwan}
\author[0000-0001-7218-7407]{Lihwai~Lin}
\affiliation{Institute of Astronomy and Astrophysics, Academia Sinica, No. 1, Section 4, Roosevelt Road, Taipei 10617, Taiwan}
\author[0000-0001-6444-9307]{Irene Cruz Gonzalez}
\affiliation{Instituto de Astronom\'ia, Universidad Nacional Autonoma de M\'exico, Circuito Exterior, Ciudad Universitaria, Ciudad de M\'exico 04510, Mexico}
\author[0000-0001-6444-9307]{Sebasti\'an~F.~S\'anchez}
\affiliation{Instituto de Astronom\'ia, Universidad Nacional Autonoma de M\'exico, Circuito Exterior, Ciudad Universitaria, Ciudad de M\'exico 04510, Mexico}
\author[0000-0002-1370-6964]{Hsi-An Pan}
\affiliation{Department of Physics, Tamkang University, No.151, Yingzhuan Road, Tamsui District, New Taipei City 251301, Taiwan}
\author[0000-0001-6444-9307]{J. K. Barrera-Ballesteros}
\affiliation{Instituto de Astronom\'ia, Universidad Nacional Autonoma de M\'exico, Circuito Exterior, Ciudad Universitaria, Ciudad de M\'exico 04510, Mexico}
\affiliation{Institute of Astronomy and Astrophysics, Academia Sinica, No. 1, Section 4, Roosevelt Road, Taipei 10617, Taiwan}
\author[0000-0001-5615-4904]{Bau-Ching Hsieh}

\begin{abstract}
Noncircular motions  have been observed across various spatial scales in disk galaxies, yet the physical properties of the gas involved in these motions remain poorly constrained. Using data from 19 galaxies from the PHANGS-MUSE sample, we investigated the prevalence of noncircular flows at spatial resolutions of tens of parsecs. We developed a new tool for 3D kinematic modeling of data cubes and applied to the PHANGS-MUSE \ha~spectral lines to recover the underlying circular, noncircular motions, as well as the intrinsic velocity dispersion in these objects. The PHANGS-MUSE galaxies exhibit rotation supported disks with $V_\mathrm{rot}/\sigma_\mathrm{intrin}$ ratios { $\gtrsim$ 5}.
Our analysis revealed ionized gas exhibiting noncircular motions at different amplitudes, with  low velocity amplitudes of about $5\kms$ associated with the axisymmetric rotation component,  deviations of $\sim10\kms$  primarily linked to interarm and spiral arms, and larger deviations ($>20\kms$), found in the central and bar regions. We found that the velocity dispersion and the strength of ionization correlate with the amplitude of noncircular motions, suggesting that the underlying dynamics of the warm gas are closely tied to its physical properties.
\end{abstract}



\section{Introduction}
Understanding the  processes that shape the velocity field of galaxies is important to characterize the dynamics of both their structural components and that of the interstellar medium (ISM).
Often, the presence of kinematic disturbances in velocity fields have been associated with noncircular (NC) flows occurring in and out of the disk plane. These have been observed in gas observations from neutral to ionized phases \citep[e.g.,][]{Bosma1978PhD, Visser1980, Marcelin1985,2024arXiv241021147L}, and also in the stellar velocity \citep[][]{LopezCoba2022}.

Under the common assumption that gas and stars follow circular rotation only, it has been possible to explain,  to certain extent, the overall shape of velocity fields. \citet{Begeman1989} was one of the first to introduce the guidelines for describing, and modeling the dynamics of galaxies by adopting simple assumptions about the gas rotation on disk galaxies. Their kinematics description based on tilted rings has long served, and continuous to serve, as a foundation for many algorithms that model the dynamics of disk galaxies. These include  2D approaches, which aim to model the velocity field of galaxies \cite[e.g.,][]{kinemetry,DiskFit,XookSuut}, and 3D approaches, which aim to model the line-of-sight (LOS) velocities from spectral line cube observations while correcting for artificial broadening due to instrumental and spatial resolution effects \citep[e.g.,][]{BBarolo,GalPak,Price2021}.

An immediate result from these models, regardless of the method used, is that circular rotation alone cannot fully replicate the observed kinematics, revealing imprints of both small- and large-scale NC motions. This has been verified in observations at various angular and spectral resolutions, ISM gas phases, and redshifts \citep[e.g.,][]{Genzel2023}.

NC motions however, have been challenging to model using either 2D or 3D algorithms. This difficulty arises primarily from our incomplete understanding of the mechanisms governing gas and stellar flows across multiple scales in galaxies. Additionally, reproducing the observed LOS velocities from models is complicated by resolution effects, which often diminish the contribution of NC rotation.

These challenges have prevented detailed explorations of the physical processes governing gas involved in NC rotation in galaxies. Nevertheless, numerous studies have suggested a relationship between kinematics and the physical properties of the ISM \citep[e.g.,][]{Ho2014,AMUSING++, Yung-Chau2022, WHaD, Barrera-Ballesteros2023,2024arXiv241021147L,2024arXiv241020583K}. Such relationships are apparently linked to global and local perturbations.
For instance, non-axisymmetric structures such spiral arms or stellar bars induce large scale  perturbations in the gas and stars, affecting the dynamics and physical conditions of the ISM through shocks, inflowing motions and star-formation \citep[e.g.,][]{Roberts1969, Mundell1999, Riffel2008, Sormani2023, Kim2024}. In the other extreme, local perturbations to the ISM such as supernovae explosions and active galactic nuclei are able to drive winds that contribute to the observed velocity field. In either case, the amplitude of the NC motions induced can be as large as some fractions of the local circular velocities \citep[e.g.,][]{Erroz-Ferrer2015, clc2018, Krieger2019}.

Thanks to advanced observational techniques that allow access to multiple emission-lines, as is the case of the integral field spectroscopy IFS, we are able to gain insights about the dynamics of the ISM and stars, and about the properties of the underlying gas that shape the velocity fields.
The Physics at High Angular resolution in Nearby GalaxieS, PHANGS-MUSE \citep[e.g.,][]{Emsellem2022}, is a galaxy survey that used the MUSE/VLT instrument to map 19 galaxies from the nearby Universe ($D<20$~Mpc) covering fields of view larger than or approximately  to 2 times the effective radii.
These data have disentangled the properties of the warm ISM and the stellar populations at an unprecedented angular resolution of $\sim50$ pc at the average distance  \citep[][among others]{Pessa2023,Groves2023,Belfiore2023B}.

In this work, we utilize the exquisite data from PHANG-MUSE to trace the noncircular motions of ionized gas and stars, and to uncover relationships between these motions and the physical properties of the gas. As part of this analysis, we developed a software tool for the kinematic modeling of spectral-line-cube observations, that can be applied to most of the 3-Dimensional observations.

The paper is organized as follows: Section 2 describes the stellar population analysis used to recover the ionized gas properties, while Section 3 focuses on the 3D kinematic modeling of spectral lines. Section 4 explores the physical properties of gas in NC rotation. The results for the 19 PHANGS-MUSE objects are presented in Section 5, followed by a discussion in Section 6, and conclusions in Section 7. The appendix includes individual figures for each of the 19 objects analyzed.

\begin{figure*}[t!]
  \centering
     \includegraphics[width = \textwidth]{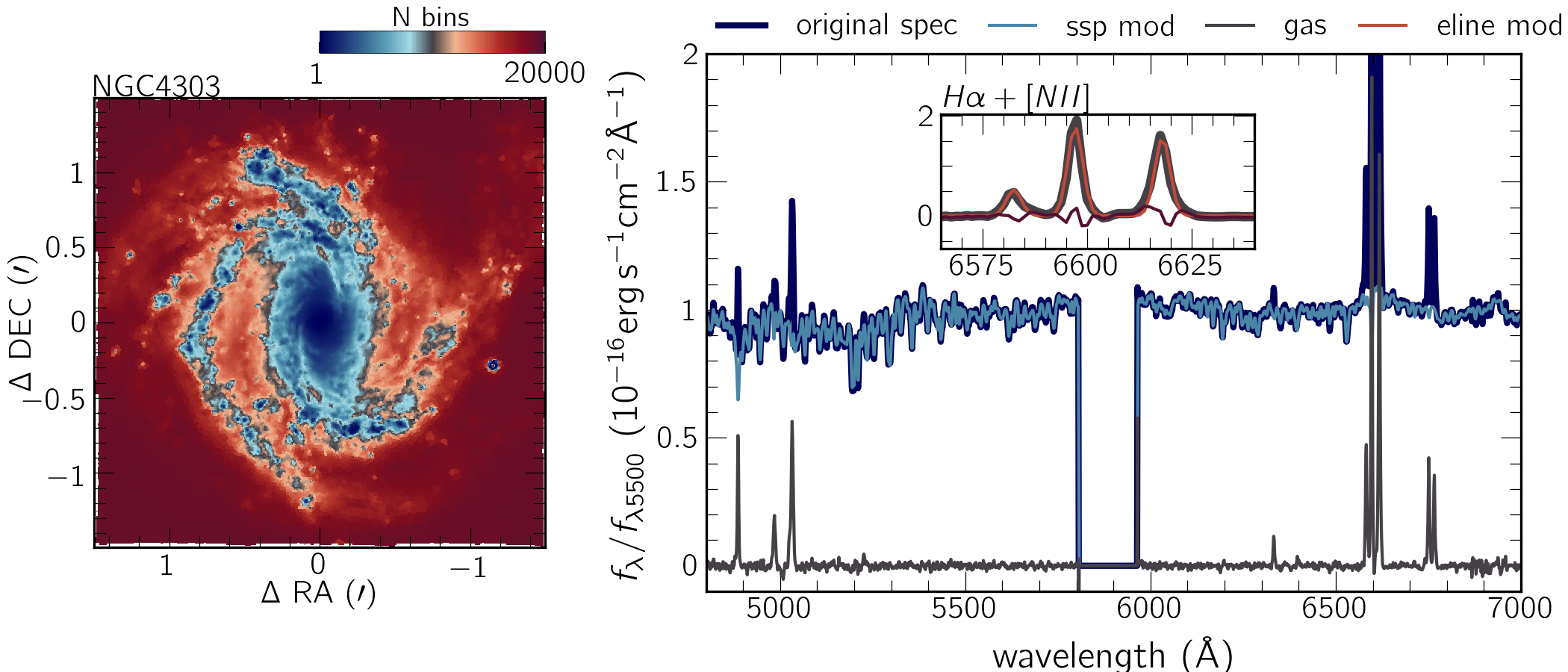}
  \caption{ Example of the spectral fitting in one object from the PHANGS-MUSE sample. {\it Left panel}: Voronoi map of NGC\,4303. The color bar indicates the bin number, each comprising one or more spaxels. {\it Right panel:} Implementation of the single stellar population analysis to an arbitrary spectrum within the galaxy. The original spectrum is shown in dark blue while the best SSP model is shown in pale blue; the gas spectrum (free of stellar continuum) is shown in gray, the emission line model in red, and the residuals in dark red. Additionally, a zoom in around the \ha~line is shown for better appreciation.
  }
  \label{fig:pipe3d}
\end{figure*}

\begin{deluxetable*}{llcccccccccccc}
\tablecaption{Main properties of the PHANGS/MUSE objects. \label{Tab:main_props}}
\tablewidth{0pt}
\tablecolumns{13}
\tablehead{
\colhead{ID} & \colhead{Object} \vspace{-0.2cm} & \colhead{d} & \colhead{scale} &\colhead{$r_e$} & \colhead{$\log M_{\star}$} & \colhead{$\log SFR$} & \colhead{$\phi$} & \colhead{$i$} & \colhead{$v_\mathrm{sys}$} & \colhead{dK01} & \colhead{EW(\ha)} & \colhead{$V_t/\sigma_0$} \\
\colhead{} & \colhead{} & \colhead{(Mpc)} &\colhead{(pc/$^{\prime\prime}$)} &\colhead{(kpc)} & \colhead{($M_{\sun}$)} & \colhead{($M_{\odot} yr^{-1}$)} & \colhead{($^{\circ}$)} & \colhead{($^{\circ}$)} & \colhead{($\kms$)} & \colhead{(dex)} & (\AA) & \colhead{}
}
\startdata
1& NGC1087& 15.8& 77& 3.2& 9.93& -0.3& 359& 42& 1528& -0.2& 54& 4.1\\
2& NGC1300& 19.0& 92& 6.5& 10.62& -0.54& 273& 41& 1576& 0.1& 10& 4.8\\
3& NGC1385& 17.2& 83& 3.4& 9.98& -0.12& 179& 40& 1503& -0.2& 65& 3.6\\
4& NGC1433& 18.6& 90& 4.3& 10.87& 0.0& 199& 27& 1081& 0.5& 3& 7.0\\
5& NGC1512& 18.8& 91& 4.8& 10.71& -0.76& 266& 40& 899& 0.2& 7& 5.8\\
6& NGC1566& 17.7& 86& 3.2& 10.78& 0.03& 217& 31& 1512& 0.5& 4& 5.0\\
7& NGC1672& 19.4& 94& 3.4& 10.73& 0.23& 131& 33& 1344& -0.0& 49& 4.6\\
8& NGC2835& 12.2& 59& 3.3& 10.0& 0.53& 1& 51& 896& -0.1& 5& 3.7\\
9& NGC3351& 10.0& 48& 3.0& 10.36& -0.81& 193& 36& 783& -0.2& 24& 6.5\\
10& NGC3627& 11.3& 55& 3.6& 10.83& 0.07& 172& 59& 716& 0.7& 1& 4.6\\
11& NGC4254& 13.1& 64& 2.4& 10.42& 0.24& 69& 35& 2411& -0.1& 12& 5.1\\
12& NGC4303& 17.0& 82& 3.4& 10.52& 0.57& 313& 23& 1569& 0.2& 13& 5.0\\
13& NGC4321& 15.2& 74& 5.5& 10.75& -0.04& 157& 35& 1582& -0.1& 20& 5.8\\
14& NGC4535& 15.8& 76& 6.3& 10.53& -0.36& 178& 37& 1965& -0.0& 16& 5.7\\
15& NGC7496& 18.7& 91& 3.8& 10.0& -0.57& 198& 29& 1655& -0.1& 35& 3.5\\
16& IC5332& 9.0& 44& 3.6& 9.67& -1.69& 66& 35& 692& -0.0& 7& 1.5\\
17& NGC628& 9.8& 48& 3.9& 10.34& -0.24& 17& 33& 659& 0.0& 2& 1.2\\
18& NGC5068& 5.2& 25& 2.0& 9.4& -0.97& 340& 29& 666& -0.2& 9& 1.1\\
19& NGC1365& 19.6& 95& 2.8& 10.99& 0.59& 212& 53& 1628& 0.1& 53& 6.5\\
\hline
\enddata
\tablecomments{From left to right columns: Object's ID; Object's name; distance to the source obtained from \citet{Leroy2021a}; spatial scale; effective radius $r_e$ obtained from \citet{Leroy2021a}; stellar mass obtained from \citet{Leroy2021a};
integrated star formation rate estimated from \ha; kinematic position angle ($\phi$); disk inclination $i$ from 3D kinematic modelling; systemic velocity; perpendicular distance  from the \citet{Kewley2001} demarcation, with positive (negative) values indicating positions above (below) this curve; equivalent width of \ha, EW(\ha); average ratio  of the circular rotation to the intrinsic velocity dispersion. Both $d_\mathrm{K01}$ and EW(\ha) were extracted over a 2\arcsec-radius circular aperture around the kinematic centre. IC\,5332, NGC\,5068, and NGC\,628  are nearly face-on and therefore their LOS velocity does not provide enough information to confidently model the rotational curves.}
\end{deluxetable*}

\section{Data and Data analysis}

In this work we used the fully calibrated cubes from PHANGS-MUSE \citep[][]{phangs_eso_data}, obtained from the ESO Science Archive Facility\footnote{\url{http://archive.eso.org/cms.html}}; specifically, we use the mosaicked datacubes, with each mosaic containing between 3 and 15 individual MUSE observations each of $1\arcmin\times1\arcmin$ field-of-view (FOV).
The MUSE instrument provides spatially resolved spectra covering the optical spectral range (4700-9300\,\AA) with a wavelength-dependent spectral resolution \citep[e.g.,][]{Bacon2017}, approximately $\sigma_\mathrm{instr} \sim 1.3$\,\AA~in the blue part and 1.1\,\AA~in the red. The data cubes are sampled 1.25\,\AA~in the spectral direction and 0\farcs{2}/pixel in the spatial direction.

The full width at half maximum resolution (FWHM) of the PHANGS-MUSE mosaics is estimated to be better than 0\farcs{8} across the majority of objects \citep[e.g.,][]{Emsellem2022}. This FWHM resolution translates to physical scales ranging between 20 pc to 70 pc, for the considered distance of the objects.
 
The mosaicked cubes were analyzed with the Python language version of the {\tt pipe3d} pipeline \citep[e.g,][]{Pipe3D_I}, originally written in Perl language. This new version named {\tt pyPipe3D} \citep[e.g.,][]{Lacerda2022} share the same routines from its predecessor. The {\tt pipe3d} routines have successfully recovered the main properties of the ionized gas and the stellar continuum in a wide variety of integral field unit datasets \citep{2022ApJS..262...36S}. Early results on the PHANGS-MUSE objects using this pipeline were presented in \cite{N1087}.

Briefly, {\tt pyPipe3D} performs a decomposition of the observed spectra into multiple simple stellar populations (SSPs) each with different age and metallicities.
In alignment with the analysis from \cite{Emsellem2022}, we use the same stellar libraries based on the extended MILES project \citep[E-MILES,][]{EMILES}, MILES/FWHM $=2.5$\AA. We adopted 13 different ages (Age/Gyr = 0.03, 0.05, 0.08, 0.15, 0.25, 0.4, 0.6, 1, 1.75, 3, 5, 8.5, 13.5), and 6 metallicities ($\mathrm{[Z/H]} = -1.49, -0.96, -0.35, +0.06, +0.26, +0.40$), for a total of 78 templates.

Since MUSE exhibits a lower spectral resolution in the blue region compared to E-MILES, we convolved the stellar libraries with the MUSE line-spread function \citep[e.g.,][]{Bacon2017} in those regions  where $\sigma_\mathrm{instr}>\sigma_\mathrm{EMILES}$; that is, for $\lambda < 6850$\,\AA.

Before proceeding with the SSP analysis, the cubes were corrected for Galactic extinction with the coordinates recorded in the header, and flux units were scaled to the ones required by {\tt pyPipe3D}, namely $\times10^{-16}\mathrm{erg\,s^{-1}\,cm^{-2}\,\AA^{-1}}$. Additionally, to avoid contamination of sky lines in the red part of the spectra we restrict the SSP analysis to $\lambda<7000$\AA.

To increase the signal-to-noise (SN) of the stellar continuum in all spatial pixels (spaxels), we performed a Voronoi binning-segmentation \citep[][]{cappellari03}  on a 2D-slice of the mosaicked cubes centered around 5700\,\AA, and integrated over a $\pm$50\,\AA~spectral window, ensuring bin-sizes of the order of the PHANGS-MUSE/FWHM resolution and SN\,$\sim50$. This binned map will serve to compute different properties of the underlying stellar continuum, while the ionized gas properties are analyzed in a spaxel-wise sense. The emission-line properties such as flux, velocity and velocity dispersion are derived through Gaussian fitting to the strongest emission-lines in the spectra (\hb, \oiii, \nii, \ha, \sii).  In this work we will not use directly the {\tt pyPipe3D} velocity maps, but they will serve as reference for comparing our analysis. An example of the spectral and emission line fitting is shown in Figure~\ref{fig:pipe3d}.

The data-products from the {\tt pyPipe3D} pipeline are a set of maps comprising information about the stellar populations (stellar velocity, stellar mass, among others products), and the ionized gas (emission-line fluxes, emission-line velocities, equivalent widths etc). We refer to \cite{Lacerda2022} for a thorough description of the analysis.

\section{Kinematic modeling}
\subsection{3D fitting technique}
For modeling the gas kinematics, we developed a modified version of the \XS~package \citep[e.g.,][]{XookSuut}, building on the existing algorithm. \XS~was originally designed to model the kinematics of galaxies using  velocity or first moment  maps.
In this new approach we model a datacube containing the spectral line of interest, accounting for the instrumental configurations of the observation; that is, the spatial resolution and the spectral broadening.

As shown in Figure~\ref{fig:pipe3d}, the observed MUSE cube consist of two spatial axes  (R.A., Dec.) or ($x$, $y$) on sky coordinates and one spectral axis along the $z$ axis. Each 2D slice on a cube is called channel.
We proceed by isolating a single spectral line from a cube free of stellar continuum. This is achieved by removing the best SSP model cube from the original MUSE cube as described in \cite{Lacerda2022}. The result of this procedure is a cube containing only the ionized gas plus imperfections in the SSP modeling. Since the MUSE spectral range covers a wide set of emission-lines, we separate one spectral line, \ha~in this case, by trimming the gas cube along the spectral axis.
The latter is achieved by selecting a $\pm500\kms$ window around \ha~after considering the \ha~systemic velocity estimated from the SSP analysis. Given the closeness of the Nitrogen lines \nii$\lambda\lambda6548,6584$, we inspected the \ha~narrow band subcubes to avoid contamination of these lines and reduced the velocity window when necessary.

The following procedure is valid for any 3D-observation, without limiting to wavelength sampled data cubes. First, we assume that any given spectrum in the cube containing the spectral line of interest, called observed cube from now on, can be described as a Gaussian function, having an intrinsic velocity $v_0$ given by the gravitational potential, intensity $f_0$, and velocity dispersion $\sigma_0$. Thus, $\sigma_0$ is the intrinsic dispersion ($\sigma_\mathrm{intrin}$) of the line caused by the natural width, thermal broadening, sound speed, turbulence, or any line-of-sight (LOS) projected velocity dispersion. A spectral line in this cube is modeled as:
\begin{equation}
\label{Eq:Sintrin}
 S_\mathrm{intrin}(f_0,v_0,\sigma_0)= f_0\exp(-\Delta v^2/2\sigma_0^2).
\end{equation}
Note that we transformed the spectral axis $z$ to velocity following standard transformations from $\lambda$ and $\nu$ to $v$\footnote{For wavelength-based spectral axes, such as in MUSE, the optical definition of the Dopppler velocity is adopted:  $v = c(\lambda/\lambda_\mathrm{rest}-1)$. For frequency-based spectral axes, the radio definition is commonly used:  $v=c(1-\nu/\nu_\mathrm{rest})$, where $c$ is the speed of light, and $\lambda_\mathrm{rest}$ ($\nu_\mathrm{rest}$) is the rest wavelength (frequency) of the spectral-line.}. $\Delta v$ represents spectral shifts with respect to the central velocity $v_0$.

A Gaussian-shaped profile is commonly observed at low to intermediate spectral resolutions ($\sigma_\mathrm{instr}\gtrsim50\kms$), regardless of the gas phase. However, at very high spectral resolutions, on the order of a few km/s, spectral lines are observed to deviate from Gaussian shapes \citep[e.g.,][]{deBlok2008}.
Such cases present a significant challenge for modeling their LOS velocity distribution and require additional assumptions beyond the ones adopted here \citep[see ][]{Kabasares2022}.

The intrinsic emission-line profile in Eq.~\ref{Eq:Sintrin} is broadened as a result of the finite spectral resolution of the instrument, the line-spread-function LSF,  characterized by $\mathrm{ \sigma_{instr}}$ and assumed constant within the spectral window containing the emission-line. The second artificial broadening is caused by atmospheric conditions or imaging processing, known as point-spread-function (PSF) \footnote{The PSF is known as synthesised Beam in radio astronomy, with similar effects on the angular resolution.}.
The effects of the angular and the spectral resolution on gas kinematics have been extensively discussed in the literature, with a general consensus for being corrected to obtain a more accurate characterization of gas kinematics \citep[e.g.,][]{Begeman1989, Cresci2009, Davies2011, BBarolo, GalPak, Bekiaris2016, Levy2018, Chung2021}.

Here we assume that the PSF has constant shape through the different velocity channels, and it can be represented with a unique 2D elliptical Gaussian function, with semi-minor axis ${b_{min}}$,  semi-major axis ${ b_{maj} }$\footnote{We define the Beam/PSF major and minor FWHMs as BMAJ=$b_{maj}\sqrt{8\ln2}$ and BMIN=$b_{min}\sqrt{8\ln2}$, oriented at BPA=$b_\mathrm{pa}$ from N to E.}, and position angle $b_\mathrm{pa}$. Under these considerations, the instrumental line broadening is described by the product of the PSF and spectral broadening as follows
\begin{align}
 g(x,y,z) &= \frac{1}{2\pi b_{maj} b_{min}}  \exp \Big(  -\Big( \frac{x^2}{2b_{maj}^2} + \frac{y^2}{2b_{min}^2} \Big) \Big) \notag \\
          & \quad \times \frac{1}{\sqrt{2\pi} \sigma_\mathrm{instr} } \exp \Big( -\frac{z^2}{2\sigma^2_\mathrm{instr}} \Big).
\end{align}
This is a 3-dimensional Gaussian kernel composed by one 2-dimensional kernel which takes place on spatial coordinates  and one 1-dimensional kernel along the spectral axis. Note that although the kernel is Gaussian, it can induce to the appearance of asymmetric line profiles \citep[see][]{Kabasares2022}.

The expression to model an observed spectral line is then composed by the intrinsic Gaussian profile $S_\mathrm{intrin}$, convolved by the 3D kernel,
\begin{equation}
\label{Eq:Gaussian_model}
I(x,y,v)_\mathrm{model}=S_\mathrm{intrin} \otimes  g(x,y,v).
\end{equation}
The symbol $\otimes$ stands for convolution. $S_\mathrm{intrin}$ is the intrinsic spectral line profile free of any instrumental or observational broadening.
The right hand side of this equation can be generalized to a cube, in such case $\sigma_0=\sigma_0(x,y)$, $v_0=v_0(x,y)$, and $f_0=f_0(x,y)$. Consequently, we initialize an empty cube  of the spectral line of interest $S_\mathrm{intrin}(x,y,v)$, with equal dimensions as the observed cube; namely, $nz\times ny\times nx$ corresponding to $N$ individual pixels. The variables $\{f_0$, $v_0, \sigma_0 \}$ will be assigned to each spectrum in a subsequent procedure.

The 3D convolution in Eq.~\ref{Eq:Gaussian_model} is implemented through fast Fourier transforms (FFT). This is achieved by using the {\tt pyFFTW} library \citep{pyFFTW}, which is a pythonic wrapper around the Fastest Fourier Transform in the West  {\tt FFTW} in C-language \citep{FFTW05}.
To optimize this operation, we take advantage of two properties of the FFT. First, by applying the convolution theorem to Equation~\ref{Eq:Gaussian_model} we obtain
\begin{equation}
\label{Eq.conv_theorem}
 \mathcal{F}\{ I_\mathrm{model} \} = \mathcal{S}_\mathrm{intrin} \cdot \mathcal{G},
 \end{equation}
where $ \mathcal{S}_\mathrm{intrin} =  \mathcal{F}\{S_\mathrm{intrin} \} $,  $ \mathcal{G}  =  \mathcal{F}\{g \} $, and  $\mathcal{F}$ represents the discrete Fourier transform operator.
Then we just need to apply the inverse Fourier transform $\mathcal{F}^{-1}$ to both sides of this equation, to obtain the expression
that will allow us to model a spectral line considering the observational PSF and the spectral broadening:
\begin{equation}
\label{Eq:fft}
  I_\mathrm{model} = \operatorname{\mathbb{R}e} \{ \mathcal{F}^{-1} \{ \mathcal{S}_\mathrm{intrin} \cdot \mathcal{G} \}.
\end{equation}
Where $\operatorname{\mathbb{R}e}$ stands for real part.
Second, FFT performs more efficiently when the data dimensions are powers of 2. Hence, we apply zero padding to the nearest power of 2 along each each axis of the intrinsic and kernel cubes in Eq.~\ref{Eq:Gaussian_model}. This procedure significantly speeds up the convolution operation, thereby reducing the overall fitting time, at the cost of increasing the size of the cubes. Once completed this procedure, the model cube is reverted to its original dimensions.

The next steps consist in determine the set of parameters describing the $S_\mathrm{intrin}$ cube on Equation~\ref{Eq:Sintrin}.
The velocities $\{ v_0 \}$ and dispersions $\{ \sigma_0 \}$ of each Gaussian profile are estimated in a nonparametric way, following the same procedures described in \citet{XookSuut}, but extended here to account for the velocity dispersion and adapted for data cubes.
Briefly, this technique adopts the thin disk approximation; namely, the kinematic position angle ($\phi$), inclination ($i$) and kinematic centre are assumed constant throughout the gaseous disk.
Once defined a starting geometry for the disk, a set of $\{ v_{0,k},~\sigma_{0,k}\}$ is estimated on $k$ concentric rings from the 1st ($M_1$) and 2nd ($M_2$) moment maps of the observed cube (see moment definitions in Appendix~\ref{App:momaps}). This procedure follows an algebraic approach based on the method introduced by \citet{Barnes2003} (see details in Appendix~\ref{App:v0sigma0}).

The estimation of $\{ v_{0,k} \}$ requires the adoption of a given  model for the gas rotation. Explicitly for Eq.~\ref{Eq:Sintrin}, $\Delta v$ follows
\begin{equation}
\label{Eq:deltaV}
 \Delta v(x,y) = v_\mathrm{channel}(x,y) - v_\mathrm{model}(x,y)
\end{equation}
where $v_\mathrm{channel}$ is the observed channel velocity and $v_\mathrm{model}$ is the gas rotation model at the same channel.  Here, we extend the axisymmetric and nonaxisymmetric kinematic models presented in \citet{XookSuut}. In this work, we adopt the most simple case for the gas rotation described by circular motions,
\begin{equation}
\label{Eq:circ}
v_\mathrm{model} = v_t(r) \sin i \cos \theta + v_\mathrm{sys},\, v_t \equiv v_0.
\end{equation}
Where $v_t$ is the tangential velocity or circular rotation, $\theta$ is the azimuthal angle on the disk plane with $\theta=0$ along the line of nodes, $v_\mathrm{sys}$ is the systemic velocity, and $r$ represent distances measured on the disk plane and is related to the sky coordinates and the disk geometry.

The dispersion on the other hand is not model dependent and is assumed axisymmetric, namely $\sigma_0=\sigma_0(r)$. This is motivated on the observed dispersion profiles on bulge+disk galaxies, where dispersion is enhanced towards center, probably induced by gravitational virial motions on gas, with a gradual decrease at larger distances from the nucleus \citep[e.g.,][]{Barroso2017,Mogotsi2019}. Notice that $\sigma_0$ represents the line-of-sight dispersion of the intrinsic spectral line profile, and  does not distinguish between different broadening sources (e.g., thermal, turbulence, etc) or their projected components along the LOS \cite[see][their Equation (27)]{Cappellari2020}.

Maps of the intrinsic velocities $v_0(x,\,y)$ and $\sigma_0(x,\,y)$ are constructed by linear interpolation of $\{ v_{0,k},~\sigma_{0,k}\}$ over the considered rings.

Finally, we scale the intensity of each Gaussian $f_0$, to the integrated flux (0th moment) of the observed cube (see details in Appendix~\ref{App:v0sigma0}). { This analysis will produce an intensity map similar to the observed one.}
This approach is commonly adopted in 3D modeling of spectral lines to reduce the number of variables to fit, and to produce a non-axisymmetric model of the intensity distribution \citep[e.g.,][]{Takakuwa2012,BBarolo}.
After this step, a model cube of the spectral line $I(x,y,v)_\mathrm{model}$ has been created. Following, we proceed to estimate the best values representing the disk geometry, the intrinsic gas circular velocity, and the intrinsic velocity dispersion profiles through Least-Squares minimization adopting the {\tt LMFIT} library \citep[e.g.,][]{lmfit}.

The cost function to minimize adopts the expression
\begin{align}
\label{Eq:cost}
 J &= \sum_{x,\,y,\,v} \frac{1}{\sigma^2_\mathrm{MAD}} \Big( I_\mathrm{obs}(x,y,v)- I_\mathrm{model}(x,y,v) \Big)^2 \frac{N^2}{N_\mathrm{eff}^2} +  \notag \\
          & \quad \sum_{x,\,y} \Big( M_\mathrm{1,obs}-M_\mathrm{1,model}\Big)^2 + \Big( M_\mathrm{2,obs}-M_\mathrm{2,model}\Big)^2,
\end{align}
where $N_\mathrm{eff}$ is the number of effective (i.e., non-blanked) pixels used to create the model cube, and $\sigma_\mathrm{MAD}$ is the cube { characteristic error, estimated from the the median absolute deviation (MAD) from the negative fluxes.} The cost function compares the 3D FFT-convolved model cube, along with its first and second moments, to their observational counterparts, which are affected by the observational PSF and LSF. Note that at each iteration of Eq.~\ref{Eq:cost}, a new cube
$I_\mathrm{model}$ is created following Eq.~\ref{Eq:fft}. Thus, the cost function is evaluated only after the model cube of the spectral line is created, instead of a ring-basis evaluation.

In order to prevent from modeling very low signal to noise spectra or line-less spectra, which may cause the fitting procedure to become unstable, we implemented an automated 3D masking procedure based on the MAD error of a smoothed version of the observed cube $\sigma_\mathrm{MAD}^\mathrm{smooth}$, as detailed in \citet{Dame2011}.  The MAD error in the smoothed cube is estimated from the median absolute deviation from the negative fluxes.
In this way, pixels in the smoothed cube with fluxes lower than $\sigma_\mathrm{MAD}=6\times\sigma_\mathrm{MAD}^\mathrm{smooth}$ are blanked on the original cube. This procedure yields in general to an efficient method of line detection.

Errors on each fitted parameter are estimated from bootstrap, here after 100 realizations of Eq.~\ref{Eq:cost}. At each realization, the spectra in $I_\mathrm{obs}$ are randomly perturbed by adding Gaussian noise, with a standard deviation given by the average negative fluxes of each spectrum. Additionally, the disk position angle and inclination are perturbed around their best-fit values. Errors are obtained from the standard deviation of each parameter.

The entire procedure described above has been packed into a Python tool named {\tt XS3D}  available on zenodo \citep[e.g.,][]{xs3d} or on  Github\footnote{\url{https://github.com/CarlosCoba/XS3D}}. This tool has been generalized to handle wavelength, frequency, and velocity spectral axes, and like its 2D version {\tt XS}, it allows for modeling both circular and noncircular motions in disk systems. Moreover, since velocities are estimated non-parametrically, the analysis can be extended to gaseous disks of arbitrary scales, as is the case of protoplanetary disks, in such case $v_t(r)$ in Eq.~\ref{Eq:circ} represents Keplerian rotation \citep[e.g.,][]{Teague2021}.

\subsection{Implementation of {\tt XS3D} on PHANGS-MUSE}

\begin{figure*}[t!]
\centering
\includegraphics[width=1\textwidth,keepaspectratio]{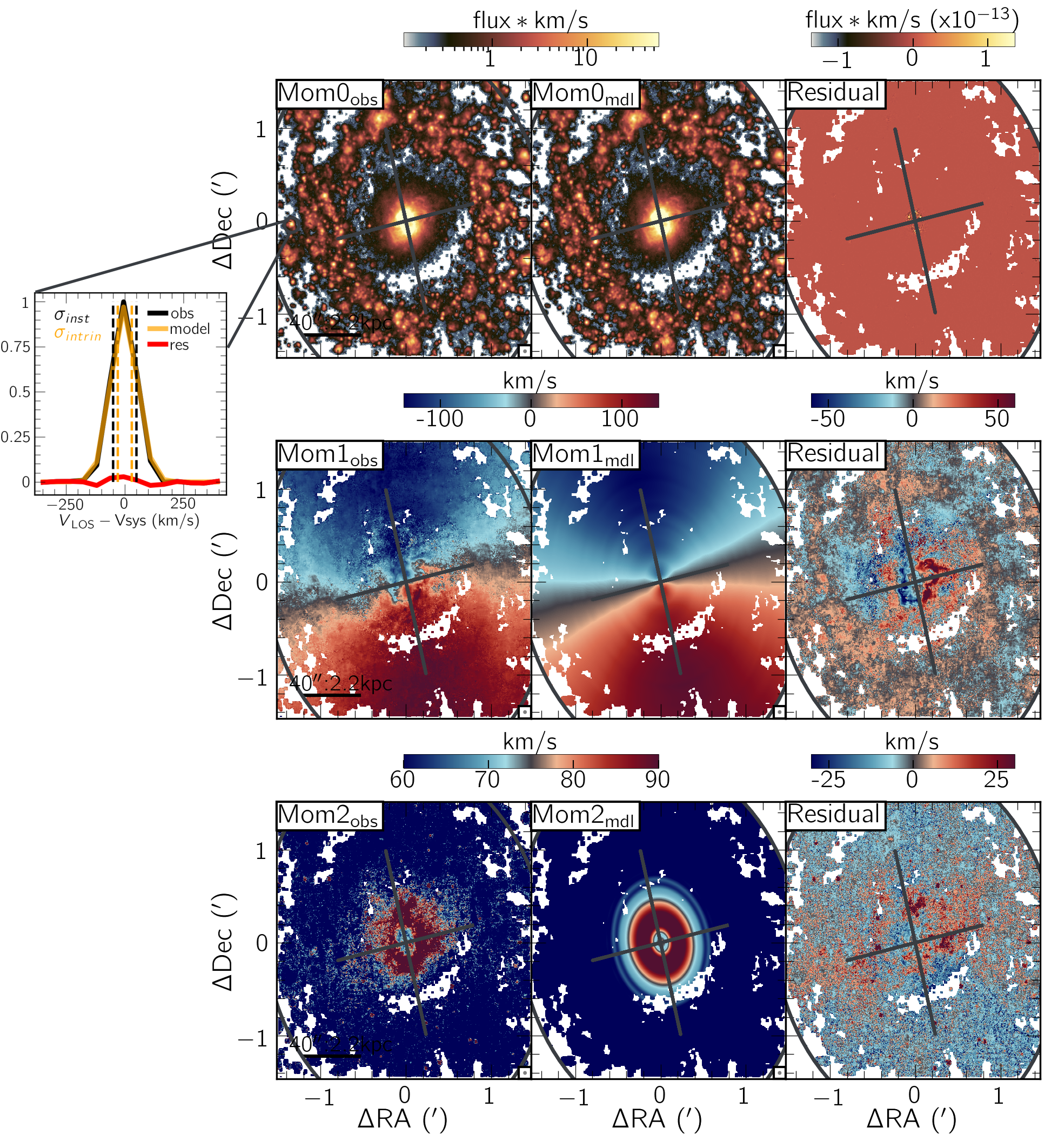}
\caption{Moment maps extracted from the observed ($I_\mathrm{obs}$) and model ($I_\mathrm{model}$, e.g., Eq.~\ref{Eq:fft}) cubes. The white patches show the blanked regions from the automated 3D masking procedure. Each column shows, from left to right, the observed moment map, the model, and the residual map (observed-model), respectively. Each  row from top to bottom shows the intensity weighted map, the velocity weighted, and the dispersion weighted (LSF and PSF affected) map. The black ellipse in each panel represents the best-fit geometry for the disk, while the central black cross shows the major and minor axes kinematic position angles. The PSF/Beam size is shown at the bottom right of each panel. The top-left inset shows an \ha~spectrum taken at an arbitrary position; the observation is shown in black, the model in orange and the residual in red. The vertical dashed lines show the instrumental and intrinsic dispersion in black and orange colors, respectively. The flux units are $\mathrm{ 10^{-16} erg\,s^{-1}\,cm^{-2}\,\AA^{-1} }$.}
\label{fig:moments}
\end{figure*}

\begin{figure*}[t!]
\centering
\includegraphics[]{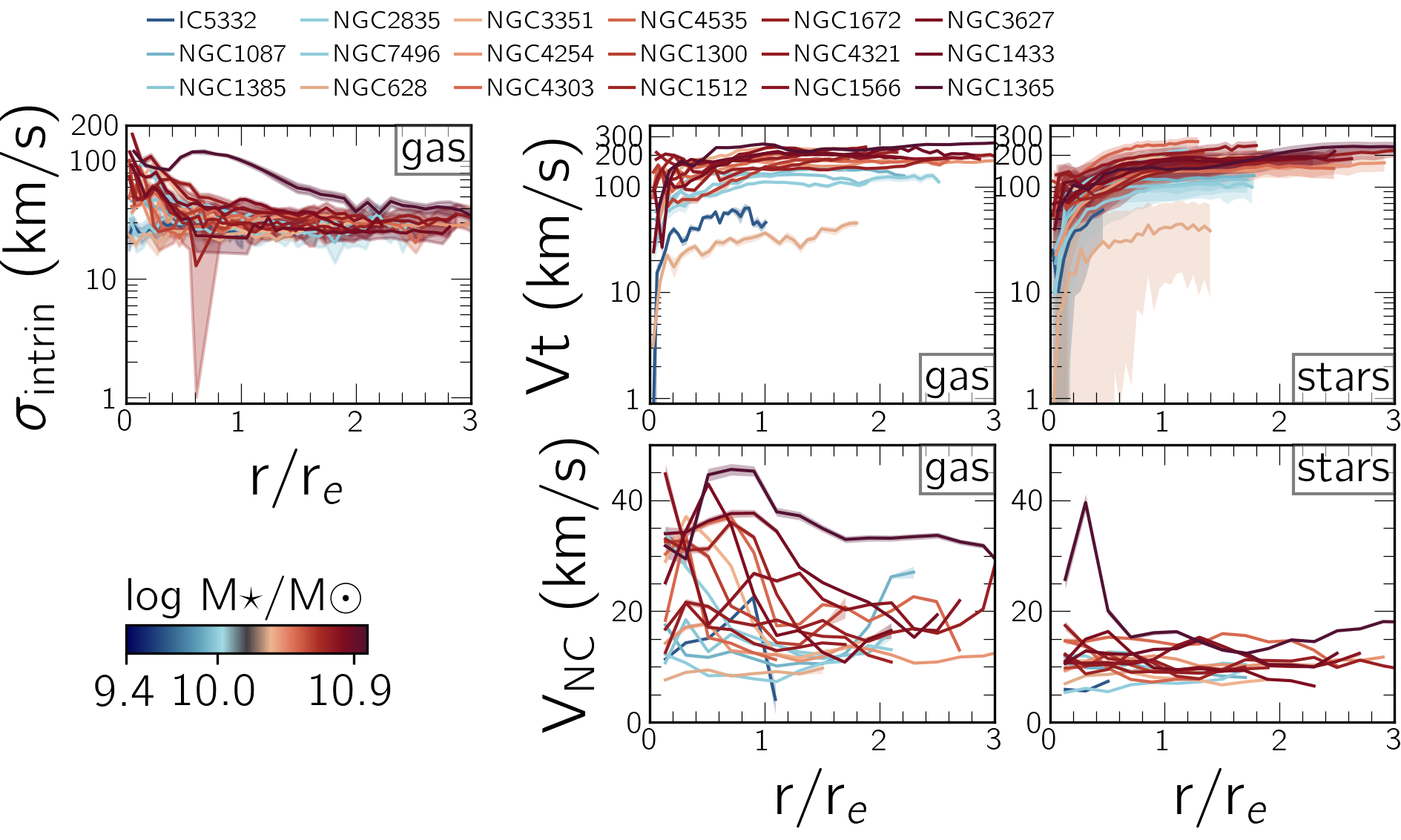}
\caption{\ha~intrinsic dispersion and rotation curves of the PHANGS-MUSE galaxies. The color of the lines indicates the stellar mass of each galaxy, with reddish tones representing the most massive and bluish tones the least massive objects.
{\it Bottom panel:} Azimuthally averaged radial profiles of noncircular motions, as defined in Eq.~\ref{Eq:residuals}. The stellar rotation curves were derived from the stellar velocity maps.
}
\label{fig:rotcurve_gas}
\end{figure*}

We proceed to model the kinematics of the \ha~line, since it is considered a good tracer of SF, and therefore a good tracer of the structure of disk galaxies. The required inputs for the pipeline are initial guess values for the disk geometry, which we took from the S4G catalogue \citep[e.g.,][]{S4G_release},  and the maximum ring radius (see Appendix~\ref{App:light_moms}).

\begin{figure*}[t!]
\centering
\includegraphics[width=0.32\textwidth]{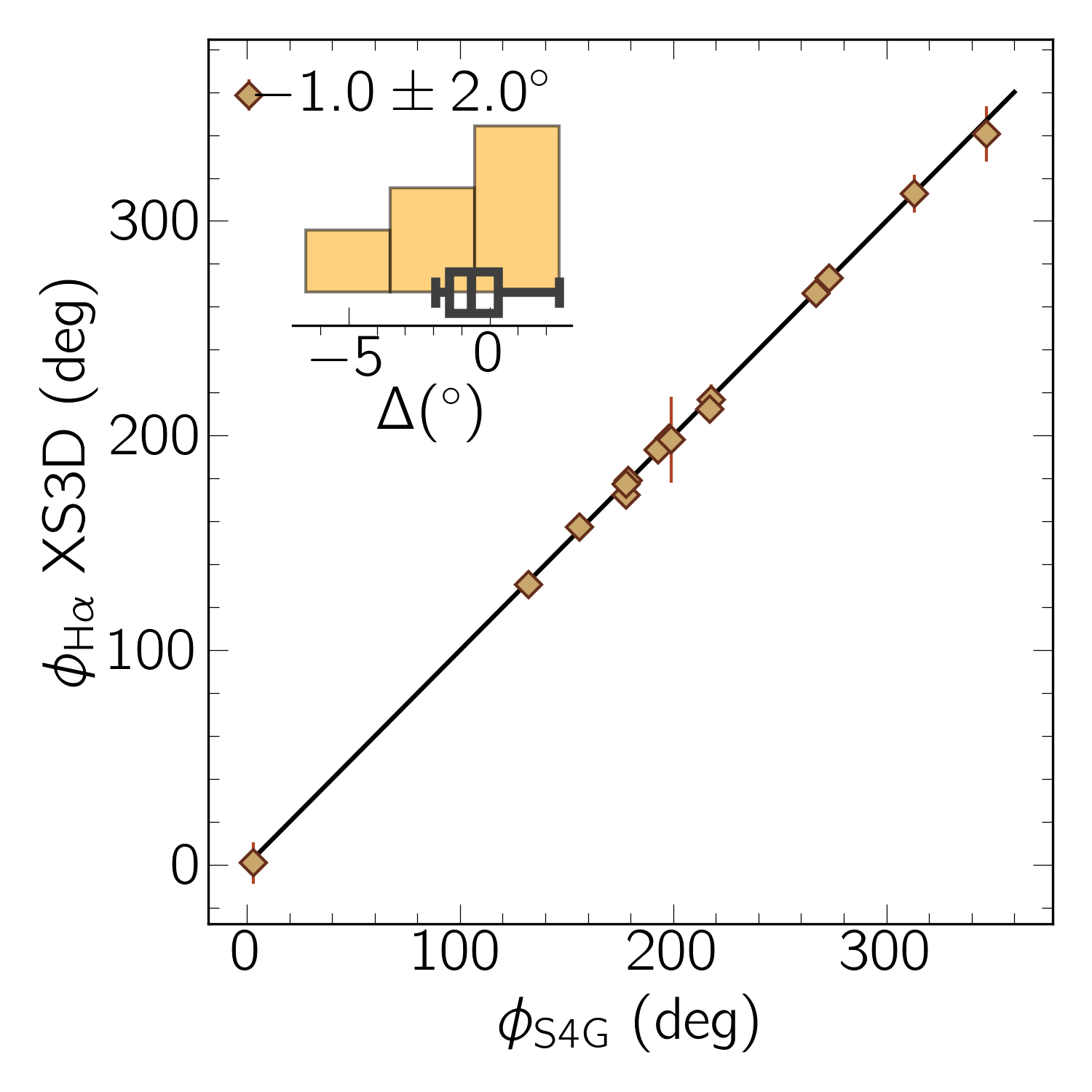}
\includegraphics[width=0.32\textwidth]{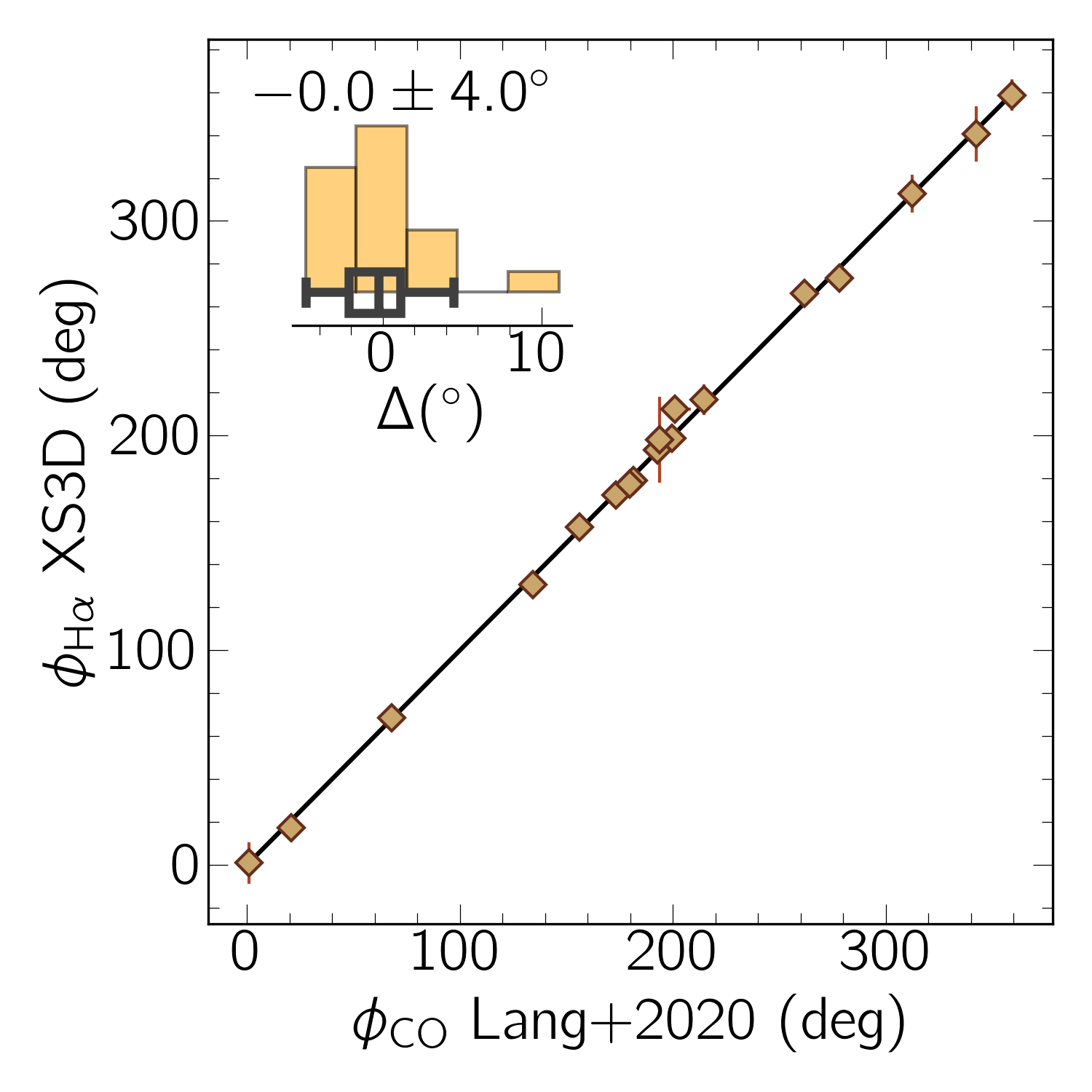}
\includegraphics[width=0.32\textwidth]{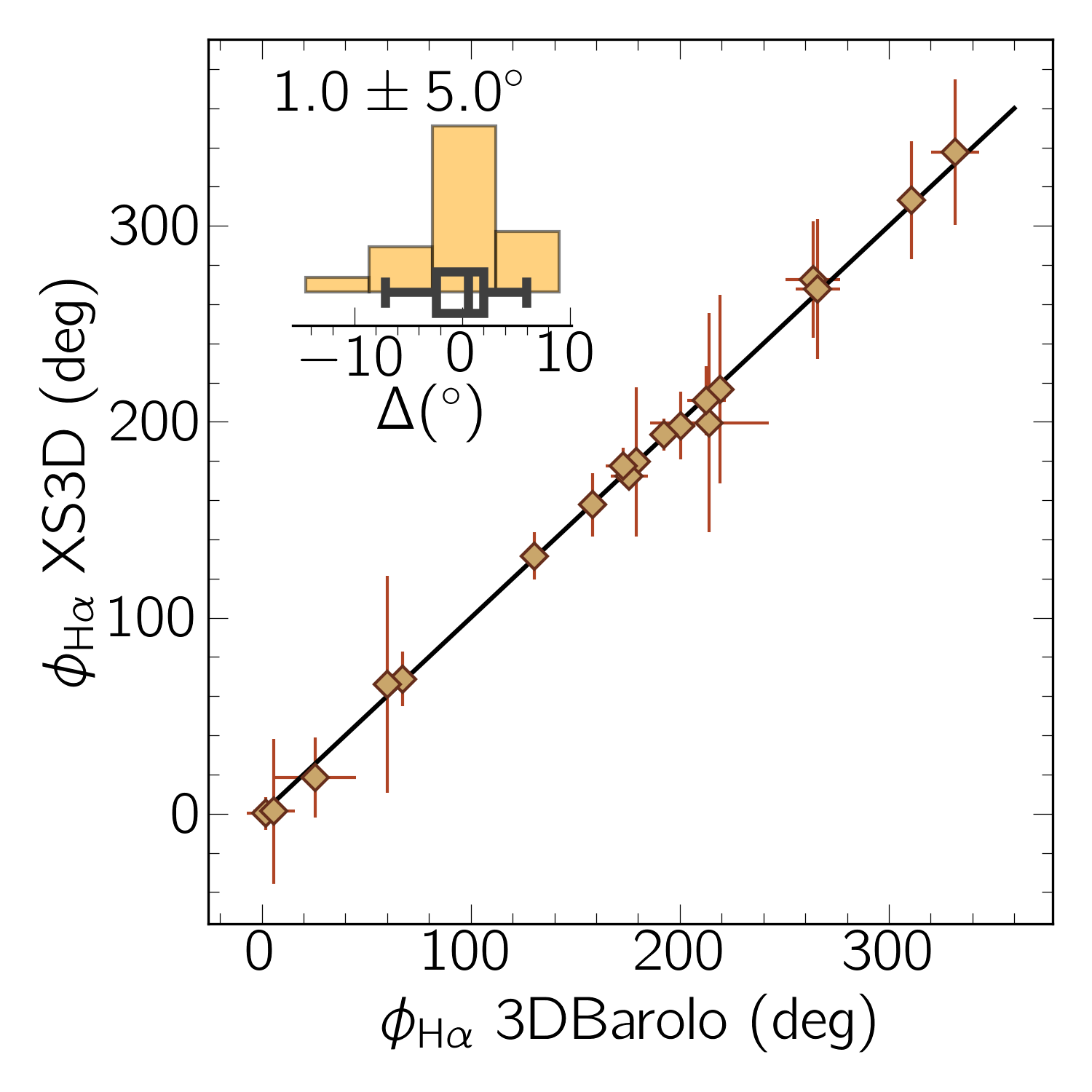}
\includegraphics[width=0.32\textwidth]{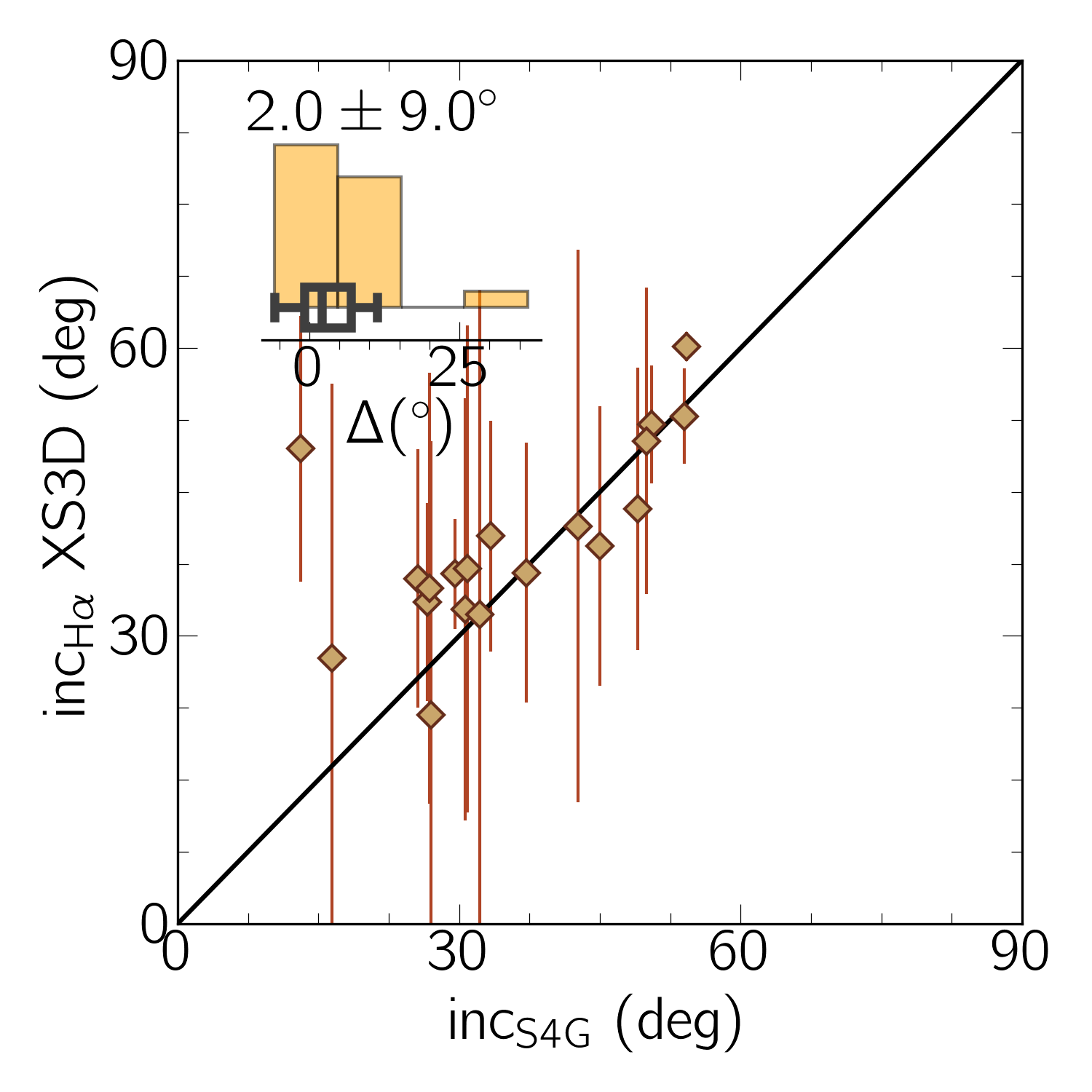}
\includegraphics[width=0.32\textwidth]{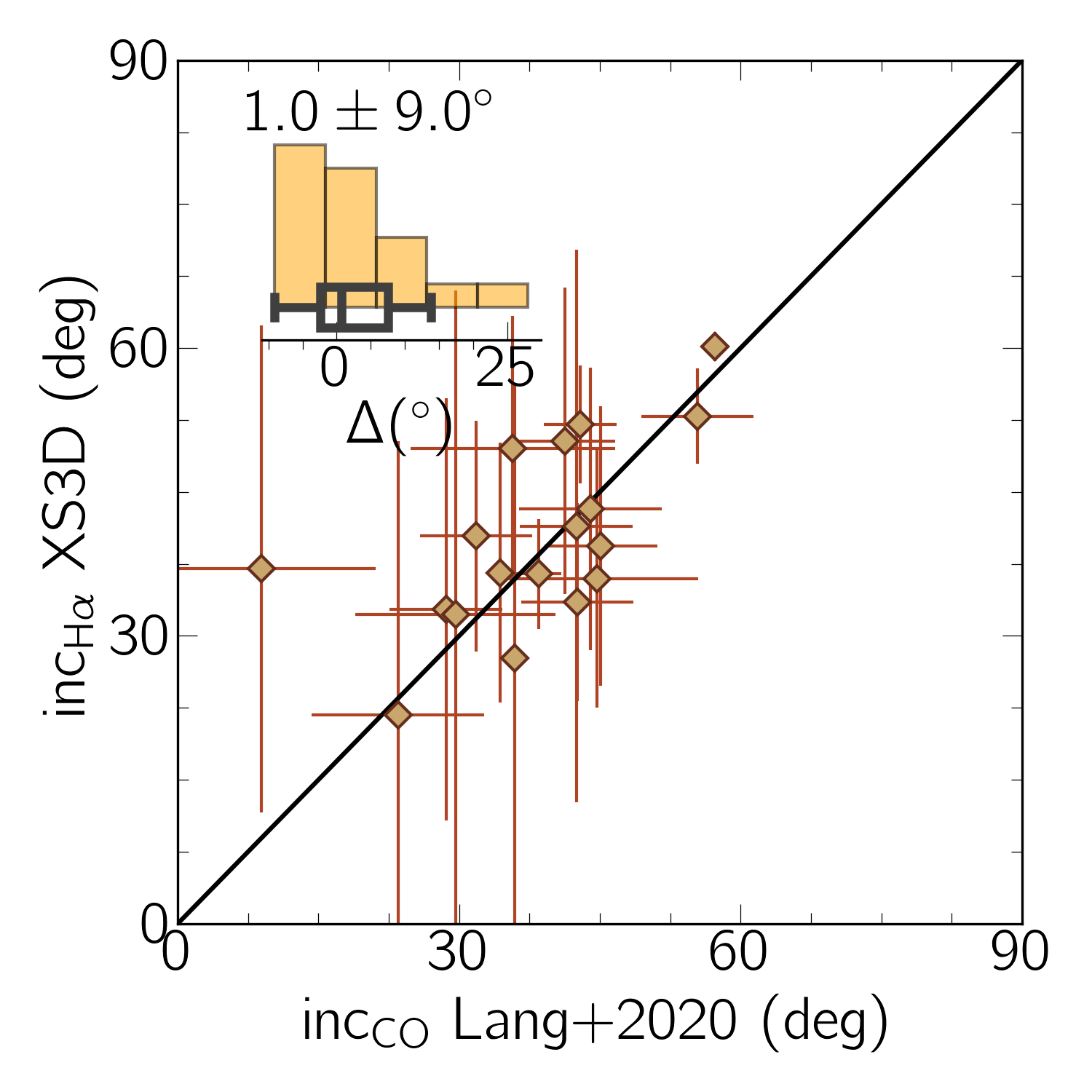}
\includegraphics[width=0.32\textwidth]{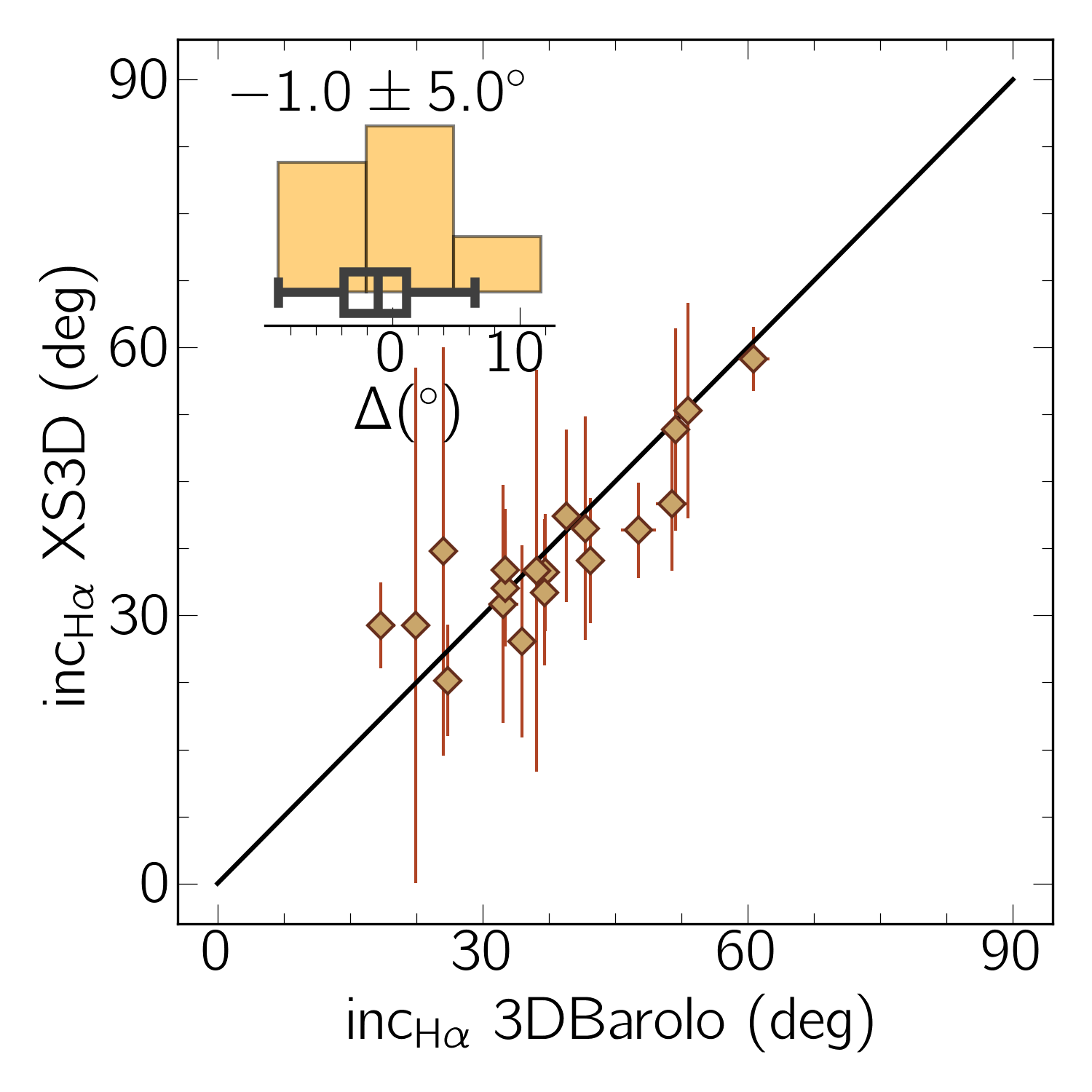}
\caption{ {\tt XS3D} results for the \ha~disk geometries for the PHANGS-MUSE objects compared to other works. Top figures from left to right show the position angle $\phi$ derived from {\tt XS3D} on the $y$-axis compared to  S4G \citep[e.g.,][]{S4G_release},  molecular gas CO from PHANGS-ALMA \citep[e.g.,][their Table~(2)]{Lang2020},  and from {\tt 3DBarolo} (see Appendix~\ref{App:3Dbarolo}), respectively. Bottom figures show similar comparison but for the disk inclination angle, with {\tt XS3D} results on the $y$-axis. The straight line represents the 1:1 relation while box plots of the difference $\Delta=y-x$ are shown in the upper-left corner of each panel, with the mean and standard deviation indicated.}
\label{fig:orientations}
\end{figure*}

The spectral resolution of MUSE is wavelength dependent \citep[e.g.,][]{Bacon2017,Emsellem2022}; at \ha\,($\lambda_\mathrm{rest}$ = 6562.68~\AA) the FWHM is $\sim 2.54$\,\AA, equivalent to $\sigma_\mathrm{instr}\sim49\kms$, which is comparable to the channel width of $57 \kms$.
At the average redshift of the sample $z\sim 0.005$, the MUSE LSF varies in less than $1\kms$ within a window of $500\kms$ around \ha. Thus, we consider it as constant.

The PSF spatial resolution of the PHANGS-MUSE objects was obtained directly from the measurements on the $R$-band (6483.58 \AA) of the mosaicked cubes from \citet{Emsellem2022} (i.e., their Table~(1)). Following these authors, we adopted the Gaussian PSF of the homogenized mosaics, which range from 0\farcs{56}--1\farcs{25} for the sample.

Results from {\tt XS3D} include: (i) the moment maps extracted from the observed and model cubes; (ii) the velocity profile of the circular rotation and intrinsic velocity dispersion; (iii) position velocity diagrams (PV) along the kinematic major and minor axes;  (iv) channel maps of the observed and model cubes; (v) channel maps for the residual cube ($I_\mathrm{obs}-I_\mathrm{model}$).
The first set of figures are shown in Figure~\ref{fig:moments} for NGC\,3351, while the remaining figures for this and all the PHANGS-MUSE objects are moved to the Appendix~\ref{App:xs3d_outputs} for a smooth reading.

NGC\,3351 was chosen here given the complex structure it presents:  (1) the \ha~map ($M_0$) shows many clumpy structures of ionized gas across the disk; (ii) it shows an inhomogeneous disk with a ring-like structure; (iii) patches are present across the entire disk as a result of line-less regions, or with poor signal to noise; (iv) it shows a complex velocity field ($M_1$) at local and large scales; (v) it exhibits a complex, yet axisymmetric, LOS velocity dispersion ($M_2$).

As observed in Figure~\ref{fig:moments}, the axisymmetric rotation model reproduces the large scale rotation of the disk. The residual velocity map on the right shows redshifted (positive) and blueshited (negative) values respect the systemic velocity, highlighting regions that deviate from the local bulk motion, which is dominated by circular rotation. The model dispersion map, $M_{2,\mathrm{mdl}}$, derived from the second moment of the $I_\mathrm{model}$ cube, reproduces the observed LOS velocity dispersion map,  which is affected by the MUSE LSF and PSF.

The rotational curves and dispersion profiles for the PHANGS-MUSE objects are presented in Figure~\ref{fig:rotcurve_gas} (see also  Appendix~\ref{App:xs3d_outputs} for individual cases), while the disk projection angles derived from the kinematic analysis are shown in Table~\ref{Tab:main_props}. The stellar rotation curves were modeled with the stellar velocity maps from the SSP analysis adopting \XS.
In the former figure, and for a better comparison, we normalized the galactocentric distance by the effective radius $r_e$, obtained from \cite{Leroy2021a}.

\subsection{\ha~disk geometries}
In order to asses the performance of {\tt XS3D} on PHANGS-MUSE, we first compare our results for the disk geometry with other works. Uncertainties in the disk geometry are one of the main sources of errors on rotational curves and therefore in noncircular motions.
Previous studies on PHANGS-ALMA \citep[][]{Leroy2021a}, have provided detailed analyses to estimate the cold disk orientations of 67 nearby galaxies, all of which are included in the PHANGS-MUSE sample, based on 2D modeling of CO velocity maps \citep[e.g.,][]{Lang2020}. In Figure \ref{fig:orientations}, we compare our results with those from \citet{Lang2020}. Additionally, we compare our results with those from S4G and include results from other 3D fitting algorithms, specifically {\tt 3DBarolo} \citep[e.g.,][]{BBarolo} (see Appendix~\ref{App:3Dbarolo} for details).

For the position angle $\phi$, we find good agreement between the {\tt XS3D} values and those from other studies. In all cases we find values lying along the 1:1 relation, exhibiting a mean on the differences about $0^{\circ}$ with a scatter about $\lesssim 5^{\circ}$.
The inclination angles from {\tt XS3D} are in good agreement with those from S4G, with an offset of $2^{\circ}$ on average respect the 1:1 line, and a dispersion of $9^{\circ}$. Similar results are obtained when comparing with their molecular gas counterparts.
Finally, inclination angles from {\tt 3DBarolo} and {\tt XS3D} follow the 1:1 relation with a scatter of $5^{\circ}$ around this line.

\subsection{Noncircular motions}

In this work we are interested in measuring deviations from circular rotation. Hence, we focus on the 1st moment residuals from Figure~\ref{fig:moments}.

We define noncircular motions as the absolute value of the observed and model velocities:
\begin{equation}
 \label{Eq:residuals}
 V_\mathrm{NC}  =  { |\Delta V|/ \sin i }, \mathrm{~~ with~}\Delta V = M_\mathrm{1,obs} -   M_\mathrm{1,model},
\end{equation}
where $M_1$ is the first moment map extracted from the observed and model cubes, respectively. $\Delta V$ represents the residual velocities from an axisymmetric rotating disk.
This map is commonly used to identify noncircular motions in disks, ranging from kpc scales to au scales \citep[e.g.,][]{Walsh2017}. Here, we correct $\Delta V$ by the disk inclination to measure the deprojected amplitude of those noncircular flows induced by nonaxisymmetric structures on the disk plane. Other corrections to $\Delta V$ require the adoption of specific models for the gas rotation.
From now on, we will use the terms ``amplitude" and ``strength" interchangeably to refer to the magnitude of the residual velocities as described in Equation \ref{Eq:residuals}.

In Figure~\ref{fig:rotcurve_gas}, we also show the azimuthally averaged radial profiles of $V_\mathrm{NC}$ for both the ionized gas and stellar velocities, using the disk geometry parameters listed in Table~\ref{Tab:main_props}.
These velocities are relatively low, $\lesssim 30 \kms$, as expected for rotation-supported disks. However, the azimuthal averaging of noncircular motions tends to give more weight to the smaller amplitudes \citep[e.g.,][]{Erroz-Ferrer2015}, as these typically dominate in the number of spaxels across all rings. Consequently, large NC amplitudes contribute less to the average unless a particular ring is dominated by NC motions, as is the case in NGC\,1365.
It is noteworthy that both stars and gas show similar residual radial profiles, indicating that rotational curves are similar in shape, although with different amplitudes due to the known  rotational lag of stars relative to the gas \citep[e.g.,][]{1946ApJ...104...12S,Yung-Chau2022}.

\begin{figure*}[t!]
\centering
\includegraphics[width=0.9\textwidth,keepaspectratio]{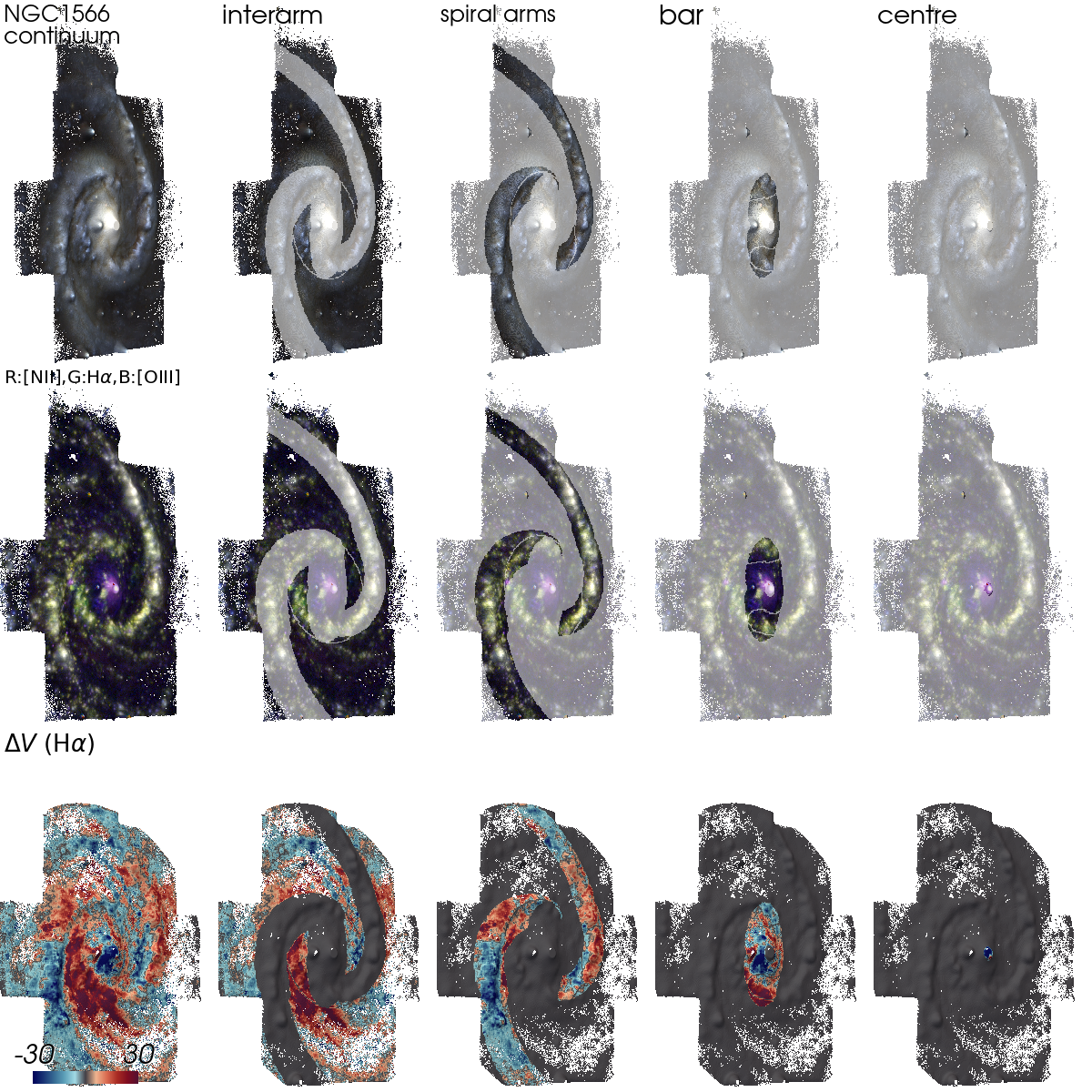}
\caption{3D visualization and environmental masking applied to various physical properties. From top to bottom, each panel shows the continuum, ionized gas and residual velocity maps, respectively. From left to right, different environmental masks are applied, where masked (i.e., blanked) regions are { shown by the desaturated areas}, and the unmasked region is indicated at the top. The first column represents the original map without masking.
The relief in each panel represents the intensity in the MUSE $r$-band. The RGB continuum image was constructed from the MUSE $gri$ bands. The RGB emission-line image shows the \nii, \ha, and \oiii~fluxes respectively. The residual velocity map shows the noncircular motions, in $\kms$, after decoupling the circular rotation.}
\label{fig:spatial_ref}
\end{figure*}
\section{Analysis}
The high angular resolution of the dataset allow us the exploration of the co-spatial dependence of the physical properties of the warm ISM, and their underlying dynamics through the gas in NC rotation. Such association between gas kinematics and their physical properties have been observed in different contexts \citep[e.g.,][]{Ho2014, Moiseev2015, AMUSING++, Sebastian_ARAA, Groves2023,2024arXiv241021147L, 2024arXiv241020583K}, which suggest that both physical properties are closely related.

\subsection{Masking procedure}\label{sec:masking}

In the following analysis we make use of the environmental masks created by \cite{Querejeta2021} based on {\it Spitzer} $3.6\mu$m images. These masks separate different structural components: spiral arms, interarm, bar, centre, { and disk (i.e., a disk without a spiral mask). The disk mask also includes photometric decomposition of near-infrared images from  \citet{S4G_release}.}
For this purpose we aligned the world-coordinate system (WCS) of both datasets and regridded the masks to the MUSE pixel size using the CASA software \citep[e.g.,][]{casa}.

Based on these environmental masks, we created one mask for each mosaicked map. These masks will subsequently be used to investigate the physical properties and noncircular motions in each environment.
The environmental masks applied to three different physical properties in NGC\,1566 are shown in Figure~\ref{fig:spatial_ref}.

\subsection{Spatial visualization of noncircular motions}

The residual maps in Figure~\ref{fig:moments} and those in Appendix~\ref{App:xs3d_outputs}, revealed that noncircular motions are present across the entire extension of optical disks; however, these maps do not provide clues that allow to associate the noncircular motions with the structural components of galaxies.
With this in mind, we proceed to transfer spatial information to each 2D map, to facilitate the identification of different noncircular driving sources. This was also motivated by the high angular resolution of the data.

We implement topographic visualization techniques through the visualization toolkit \citep[{\tt VTK},][]{vtkBook, sullivan2019pyvista}.
Here, the $X$ and $Y$ coordinates represent the sky coordinates of a given map, while the $Z$ coordinate transfer altitude by mapping some physical property of galaxies. Here we choose the continuum flux in the $r-$band to transfer relief to the maps.
Since in general the brightest structural components in disk galaxies are the spiral arms, bulge, and stellar bars, all physical properties are highlighted with respect to these nonaxisymmetric structures. We emphasize that this technique does not add a quantitative value to the maps, but rather adds qualitative spatial reference that can be important to unveil certain type of nonaxisymmetric motions \citep[see for instance,][]{2017MNRAS.465.3446G}. The above technique is already displayed in all the continuum, ionized-gas, and residual velocity maps in Figure~\ref{fig:spatial_ref}.

\subsection{Ionized gas properties}

The following analysis is to unveil possible correlations between underlying NC motions and the physical properties of the ionized gas.

\begin{figure*}[t!]
\centering
\includegraphics[width=1\textwidth,keepaspectratio]{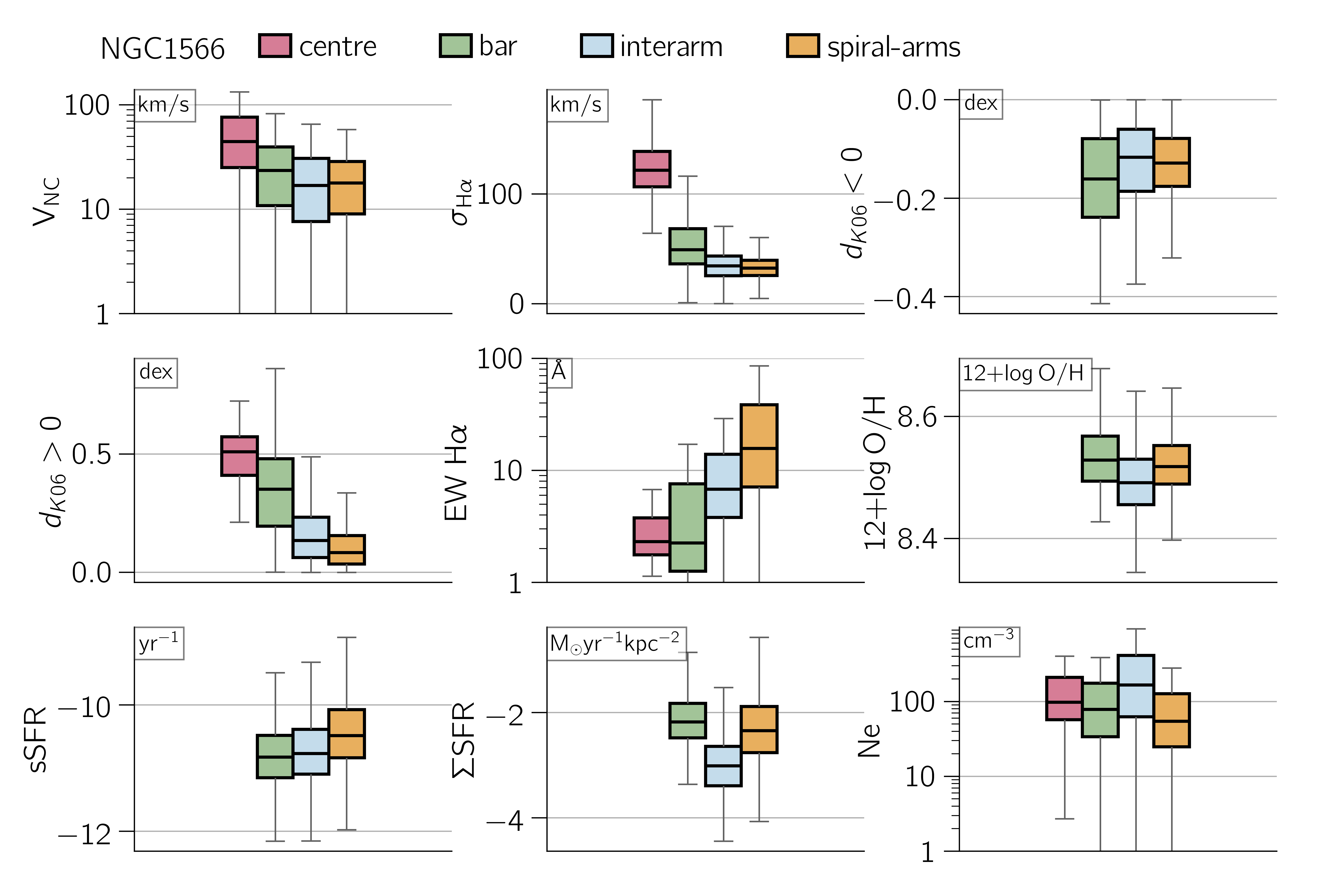}
\caption{Physical properties explored, segregated by different environments in NGC\,1566. Each panel shows the distribution of a given physical property across different galactic environments: centre, bar, interarm region, and spiral arms. From left to right and from top to bottom: Deprojected noncircular amplitude, LSF-corrected \ha~LOS dispersion, negative perpendicular distance to the K01 curve (SF-ionization), positive distance to the K01 curve (AGN-like ionization), equivalent width of \ha, specific star formation rate, star formation surface density, electron density.
Each box plot represents the distribution of spaxels within each environment. If a box plot for a given environment is not shown, it indicates that either there are no spaxels within that mask or the spaxels do not satisfy the required condition. Units are shown in the upper left of each panel.}
\label{fig:rad_profs}
\end{figure*}

\begin{enumerate}[leftmargin=*]

 \item Ionization source. The \nii/\ha~vs. \oiii/\hb\,line ratios \citep[BPT,][]{Baldwin1981}, have been extensively used to classify the ionizing source of objects at both kpc and sub-kpc scales, while the \cite{Kewley2001} curve (K01) drawn in this diagram has been commonly adopted a as reference to separate sources exhibiting spectra typically of \hii~region and those showing a harder ionization like AGNs.
 The location of a source in this diagram, specifically the perpendicular distance to the K01 curve ($ d_\mathrm{K01}$), with positive values indicating AGN-like ionization and negative indicating SF-like ionization, can be used as a proxy for the hardness of the ionization source \citep[for instance,][]{2018ApJ...861...50B}, even though its exact location in this diagram depends on the electron density, gas phase metallicity among other parameters.
We adopt this distance to map the ionization state of the gas in NC rotation. ${d_\mathrm{K01}}$  is obtained from minimizing the expression
 \begin{equation}
  d_\mathrm{K01}[\mathrm{dex}] = \sqrt{(x-x_i)^2 + (y_\mathrm{K01}(x)-y_i)^2},
 \end{equation}
 where $x$ is obtained for each spaxel from solving the equation
 \begin{equation*}
   \big(y_\mathrm{K01}(x)-y_i\big)y_\mathrm{K01}^{\prime}(x) + (x-x_i)  = 0.
 \end{equation*}
In these two equations, the pair $(x_i, y_i)$ represents the $\log$(\nii/\ha)~and $\log$(\oiii/\hb) ratios at a given location in the mosaicked images. The term  $y_\mathrm{K01} = y_\mathrm{K01}(x)$ denotes the $\log$(\oiii/\hb)~value from the K01 curve at a given $x$, while $y_\mathrm{K01}^{\prime}$ is the derivative of the K01 curve. The $d_\mathrm{K01}$ values for the central spaxels of each galaxy are listed in Table~\ref{Tab:main_props}.
 \item Equivalent width of \ha~EW(\ha).
 Together with the BPT diagram, the EW(\ha)~has been used to classify the emission line spectra of galaxies \cite[e.g.,][]{Stasinska2008}, as well as in spatially resolved regions of galaxies \citep[e.g.,][]{Levy2019, Sebastian_ARAA, Lacerda2020, Kalinova2021, Belfiore2022}. Multiple demarcations have been proposed according to the magnitude of this parameter \citep[e.g.,][]{Cid2011}, with general agreement that EW(\ha) values below 3\,\AA\ are associated with evolved stars (post-AGB), while larger values may indicate either star formation processes (EW(\ha) $>$ 3\,\AA) or an AGN (EW(\ha) $>$ 10\,\AA). We use the EW(\ha) maps from the {\tt pyPipe3D} products. The EW(\ha)  for the central spaxels of each galaxy are shown in Table~\ref{Tab:main_props}.

\item \ha~velocity dispersion. The ionized gas velocity dispersion is a measure of the turbulence of the warm ISM, and it has been related to the physical properties of the ionized gas \citep[e.g.,][]{WHaD, AMUSING++}. Here, we use the second-moment, LSF-corrected map derived from the observed cubes to account for non-axisymmetric features in the velocity dispersion and to measure the local gas dispersion.
\item Gas phase metallicity (12 + $\log$ (O/H)). For computing this parameter we have adopted the \cite{Pilyugin2016} calibrator based on strong emission lines, namely, \nii6548,6584, \sii6717,6731, \oiii5007,4959, \ha~and~\hb. All these lines are accessible within the spectral range covered by MUSE. We did not impose additional criteria on the emission lines other than $\sn>1$ and negative $d\mathrm{_\mathrm{K01}}$ values.
\item Specific star formation rate (sSFR). The star formation rate (SFR) was computed from the \ha~luminosity, adopting the distances reported in Table~\ref{Tab:main_props}. We correct the \ha~flux from dust attenuation adopting the \cite{Cardelli1989} extinction curve with $R_V = 3.1$, and case B of recombination \citep[e.g.,][]{osterbrock89}. Then we use the \cite{Kennicutt1998} relation to convert the \ha~luminosity to SFR. The latter map was transformed to surface density $\Sigma$SFR, by dividing it by the physical size of the pixel area.
Finally, the specific SFR  is computed from the ratio between the star formation surface density and the stellar mass surface density obtained from the SSP analysis, sSFR = $\mathrm{\Sigma SFR/\Sigma M_\star}$. We correct the surface density quantities by multiplying by the cosine of the inclination derived from the kinematic analysis. The SFR is computed only for spaxels where $d{_\mathrm{K01}}<0$.
\item Electron density ($\mathrm{N_e}$). Some NC motions may result from violent interactions of gas flows, leading to changes in the density distribution of the ISM.
We adopt the parameterization from \cite{Sanders2016} derived for a 5-level atom approximation of the \sii~ion, given by $\mathrm{ N_e(R_{[SII]})} = (c\mathrm{R_{[SII]}}-ab) / (a-\mathrm{R_{[SII]}})$, with $\mathrm{R_{[SII]}}=$\sii6717/\sii6731, $a=0.4315,\, b = 2 107,\, c = 627.1$, valid for $0.4375 <\mathrm{R_{[SII]}}<1.4484$ and $T_e=10^4$\,K. A cut of $\sn > 1$ was applied to both emission lines.
%
\end{enumerate}
The masking procedure applied to these properties for one object is shown in Fig.~\ref{fig:rad_profs}. Each column in the figure shows the resulting spaxel distribution for the considered property after applying the environmental masks.

\begin{figure*}[t!]
  \centering
     \includegraphics[width=\textwidth,keepaspectratio]{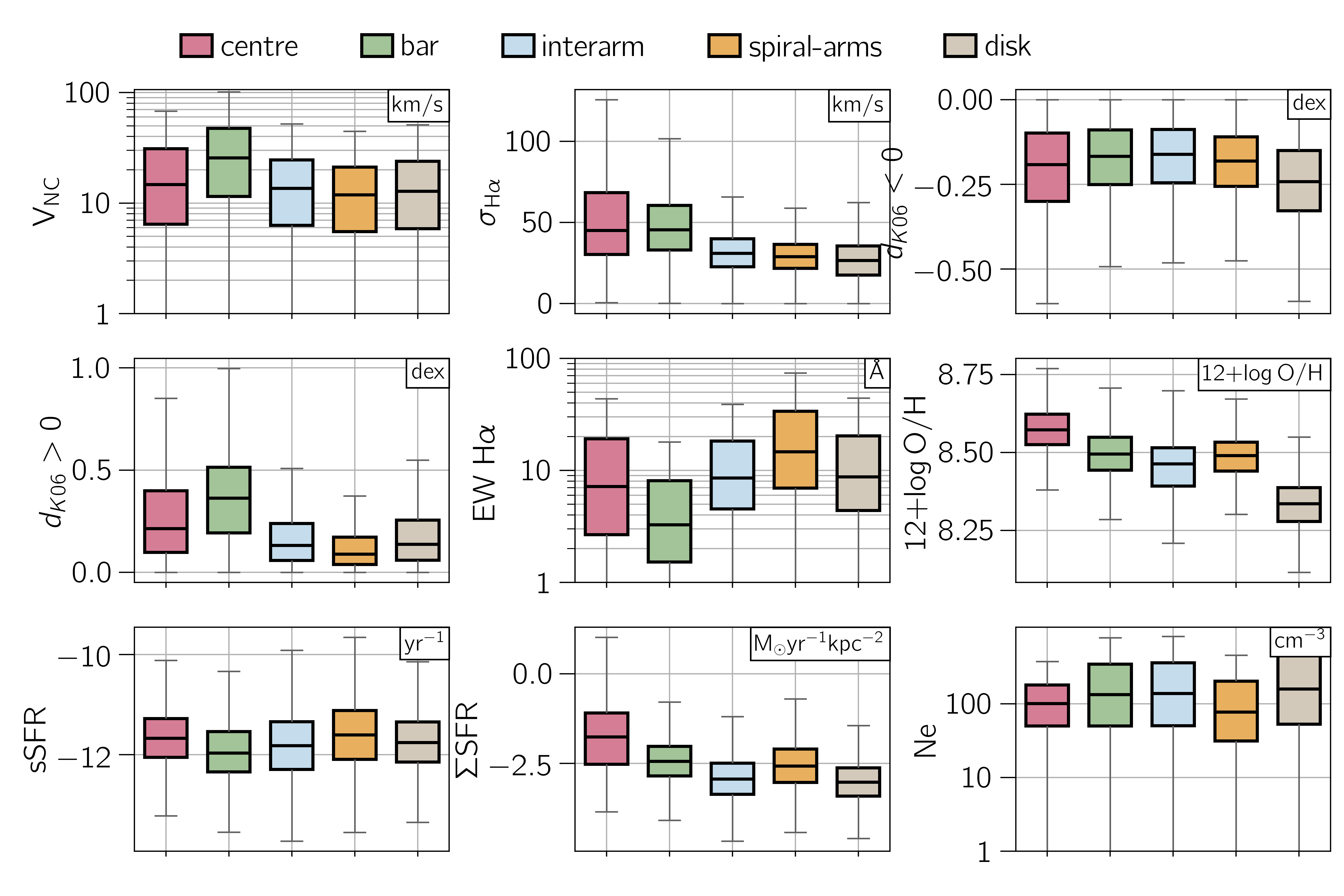}
  \caption{Similar plots as in Figure~\ref{fig:rad_profs}, but this time including the 19 objects from PHANGS-MUSE. Each panel comprises $\sim2.6\times10^6$ spaxels spanning the 5 different environments.}
  \label{fig:boxplot_all}
\end{figure*}

\section{Results}

\subsection{Ionized gas intrinsic dispersion}
The intrinsic dispersion profiles from Figure~\ref{fig:rotcurve_gas} exhibit a relatively flat behavior around $30\kms$ across most of the optical extension, with a { noticeable increase in the inner $0.6r_e$.  Since beam-smearing effects have been already accounted for in the 3D modeling, the central enhancement in the velocity dispersion cannot be attributed to instrumental effects.
In all cases, the intrinsic dispersion values exceed the thermal broadening of the Hydrogen at $T=10^4\mathrm{K}$ ($\sigma_\mathrm{therm}=9.08\kms$), but are comparable to the  dispersion observed in SF regions \citep[e.g.,][]{Reynolds1985,Moiseev2015,Law2022}. Therefore, an additional contribution due to nonthermal motions, such as gravitational effects or turbulence, might be responsible for the large dispersion values \citep[e.g.,][]{Weijmans2008}.}

An intrinsic dispersions of $30\kms$ corresponds to LOS dispersions of approximately $58\kms$, which is consistent with the values observed in the second-moment maps shown in Figures~\ref{fig:moments} and \ref{App:xs3d_outputs}.
Such dispersion and rotational curve profiles imply rotational to dispersion ratios, $\langle V_t/\sigma_\mathrm{intrin} \rangle$, of approximately 5, typical of  rotation supported systems.

The observed LOS velocity dispersions, along with the intrinsic dispersion values, are in agreement with those recently reported in PHANGS-MUSE \citep[e.g.,][]{Emsellem2022,Groves2023}.

\subsection{Relationship between physical properties and environment}

\begin{figure*}[t!]
  \centering
     \includegraphics[width=\columnwidth,keepaspectratio]{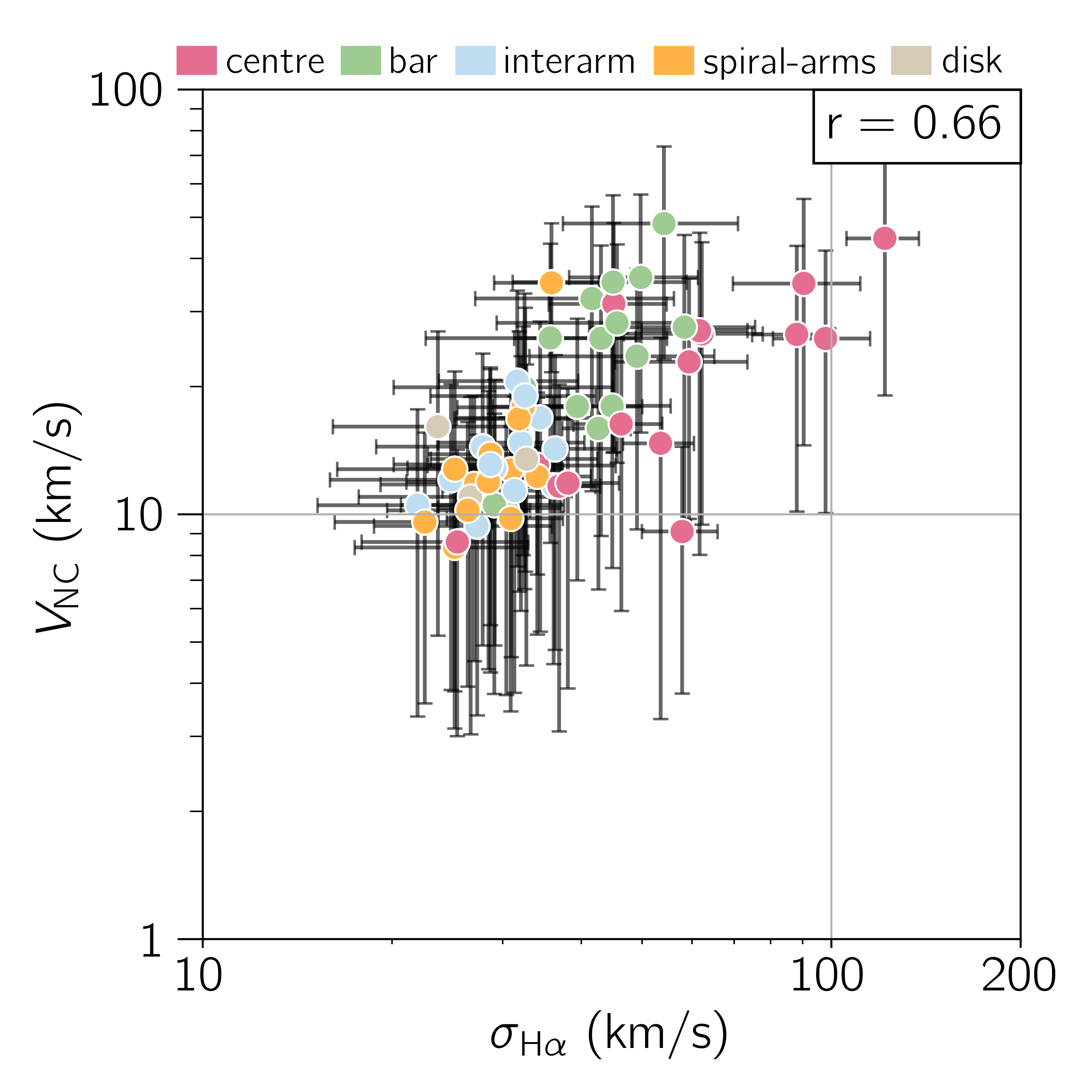}
     \includegraphics[width=\columnwidth,keepaspectratio]{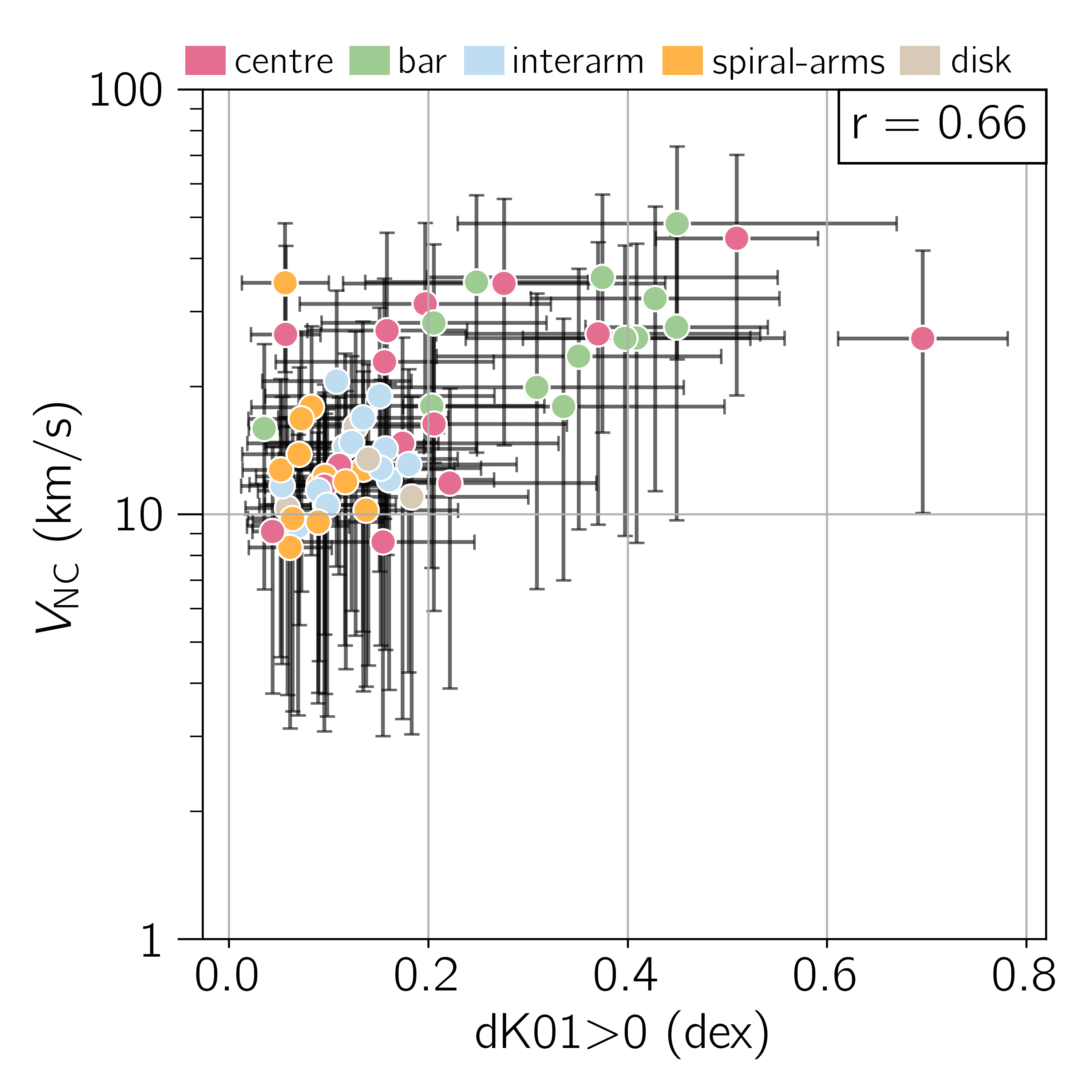}
  \caption{ Relationship between the amplitude of non-circular (NC) motions and ionized gas properties across different galactic environments in the PHANGS-MUSE sample. Left: Line-of-sight (LOS) LSF-corrected \ha~velocity dispersion. Right: Positive offset from the \cite{Kewley2001} demarcation, indicating AGN-like ionization. Each point represents the median value within each environment, and error bars indicate the standard deviation within each environment. The relationship is characterized by the Pearson correlation coefficient shown in the upper right corner of each panel.  }
  \label{fig:box}
\end{figure*}

Figure~\ref{fig:rad_profs} presents the results of the environmental masking procedure applied to the showcase example shown in Figure~\ref{fig:spatial_ref}.
Based on the median of the distributions, we note that some physical properties exhibit a { progressively decrease, with the highest values in the center, followed by successively lower values toward the bar, interarm, spiral arm, and disk regions. This trend is mostly observed in the amplitude of noncircular motions, velocity dispersion, AGN-like ionization. In contrast, properties such as the sSFR and the EW(\ha) show the opposite behavior, while the electron density and gas metallicity do not exhibit clear trends (see these trends for all objects in Appendix~\ref{App:boxplot_all}). These trends, in particular the ones referring to the amplitude of NC motions and dispersion are observed in about 70\% of the sample, while in the remaining fraction there seems to be local variations, such as in NGC\,4303.
Although different environments may overlap at similar galactocentric distances, the trend observed in Figure~\ref{fig:rad_profs} closely resembles that reported in studies of radial profiles of physical properties, where inside-out growth is commonly identified \citep[e.g.,][]{Barrera-Ballesteros2023,Groves2023}. }

{ In NGC\,1566}, there does not seem to be a particular trend in the case of spaxels dominated by SF. However, those associated with an AGN-like ionization (i.e., where $d_{K06}>0$) exhibit harder ionization in the central region, followed by the bar-region probably induced by shocks, and the interarm region and spiral arms. Such behavior in the ionization source is consistent with the trend observed in the EW(\ha), where the lowest values $\sim3$\AA, are located in the centre probably induced by post-AGB stars or HOLMES, while the spiral arms show characteristic EWs of SF ionization.

In order to explore the global behavior of these properties across the sample, we staked all properties in the different environments as shown in Figure~\ref{fig:boxplot_all}. { Based on the median and the interquartile range  of the distributions}, the higher amplitudes of noncircular motions are located in the centre and bar, while the lower amplitudes in the { spiral arms, interarms, and disk}. However, even in the latter environments, noncircular velocities are noneligible, with amplitudes reaching the $20\kms$.
A similar case is observed in the velocity dispersion, where the lowest values are located in the spiral arms, interarms, and disk.
AGN-like ratios ($d_{K06}>0$) are observed across all environments, including the spiral arms, with the central region and bar showing the hardest ionization. However, the presence of such line-rations (i.e., large \oiii/\hb~and \nii/\ha~ratios) in the interarm and spiral arm regions suggest that other sources different to AGN-ionization are  contributing significantly in those regions, such as low velocity shocks or hot low-mass evolved stars (HOLMES).

The amplitude of the NC motions and the LOS velocity dispersion are two quantities that describe the dynamical state of the gas. In Figure~\ref{fig:box} we plot both parameters including all environments and we observe a clear trend of increasing noncircular amplitudes with higher gas velocity dispersion. This relation shows a moderate correlation coefficient of 0.66.  While previous studies have suggested correlations between ionized gas properties and the velocity dispersion, \citep[e.g.,][]{Monreal2010, Ho2014, AMUSING++, Johnston2023, WHaD}, here we observe that these two dynamical gas properties are closely linked. A similar trend is observed for the strength of ionization, shown on the right side of the figure.

Although the SFR surface density shows a similar increase with the amplitude of noncircular motions, as also seen in Figure~\ref{fig:boxplot_all}, this trend disappears when normalized by the stellar mass.

\subsection{Source of NC motions in the PHANGS-MUSE objects}

The 3D kinematic modeling of the PHANGS-MUSE objects showed that NC motions are ubiquitous in the ionized gas phase of disk galaxies.
There are at least two major drivers of NC motions in galaxies: (i) those induced by local  processes, like star-formation related and AGN activity, and (ii) those related to global perturbations in the gravitational potential.

It is evident from the surface maps from Figure~\ref{fig:spatial_ref}, that nonaxisymmetric structures introduce perturbations in the velocity field.

Spiral flows are often observed in the range of $\sim 20\kms$ when deprojected \citep[e.g.,][]{vandeVen2010,Schmidt2016, DiTeodoro2021}. Evidence of NC motions likely induced by spiral arms can be observed in NGC\,1672, NGC\,3627, NGC\,4321 among others. Most of the ionized gas in the spiral arms show velocities ranging from $5-20\kms$ in Figure~\ref{fig:boxplot_all}. This range is also observed in the disk where \hii~regions are often observed \citep[e.g.,][]{sfs2015, Santoro2022A, Groves2023, Alejandra2024}.

Perturbations induced by stellar bars are notorious in most of the sample, including NGC\,1087, NGC\,1365, NGC\,3627, NGC\,4535. While bar-perturbations induce itself noncircular motions on gas and stars \citep[e.g.,][]{LopezCoba2022}, the ionized gas in those regions show a combination of SF and a harder ionization source probably induced by shocks occurring along or near the bar-dust lanes \citep[e.g.,][]{Athanassoula1992, Regan1999, Mundell1999, vandeVen2010,Schinnerer2023A}, or by gas inflows \citep[e.g.,][]{Kim2024}. However, searching for multiple kinematic components in the spectra is beyond the scope of this paper.

The central or nuclear region is the second morphological structure in Figure~\ref{fig:boxplot_all} with the largest noncircular motions, with amplitudes around $\sim 20\kms$. The observed range of EWs(\ha) observed $\sim$ 3--11\AA, along with the presence of spaxels both below and above the K01 curve, suggests that a combination of HOLMES and AGN ionization is partially responsible of the observed motions.
The latter EWs are also found in inter-arm regions, likely associated with a diffuse ionized gas component, or with low-ionization emission-line regions (LIERs) and HOLMES, as recently observed in the PHANGS-MUSE galaxies \citep[e.g.,][]{Belfiore2022}. The systematic deviations from circular rotation in the diffuse gas may indicate the presence of a distinct kinematic component relative to the bulk motion of the ionized gas.
Recently, \cite{denBrok2020} found that diffuse ionized gas exhibits slower rotation and a higher dispersion compared to gas compatible with SF. This is consistent with the increased velocity dispersion observed in Figure~\ref{fig:boxplot_all}.

In addition to nonaxisymmetric motions, vertical perturbations could be contributing to the large scale residuals observed in the residual velocity maps.
There are two cases where the largest noncircular motions, of approximately $30\kms$, are located at the outskirt of the optical disk, these are NGC\,3627 and NGC\,1365.
These amplitudes are consistent with vertical motions of gas and stars out of the mid-disk plane, as observed in hydrodynamical simulation \citep[e.g.,][]{2016MNRAS.456.2779G, 2017MNRAS.465.3446G}. As noted in our kinematic modeling, we do account for contributions to the LOS velocity of off-plane flows.

Although the 3D modelling technique presented here provides a more accurate description of the kinematics of gaseous disks, contributions from gas random motions, and imperfections in the modelling still cannot be interpreted as deviations from circular rotation.
To quantify these motions, we estimate the error in the emission-line centroids by randomly perturbing the spectra, using
the average negative fluxes in each spectrum as the dispersion value of a normal distribution. For each object, we performed 100 realizations: in 50 realizations, the centroids were estimated using $M_1$, while in the other 50, a second-order polynomial was fitted to the peak flux and its two adjacent fluxes. The standard deviation for the entire sample was found in  $\sigma^\mathrm{centroid}_\mathrm{H\alpha} \sim 3 \kms$, which is inferior to the noncircular motions observed in the disk and spiral arms.

\section{Discussion}

Noncircular motions seem to be related with other physical properties of gas such as velocity dispersion and strength of ionization. { Although the spread in Figure~\ref{fig:box} is large, the global relation (and global correlation coefficients) between the noncircular motions with the gas dispersion and ionization strength are fairly similar. Clearly, the centre and bar regions are the two locations where the largest velocity dispersion and hardest ionization are observed.
These two environments show individual correlations coefficients in the $V_\mathrm{NC}$ vs. $\sigma$ diagram of 0.81 and 0.63, respectively. Meanwhile the interarm, spiral arms, and disk, show individual correlation coefficients of 0.44, 0.67, and -0.45 respectively.

While a global correlation is observed, an environmental dependence is notable in the bar and centre; however, this is less clear in the interarm, spiral arms, and disk. These environments are affected by the low dynamical range in both the amplitude of NC motions and the velocity dispersion, as well as by the limited statistics in the disk environment. The $\sim116\kms$ velocity resolution of MUSE  might play a role in preventing the estimation of NC flows and dispersion velocities with low amplitudes.

The moderate correlation coefficient in the interarm may suggest a possible connection between the NC motions in the spiral arms and the gas dispersion. However, the amplitude of these motions cannot be distinguished from those observed in the interarm and disk, as also shown in Figure~\ref{fig:boxplot_all}. }

Oval potentials, such as those associated with bars and bulges, strongly influence the conditions of the ISM, with noncircular motions expected to arise as part of the streaming flows induced by these structures.
The overlapping of LOS velocities from gas moving at distinct velocities and orbits is a plausible explanation of the large dispersion observed in the bar and centre. Such multiple kinematic components within bars have been observed in the cold, quiescent molecular gas phase \citep[e.g.,][]{Schinnerer2023A}, typically observed at  much higher spectral resolutions, on the order of a few kilometer per second.

The ionization state of the gas in the bar and nucleus is primarily dominated by non-star-formation sources, although in some cases, proximity to the K01 demarcation suggests a possible contribution from star formation. { The significant offset from the K01 curve observed in the centre and bar environments, with individual correlation coefficients of 0.55 and 0.63 in the $V_\mathrm{NC}$ vs. $d_\mathrm{K01}$ diagram, implies that shocks may be substantially contributing to the measured line ratios.  In contrast, the interarm, spiral arms, and disk show lower correlations, around 0.33. Among these, the interarm show the largest distance from the K01 curve, $\sim0.2$ dex, towards the AGN-like region.}

A kinematic component with amplitudes around 3–5$\kms$ is likely associated with the bulk rotation of the galaxy. This component represents the axisymmetric rotation, distinguishing it from the streaming flows observed in { the spiral arms, disk, and interarm. As shown in Figure~\ref{fig:box}, these structures exhibit NC amplitudes on the order of 10--15$\kms$. While spiral-arm perturbations have been suggested to drive NC amplitudes of this magnitude \citep[e.g.,][]{Wong2004, vandeVen2010}, the fact that large-scale spiral-arm motions are not clearly distinguishable from those in the disk and interarm regions complicates their identification in observations, despite their prediction in hydrodynamical simulations. High-resolution datasets such as PHANGS-ALMA, which trace the quiescent gas phase with improved velocity resolution, may help reveal the existence of spiral-driven flows.
Other related motions, including gas inflows might be contributing to the spread in the observed amplitude in the spiral-arm, disk and interarm environments \cite[see e.g.,][]{Davies2009}.}


\subsection{Caveats: mass dependence on NC amplitudes}

It is expected that the noncircular amplitudes observed in massive galaxies are not reached in low-mass galaxies. Therefore, the amplitude of noncircular motions, as derived from the residual velocity maps, may depend on stellar mass \citep[e.g.,][]{2024arXiv241021147L}, particularly for motions associated with non-axisymmetric structures.
Stacking all velocities in different morphological structures as in Figure~\ref{fig:boxplot_all} suppose there is no mass dependence on residual velocities. The PHANGS-MUSE galaxies however, do not provide a wide range of stellar masses, nor a large statistical sample that allow to unveil a mass dependence on $V_\mathrm{NC}$, if any.

On the other hand, in nonaxisymmetric structures, like bars, the viewing angle plays a role in the observed amplitude of noncircular motions. It is known that bar-motions are observed larger, when the bar departs from the disk major axis \citep[e.g.,][]{Holmes2015, LopezCoba2022}.

\section{Summary and conclusion}
In this work we addressed the physical properties of gas departing from circular rotation by taking advantage of the exquisite spatial resolution provided by the PHANGS-MUSE galaxies.
For this analysis, we built upon our previous kinematic packages for extending the analysis to spectral-line cube observations, without limiting to wavelength type observations such as MUSE/VLT. This tool named {\tt XS3D}, allowed us to construct cube observations of the \ha~spectral line to study the kinematics of the PHANGS-MUSE galaxies.

We find that the PHANGS-MUSE galaxies are characterized by a roughly constant intrinsic velocity dispersion of about $30\kms$ on average, except in regions dominated by strong noncircular motions. These values result in circular to dispersion ratios $\langle V_t/\sigma_\mathrm{intrin} \rangle$  greater than 5 for most of the sample. Exceptions include  NGC\,628, IC\,5332, and NGC\,5068 where low inclination angles, combined with the $116\kms$ resolution of MUSE, prevent a reliable estimation of  $V_t(r)$.

We measured noncircular motions in these objects through deprojected residual maps from the first moment maps of an observed and model cube.
We adopted environmental masks to separate from centre, bar, interarm, spiral-arms and disk to explore the noncircular motions, and the gas physical properties in the PHANGS-MUSE objects.

We find that { physical properties such as the gas dispersion, strength of ionization and amplitude of NC motions show a progressively increase towards the centre and bar, while decreasing values are observed in the bar, interarm regions, and spiral arms}. The gas velocity dispersion is enhanced in the central regions and bars, likely driven by a combination of virial motions, non-axisymmetric flows, and hard ionization sources such as AGN or shocks.

We find the bar and centre to be the regions where the largest amplitude of noncircular motions are observed about 20--50$\kms$, meanwhile the lowest amplitudes about 5--15$\kms$ are found in the interarm, spiral-arms, and disk. { Unlike the flows associated with oval potentials, we did not find clear evidence of spiral-arm-related perturbation flows.}

Finally, we find that the amplitude of noncircular motions is closely linked to the physical properties of the ionized gas, particularly the velocity dispersion and the strength of ionization. Both parameters show moderate correlations with the amplitude of noncircular motions, demonstrating its importance as a key parameter for characterizing the physical conditions of the ionized gas.

\section*{Acknowledgment}
We thank the anonymous referee for their valuable suggestions, which significantly improved the clarity and content of this article.

C.L.C. acknowledges support from Academia Sinica Institute of Astronomy and Astrophysics.
L.L. thanks the support  by the Ministry of Science \& Technology of Taiwan under the grants NSTC 113-2112-M-001-006 - and NSTC 114-2112-M-001 -041 -MY3.
I.C.G. acknowledges financial support from DGAPA-UNAM grant IN-119123 and CONAHCYT grant CF-2023-G-100.
H.A.P. acknowledges support from the National Science and Technology Council of Taiwan under grant 110-2112-M-032-020-MY3.
S.F.S. thanks the PAPIIT-DGAPA AG100622 project and CONACYT grant CF19-39578.
J.B-B acknowledges funding from the grant IA-101522 (DGAPA-PAPIIT, UNAM) and support from the DGAPA-PASPA 2025 fellowship (UNAM).

The SuMIRe Cluster is primarily funded by the Academia Sinica Investigator Award (grants AS-IA-107-M01 and AS-IA-112-M04) and the National Science and Technology Council, Taiwan (grant NSTC 112-2112-M-001-027-MY3).

\software{Astropy \citep{astropy}, Matplotlib \citep{Matplotlib}, Numpy \citep{Numpy}, pyPipe3D \citep{Lacerda2022}, XS3D \citep{xs3d}, 3DBarolo \citep{BBarolo}, CASA \citep{casa}.}

\section*{Data availability}
The PHANGS-MUSE data cubes \citep[e.g.,][]{phangs_eso_data} used in this study are publicly available through the ESO  Science Archive Facility and can be accessed from \cite{phangs_eso_data}. Additional products from this study may be provided upon request by the authors.

\appendix
\section{3D modeling of gaseous disks}
\label{App:3dmodel}

\subsection{Moment maps}
\label{App:momaps}

Moment maps are computed following the standard definition for discontinuous datasets. These are generated from a cleaned cube after applying a 3D mask of valid data \citep{Dame2011}. The first 3 moments of a spectral line are defined as below.

The zeroth moment $M_0$, or intensity weighted:
\begin{equation}
 M_0[\mathrm{flux\,\times\,km/s}] = \sum_v I_v \delta v,
\end{equation}
where $I_v$ is the flux and $\delta v$ is the channel width in $\kms$. The sum goes over every channel from the cube.

The first moment $M_1$, is the centroid of the spectral line defined as:
\begin{equation}
 M_1[\mathrm{km/s}] =  \frac{\sum_v I_v \delta v}{\sum_v I_v}.
\end{equation}
Alternatively, {\tt XS3D} allows to compute the centroid of the line through a second-order polynomial fit to the intensity peak and the two adjacent points.

Finally, the second moment $M_2$, is the intensity weighted root mean square width of the spectral line. For a Gaussian line profile this is the velocity dispersion,

\begin{equation}
 M_2[\mathrm{km/s}] =  \Big( \frac{\sum_v I_v (v-M_1)^2}{\sum_{v} I_v} \Big)^{1/2}.
\end{equation}

\subsubsection{Light moments}
\label{App:light_moms}
If the initial disk geometry is not provided or the maximum ring is not specified, these are estimated using weighted light moments. These geometric moments follow the definition of \citet{Stobie1980} and are computed from the zeroth moment map, $M_0$.

The outermost ring included in the fitting is determined by minimizing the Gini coefficient \citep[e.g.,][]{Florian2016} computed from the moment-zero image. This approach helps exclude isolated pixels, which are likely due to spurious detections or low signal-to-noise regions. At each iteration a new set of geometric parameters are estimated.

\subsection{Initial values of $v_0$ and $\sigma_0$}
\label{App:v0sigma0}
Initial values for the velocity and dispersion are required to start the Least-Square procedure. These are obtained
from the observed moment maps by simple algebraic solution over $k$ concentric rings oriented at a position angle $\phi$ and ellipticity given by $\varepsilon=1-\cos i$. Following \cite{XookSuut} (their Equation~24.) and also \cite{Barnes2003},

\begin{equation}
 v_{0,k}^\mathrm{model}= \sum_{x,y\,\in k} \mathcal{D}_k(x,y)w_k^\mathrm{model}(x,y)/\sum_{x,y\,\in k} \big( w_k^\mathrm{model}(x,y) \big)^2;\, \mathcal{D}_k=(M_1-v_{sys})_k.
\end{equation}
The term $w_k$ is a set of weights that depend on the kinematic model adopted. For the circular rotation case, $w_k^\mathrm{circ}=\sin i \cos \theta_k$. Other kinematic models and weight functions can be found in \citet{XookSuut}.

The velocity dispersion follows a similar expression as before,
\begin{equation}
 \sigma_{0,k}=\sum_{x,y\,\in k} \mathcal{D}_k(x,y) w_k(x,y)/\sum_{x,y\,\in k}\big( w_k(x,y) \big)^2,\, \mathcal{D}_k=\sqrt{ (M_2-\sigma_\mathrm{instr})_k^2}, \, w_k=1~\forall~k.
\end{equation}

These operations will retrieve a set of $\{v_{0,k}, \sigma_{0,k} \}$ for each ring. Then these velocities are linearly interpolated to create maps of the velocity and intrinsic dispersion, which are passed to $S_\mathrm{intrin}$ to create an initial model cube in Equation~\ref{Eq:Sintrin}. At each iteration of Equation~\ref{Eq:cost} a new set of velocities $\{v_{0,k}, \sigma_{0,k} \}$ and $v_\mathrm{sys}$ are estimated, as well as a new disk geometry.

The intensity of each Gaussian is obtained by rescaling the fluxes of the model cube to the observed one.
Once estimated $\sigma_0$ and $v_0$, an unnormalized $S_\mathrm{model}^{\prime}$ is created with corresponding moment zero map $M_\mathrm{0,model}^{\prime}$. Then, $f_0$ is estimated from the unnormalized and observed cubes as follows,

\begin{equation}
f_0(x,y) = \frac{M_\mathrm{0,obs}}{M_\mathrm{0,model}^{\prime}}.
\end{equation}

\section{Individual figures}
\subsection{XS3D PHANGS-MUSE output figures}
\label{App:xs3d_outputs}
\begin{figure}
\centering
\includegraphics[width=0.9\textwidth,keepaspectratio]{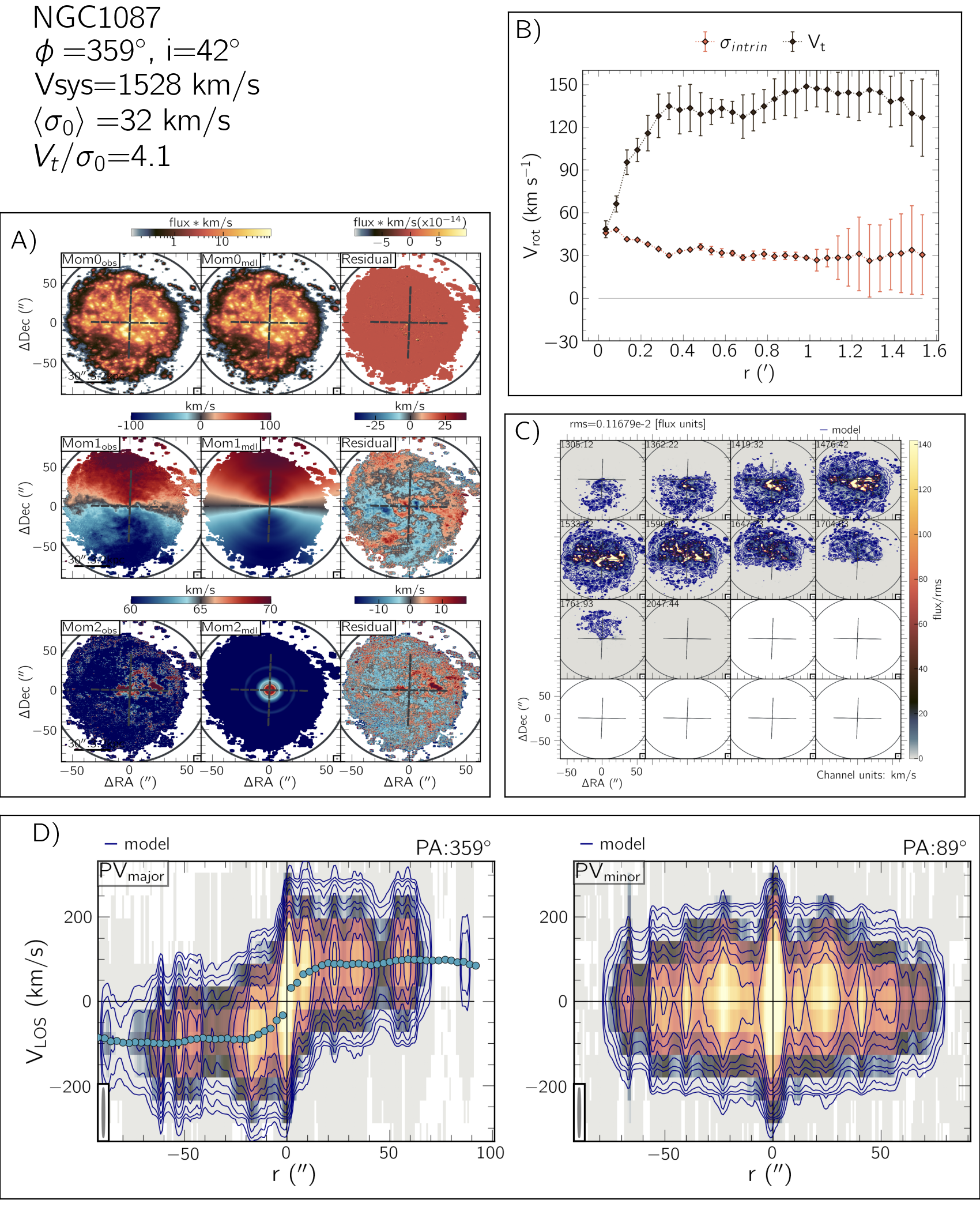}
\caption{Gallery of maps showing {\tt XS3D} outputs for NGC\,1087. Set A) from top to bottom and from left to right: (i) 0th, 1st, and 2nd moment maps extracted from the observed ($I_\mathrm{obs}$) and model ($I_\mathrm{model}$) cubes, together with the residual maps (observed-model) for each moment map. Set B) The intrinsic circular rotation $V_t$ and intrinsic velocity dispersion $\sigma_{0}$ profiles. Set C) Channel maps extracted from the observed and model cubes, with the latter shown with contours on-top. Set D) Position velocity diagram (PV) extracted over slits of $2\times$FWHM widths, positioned along the kinematic major (left) and minor axes (right). Here, the black-red-yellow colour map represents the observed flux from a broad band (the cube spectral window width), while contours on the PV maps represent the model. Contours are placed each $\sigma_\mathrm{MAD} \times 2^{n}$ for $n= -1, 0, 1, ..7$. The left-bottom ellipse represents the spatial and velocity FWHM resolutions. In all panels the large black ellipse represents the best-fit disk geometry while the black cross at the interior shows the orientation of the kinematic major and minor axes. Flux units are $\mathrm{ 10^{-16} erg\,s^{-1}\,cm^{-2}\,\AA^{-1} }$. The complete figure set (19 images) is available in the online journal.}
\label{fig:moments_all}
\end{figure}

In the online version of this article is displayed the full set of products of the {\tt XS3D} pipeline applied to the \ha~subcubes from  PHANGS-MUSE. Figure~\ref{fig:moments_all} illustrates these outputs for the first object listed in Table~\ref{Tab:main_props}. These include the best-fit disk geometry, systemic velocity, average intrinsic dispersion, rotational curve and intrinsic dispersion profile, moment maps, cube channels, and the position velocity diagrams along the major and minor axes.

\figsetstart
\figsetnum{9}
\figsettitle{XS3D outputs for the PHANGS-MUSE objects}

\figsetgrpstart
\figsetgrpnum{9.1}
\figsetgrptitle{NGC1087}
\figsetplot{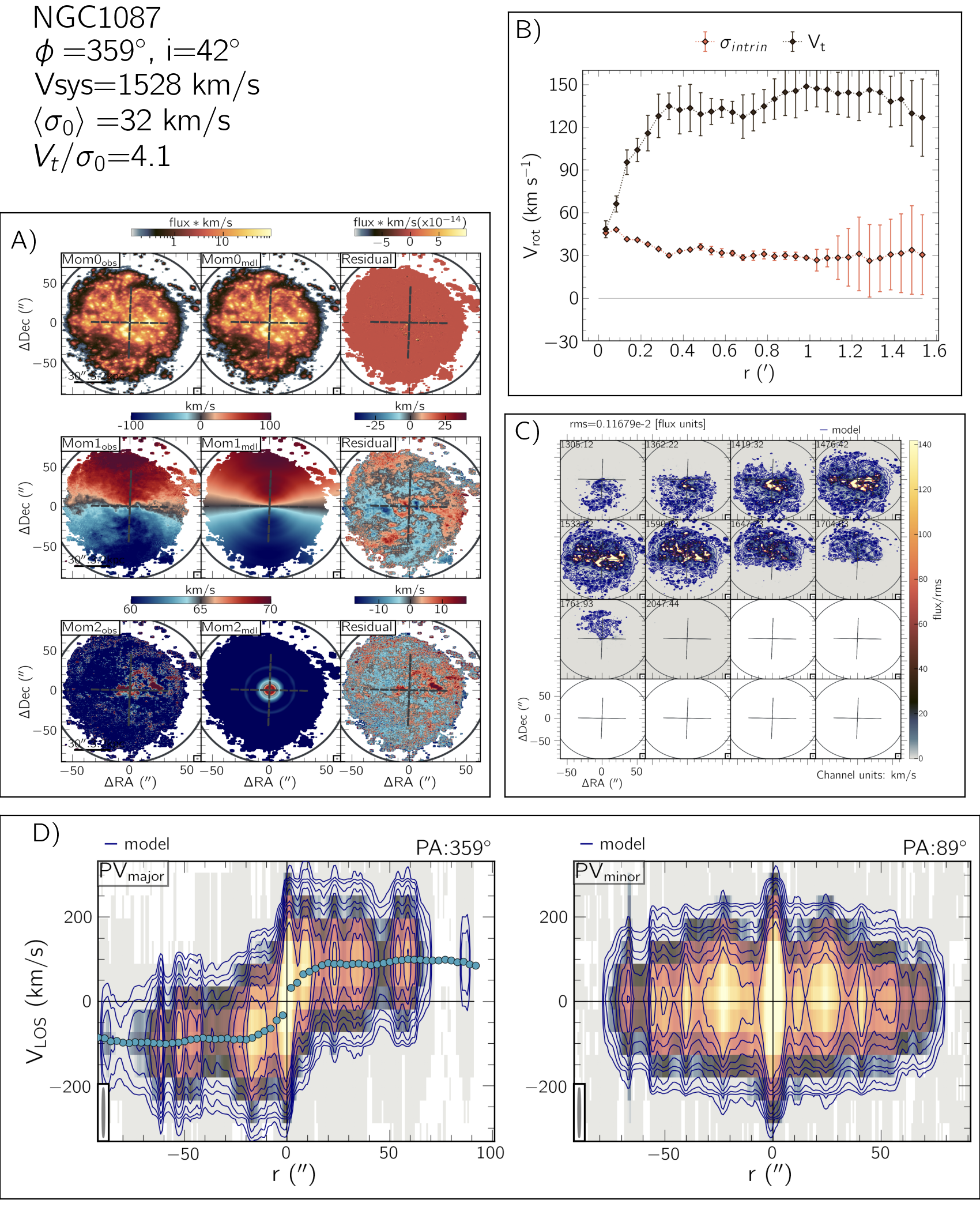}
\figsetgrpnote{Gallery of maps showing XS3D outputs on the PHANGS-MUSE objects. Set A) from top to bottom and from left to right: (i) 0th, 1st, and 2nd moment maps extracted from the observed and model cubes, together with the residual maps (observed-model) for each moment map. Set B) The intrinsic circular rotation $V_t$ and intrinsic velocity dispersion $\sigma_{0}$ profiles. Set C) Channel maps extracted from the observed and model cubes, with the latter shown with contours on-top. Set D) Position velocity diagram (PV) extracted over slits of $2\times$FWHM widths, positioned along the kinematic major (left) and minor axes (right). Here, the black-red-yellow colour map represents the observed flux from a broad band (the cube spectral window width), while contours on the PV maps represent the model. Contours are placed each $\sigma_\mathrm{MAD} \times 2^{n}$ for $n= -1, 0, 1, ..7$. The left-bottom ellipse represents the spatial and velocity FWHM resolutions. In all panels the large black ellipse represents the best-fit disk geometry while the black cross at the interior shows the orientation of the kinematic major and minor axes.}
\figsetgrpend

\figsetend

\subsection{PHANGS-MUSE 3D visualization images}
\label{App:3Dfigs}
\begin{figure*}
\centering
\includegraphics[width=0.8\textwidth,keepaspectratio]{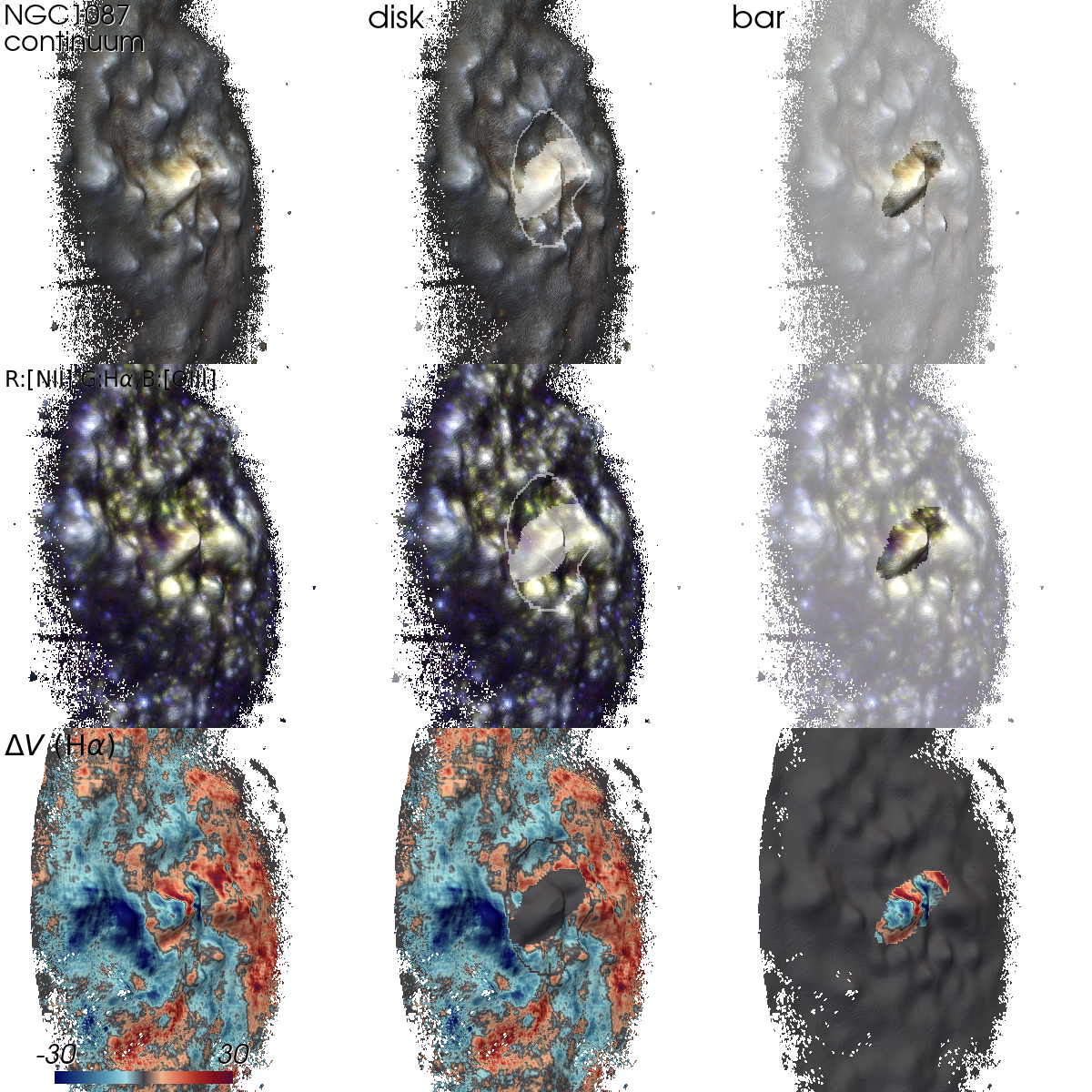}
\caption{3D visualization of the environmental masks from \cite{Querejeta2021}, applied to the PHANGS-MUSE objects. From top to bottom: $gri$ MUSE reconstructed continuum image; true color image of the ionized gas distribution with R: \nii, G: \ha, B: \oiii; and \ha~residual velocities after subtracting the circular rotation. The first column shows the original images, while the remaining columns display the corresponding masks applied, { with desaturated areas} indicating the masked (blanked) pixels in each category. Similar figures for the 19 PHANGS-MUSE objects are available in the online journal.}
\label{fig:moments_all2}
\end{figure*}

The set of figures including the environmental masks along with the  3D visualization technique, is shown in Figure~\ref{fig:moments_all2} for the first objected listed in Table~\ref{Tab:main_props}. The full set of figures for the remaining objects are displayed in the online version of this article.


\figsetstart
\figsetnum{10}
\figsettitle{Enviromental masks applied to the PHANGS-MUSE objects}

\figsetgrpstart
\figsetgrpnum{10.1}
\figsetgrptitle{NGC1087}
\figsetplot{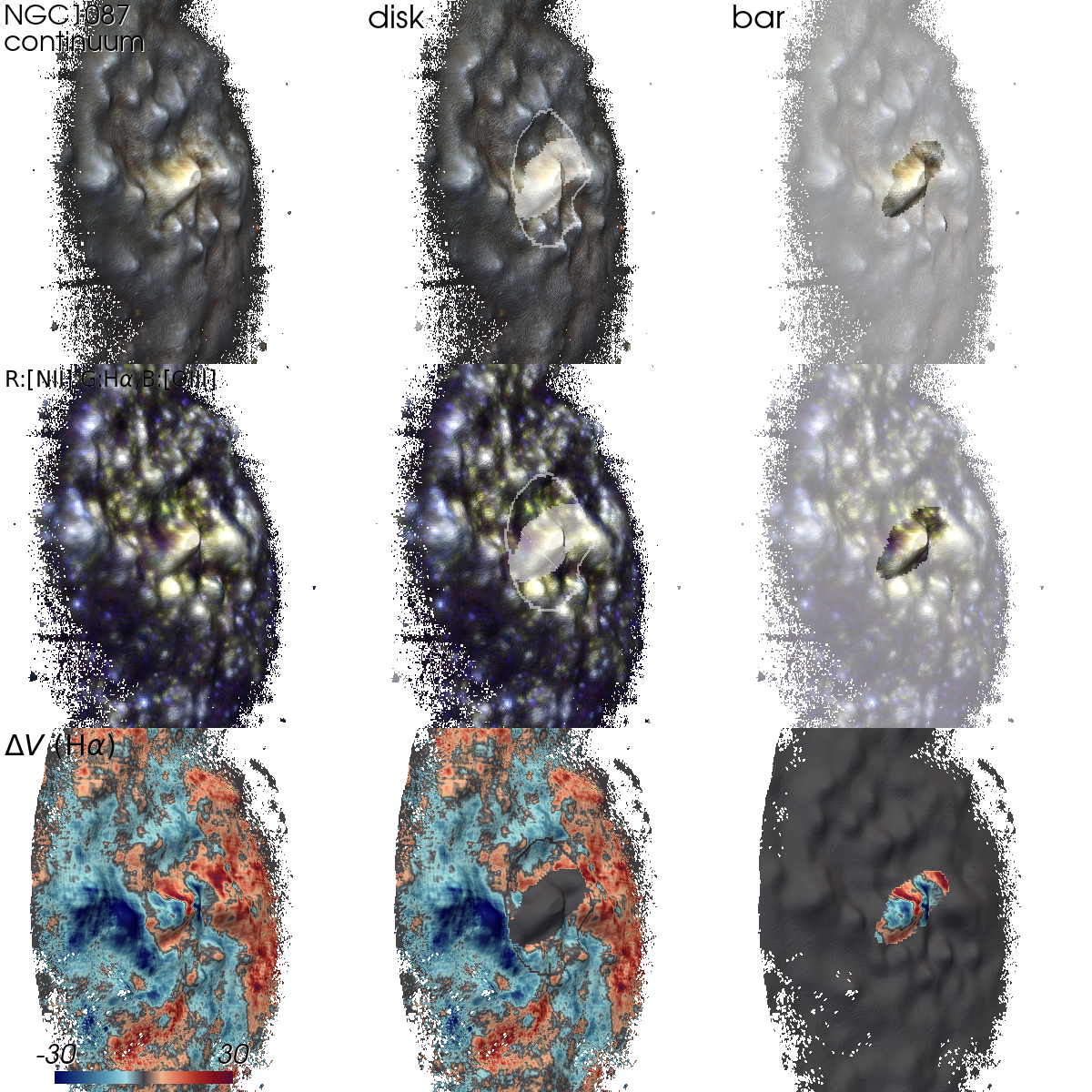}
\figsetgrpnote{3D visualization of the environmental masks applied to the PHANGS-MUSE objects. From top to bottom: $gri$ MUSE reconstructed continuum image; true color image of the ionized gas distribution with R: \nii, G: \ha, B: \oiii; and \ha~residual velocities after subtracting the circular rotation. The first column shows the original images, while the remaining columns display the corresponding masks applied, with desaturated areas indicating the masked (blanked) pixels in each category.}
\figsetgrpend

\figsetgrpstart
\figsetgrpnum{10.6}
\figsetgrptitle{NGC1566}
\figsetplot{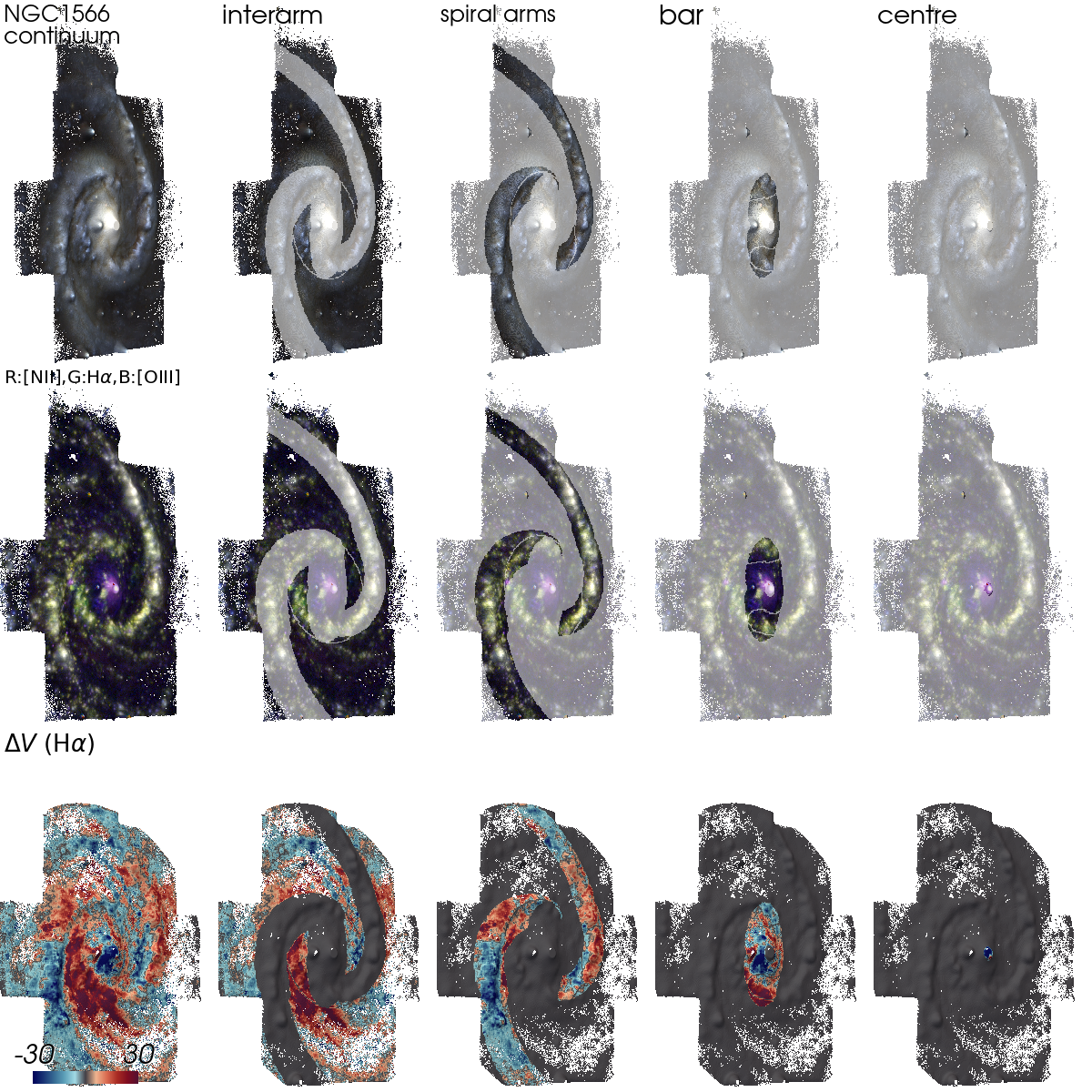}
\figsetgrpnote{3D visualization of the environmental masks applied to the PHANGS-MUSE objects. From top to bottom: $gri$ MUSE reconstructed continuum image; true color image of the ionized gas distribution with R: \nii, G: \ha, B: \oiii; and \ha~residual velocities after subtracting the circular rotation. The first column shows the original images, while the remaining columns display the corresponding masks applied, with desaturated areas indicating the masked (blanked) pixels in each category.}
\figsetgrpend

\figsetend

\subsection{Box plots of physical properties}
\label{App:boxplot_all}
\begin{figure}
\centering
\includegraphics[width=0.98\textwidth,keepaspectratio]{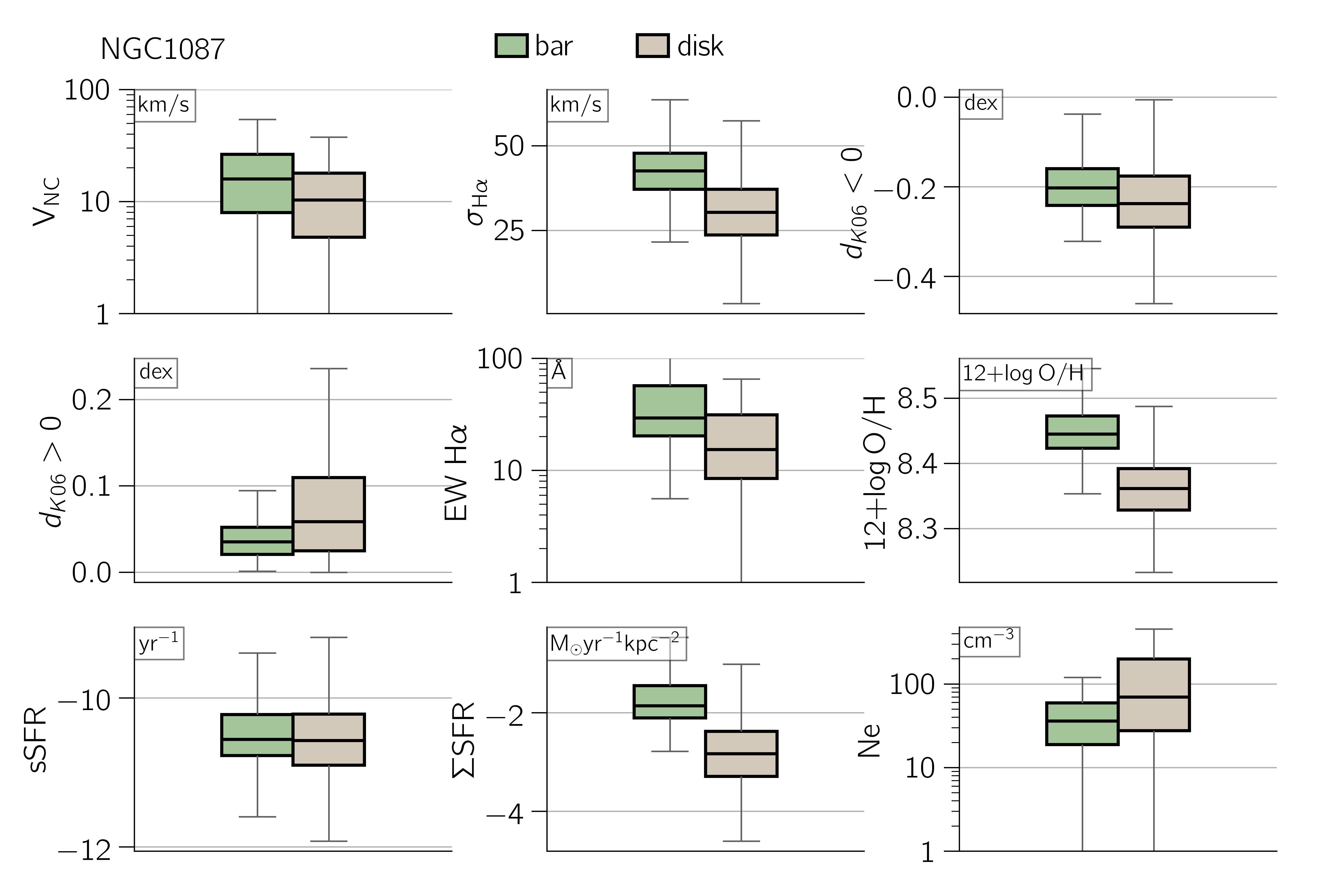}
\caption{Box plots illustrating the distribution of the physical properties analyzed in the main text across the various environments for each PHANGS-MUSE galaxy. The complete figure set (19 images) is available in the online journal. }
\label{fig:moments_all3}
\end{figure}

The set of figures including the distribution of physical properties in each environment is shown in Figure~\ref{fig:moments_all3}.



\figsetstart
\figsetnum{11}
\figsettitle{Distribution of physical properties across the different environments}

\figsetgrpstart
\figsetgrpnum{11.1}
\figsetgrptitle{NGC1087}
\figsetplot{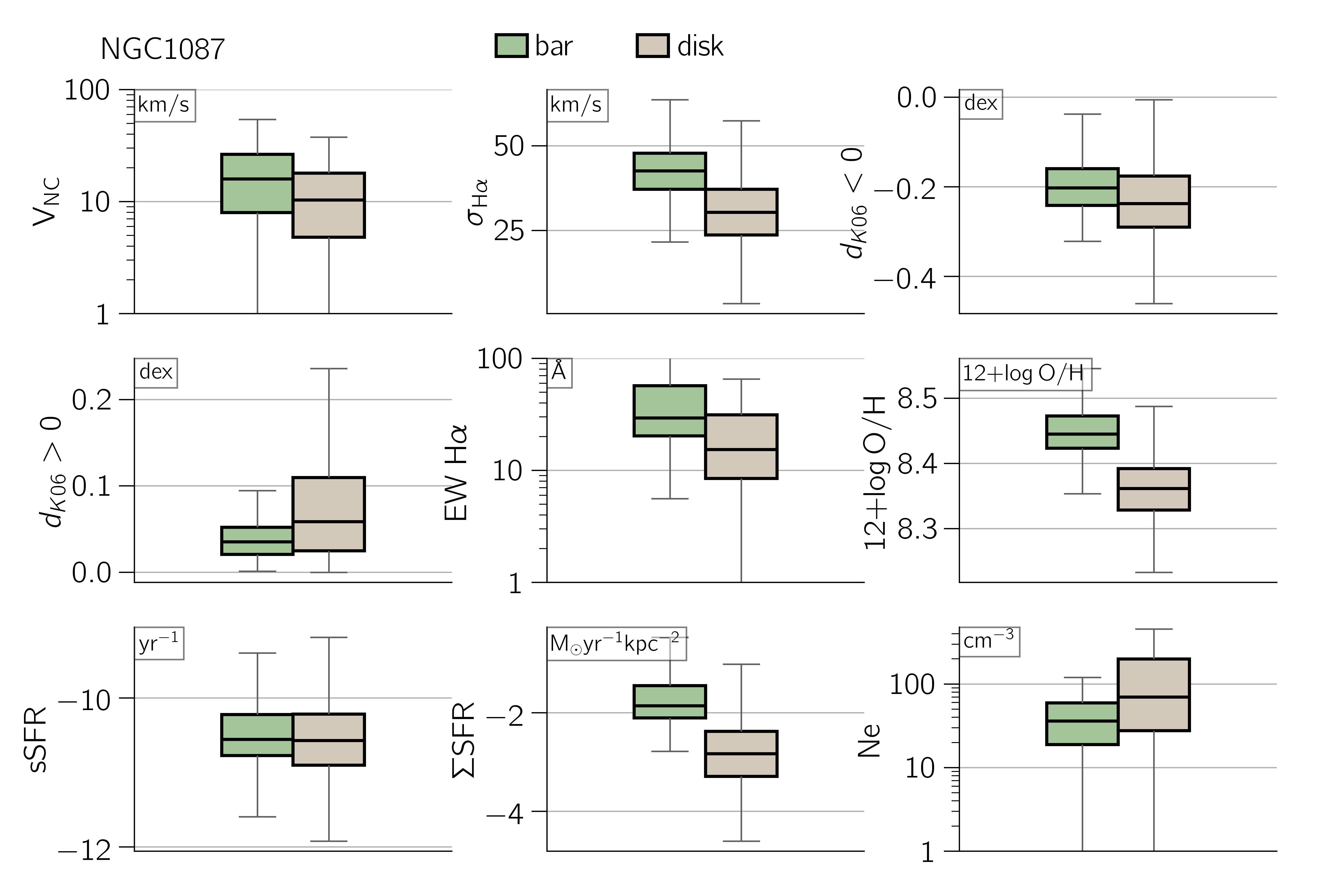}
\figsetgrpnote{Box plots illustrating the distribution of the physical properties analyzed in the main text across the various environments for each PHANGS-MUSE galaxy.}
\figsetgrpend

\figsetgrpstart
\figsetgrpnum{11.2}
\figsetgrptitle{NGC1300}
\figsetplot{boxplot_NGC1300.png}
\figsetgrpnote{Box plots illustrating the distribution of the physical properties analyzed in the main text across the various environments for each PHANGS-MUSE galaxy.}
\figsetgrpend

\figsetgrpstart
\figsetgrpnum{11.3}
\figsetgrptitle{NGC1385}
\figsetplot{boxplot_NGC1385.png}
\figsetgrpnote{Box plots illustrating the distribution of the physical properties analyzed in the main text across the various environments for each PHANGS-MUSE galaxy.}
\figsetgrpend

\figsetgrpstart
\figsetgrpnum{11.4}
\figsetgrptitle{NGC1433}
\figsetplot{boxplot_NGC1433.png}
\figsetgrpnote{Box plots illustrating the distribution of the physical properties analyzed in the main text across the various environments for each PHANGS-MUSE galaxy.}
\figsetgrpend

\figsetgrpstart
\figsetgrpnum{11.5}
\figsetgrptitle{NGC1512}
\figsetplot{boxplot_NGC1512.png}
\figsetgrpnote{Box plots illustrating the distribution of the physical properties analyzed in the main text across the various environments for each PHANGS-MUSE galaxy.}
\figsetgrpend

\figsetgrpstart
\figsetgrpnum{11.6}
\figsetgrptitle{NGC1566}
\figsetplot{boxplot_NGC1566.png}
\figsetgrpnote{Box plots illustrating the distribution of the physical properties analyzed in the main text across the various environments for each PHANGS-MUSE galaxy.}
\figsetgrpend

\figsetgrpstart
\figsetgrpnum{11.7}
\figsetgrptitle{NGC1672}
\figsetplot{boxplot_NGC1672.png}
\figsetgrpnote{Box plots illustrating the distribution of the physical properties analyzed in the main text across the various environments for each PHANGS-MUSE galaxy.}
\figsetgrpend

\figsetgrpstart
\figsetgrpnum{11.8}
\figsetgrptitle{NGC2835}
\figsetplot{boxplot_NGC2835.png}
\figsetgrpnote{Box plots illustrating the distribution of the physical properties analyzed in the main text across the various environments for each PHANGS-MUSE galaxy.}
\figsetgrpend

\figsetgrpstart
\figsetgrpnum{11.9}
\figsetgrptitle{NGC3351}
\figsetplot{boxplot_NGC3351.png}
\figsetgrpnote{Box plots illustrating the distribution of the physical properties analyzed in the main text across the various environments for each PHANGS-MUSE galaxy.}
\figsetgrpend

\figsetgrpstart
\figsetgrpnum{11.10}
\figsetgrptitle{NGC3627}
\figsetplot{boxplot_NGC3627.png}
\figsetgrpnote{Box plots illustrating the distribution of the physical properties analyzed in the main text across the various environments for each PHANGS-MUSE galaxy.}
\figsetgrpend

\figsetgrpstart
\figsetgrpnum{11.11}
\figsetgrptitle{NGC4254}
\figsetplot{boxplot_NGC4254.png}
\figsetgrpnote{Box plots illustrating the distribution of the physical properties analyzed in the main text across the various environments for each PHANGS-MUSE galaxy.}
\figsetgrpend

\figsetgrpstart
\figsetgrpnum{11.12}
\figsetgrptitle{NGC4303}
\figsetplot{boxplot_NGC4303.png}
\figsetgrpnote{Box plots illustrating the distribution of the physical properties analyzed in the main text across the various environments for each PHANGS-MUSE galaxy.}
\figsetgrpend

\figsetgrpstart
\figsetgrpnum{11.13}
\figsetgrptitle{NGC4321}
\figsetplot{boxplot_NGC4321.png}
\figsetgrpnote{Box plots illustrating the distribution of the physical properties analyzed in the main text across the various environments for each PHANGS-MUSE galaxy.}
\figsetgrpend

\figsetgrpstart
\figsetgrpnum{11.14}
\figsetgrptitle{NGC4535}
\figsetplot{boxplot_NGC4535.png}
\figsetgrpnote{Box plots illustrating the distribution of the physical properties analyzed in the main text across the various environments for each PHANGS-MUSE galaxy.}
\figsetgrpend

\figsetgrpstart
\figsetgrpnum{11.15}
\figsetgrptitle{NGC7496}
\figsetplot{boxplot_NGC7496.png}
\figsetgrpnote{Box plots illustrating the distribution of the physical properties analyzed in the main text across the various environments for each PHANGS-MUSE galaxy.}
\figsetgrpend

\figsetgrpstart
\figsetgrpnum{11.16}
\figsetgrptitle{IC5332}
\figsetplot{boxplot_IC5332.png}
\figsetgrpnote{Box plots illustrating the distribution of the physical properties analyzed in the main text across the various environments for each PHANGS-MUSE galaxy.}
\figsetgrpend

\figsetgrpstart
\figsetgrpnum{11.17}
\figsetgrptitle{NGC628}
\figsetplot{boxplot_NGC628.png}
\figsetgrpnote{Box plots illustrating the distribution of the physical properties analyzed in the main text across the various environments for each PHANGS-MUSE galaxy.}
\figsetgrpend

\figsetgrpstart
\figsetgrpnum{11.18}
\figsetgrptitle{NGC5068}
\figsetplot{boxplot_NGC5068.png}
\figsetgrpnote{Box plots illustrating the distribution of the physical properties analyzed in the main text across the various environments for each PHANGS-MUSE galaxy.}
\figsetgrpend

\figsetgrpstart
\figsetgrpnum{11.19}
\figsetgrptitle{NGC1365}
\figsetplot{boxplot_NGC1365.png}
\figsetgrpnote{Box plots illustrating the distribution of the physical properties analyzed in the main text across the various environments for each PHANGS-MUSE galaxy.}
\figsetgrpend

\figsetend

\section{Comparison with 3DBarolo}
\label{App:3Dbarolo}

We compare our results with {\tt 3DBarolo} using the same configuration adopted in {\tt XS3D}.
We run the {\tt 3DFIT} task in {\tt 3DBarolo} and performed the modeling using the same number of rings, a fixed $2.0\arcsec$ ring spacing,  and the same initial disk geometry given by S4G. We also applied local normalization for the intensity, set {\tt MASK=SEARCH}, and adopted homogeneous weighting {\tt WFUNC=0}. We adopt the default masking search and {\tt FTYPE=2}.
Radial variations were allowed in all parameters, including the disk geometry and velocities (i.e., the parameters {\tt VROT}, {\tt VDISP} and {\tt VSYS} in {\tt 3DBarolo}).
We adopted a $116\kms$/FWHM resolution equivalent to 2.03 channels ({\tt LINEAR}=0.862) and used the PSF/FWHM values reported in Table~\ref{Tab:main_props}. In both pipelines we used 4 cores for the analysis.

Results for the best-fit disk geometry obtained with {\tt 3DBarolo} were reported in Figure~\ref{fig:orientations}. For that figure, we computed the mean and standard deviation of $i(r)$ and $\phi(r)$ for each object. The agreement between both pipelines in the global inclination and position angles supports the assumption of a flat disk in {\tt XS3D} for this analysis.

\begin{figure*}[t!]
	\centering
	\includegraphics[width=0.24\textwidth,keepaspectratio]{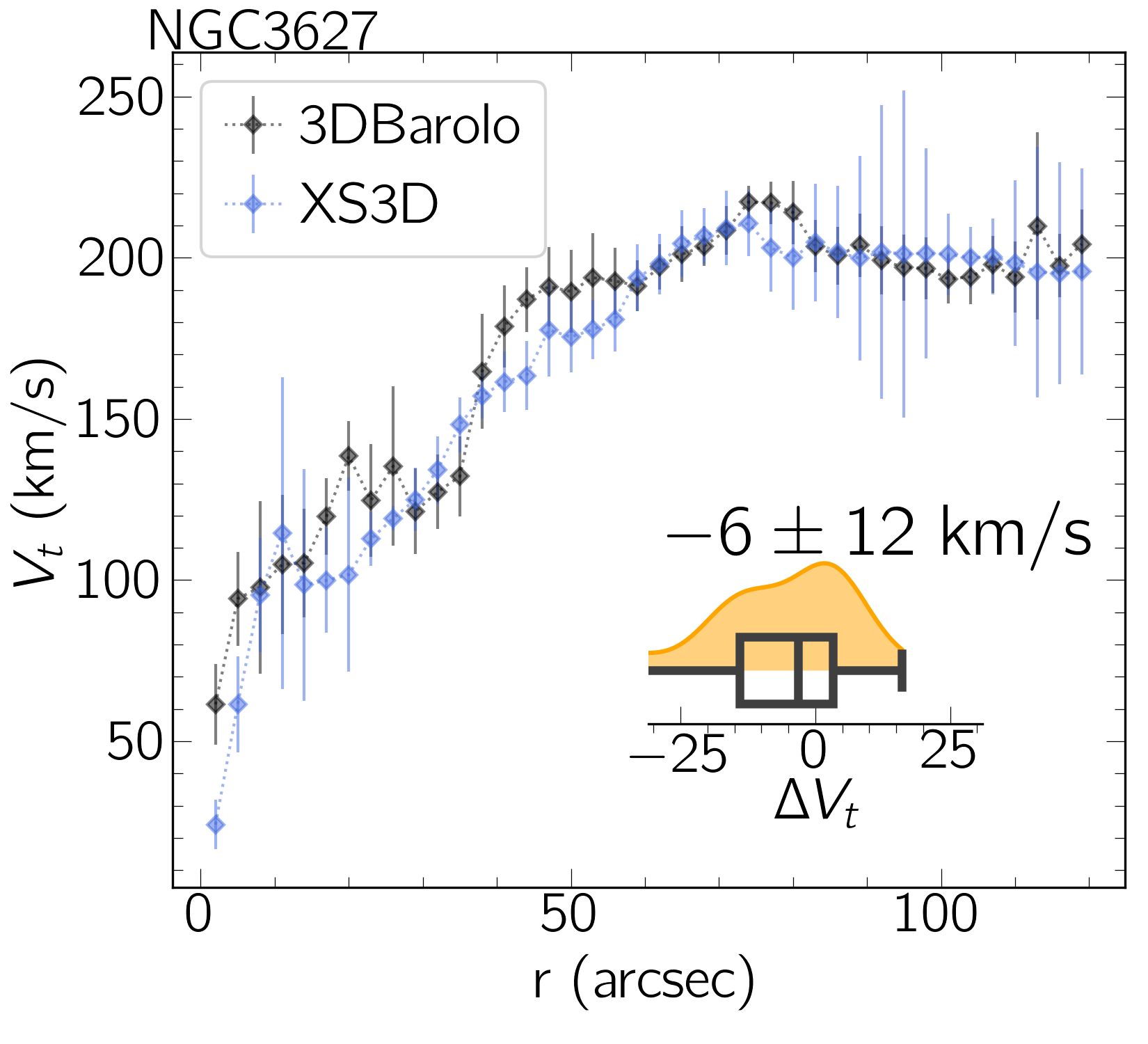}
	\includegraphics[width=0.24\textwidth,keepaspectratio]{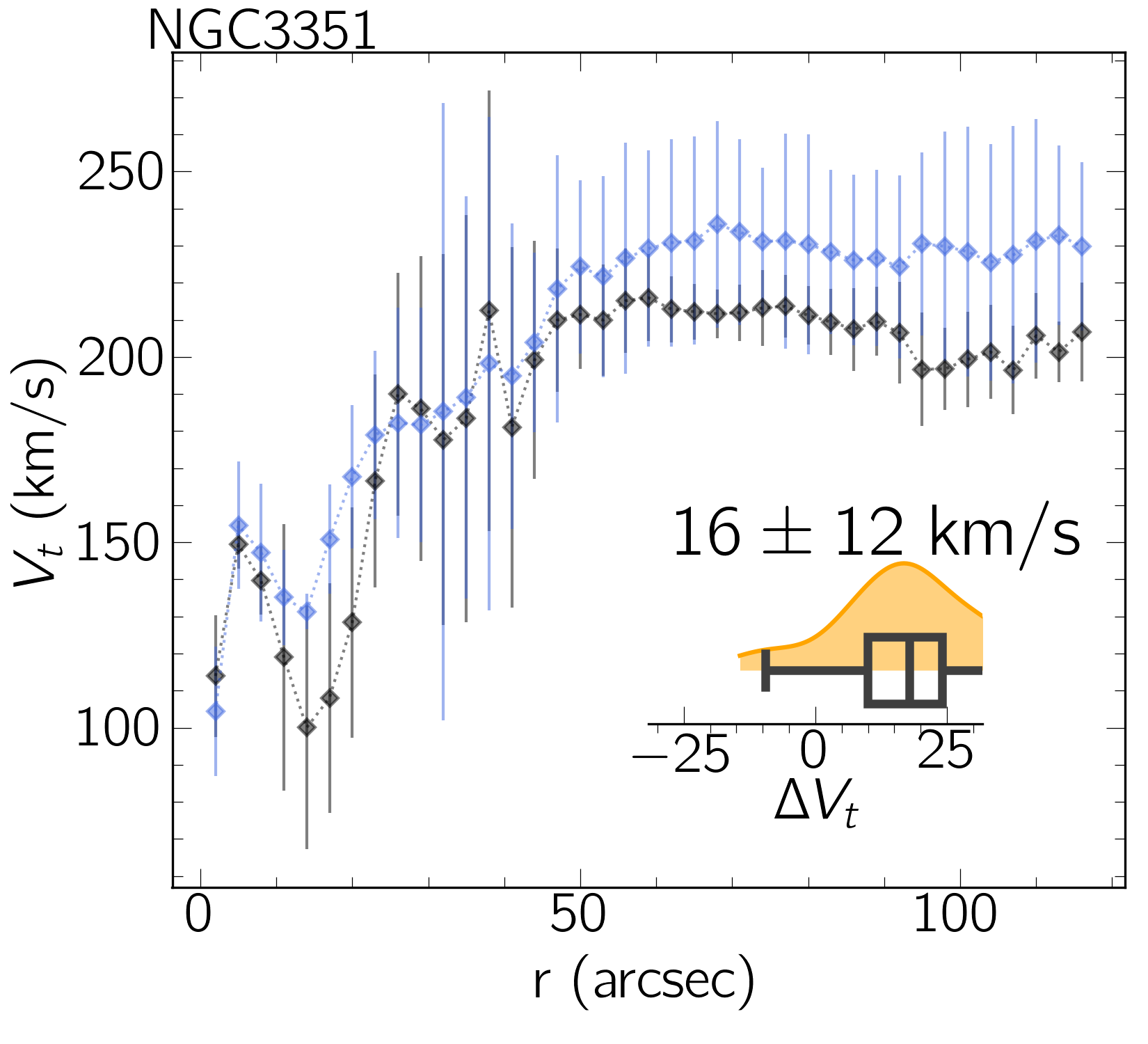}
	\includegraphics[width=0.24\textwidth,keepaspectratio]{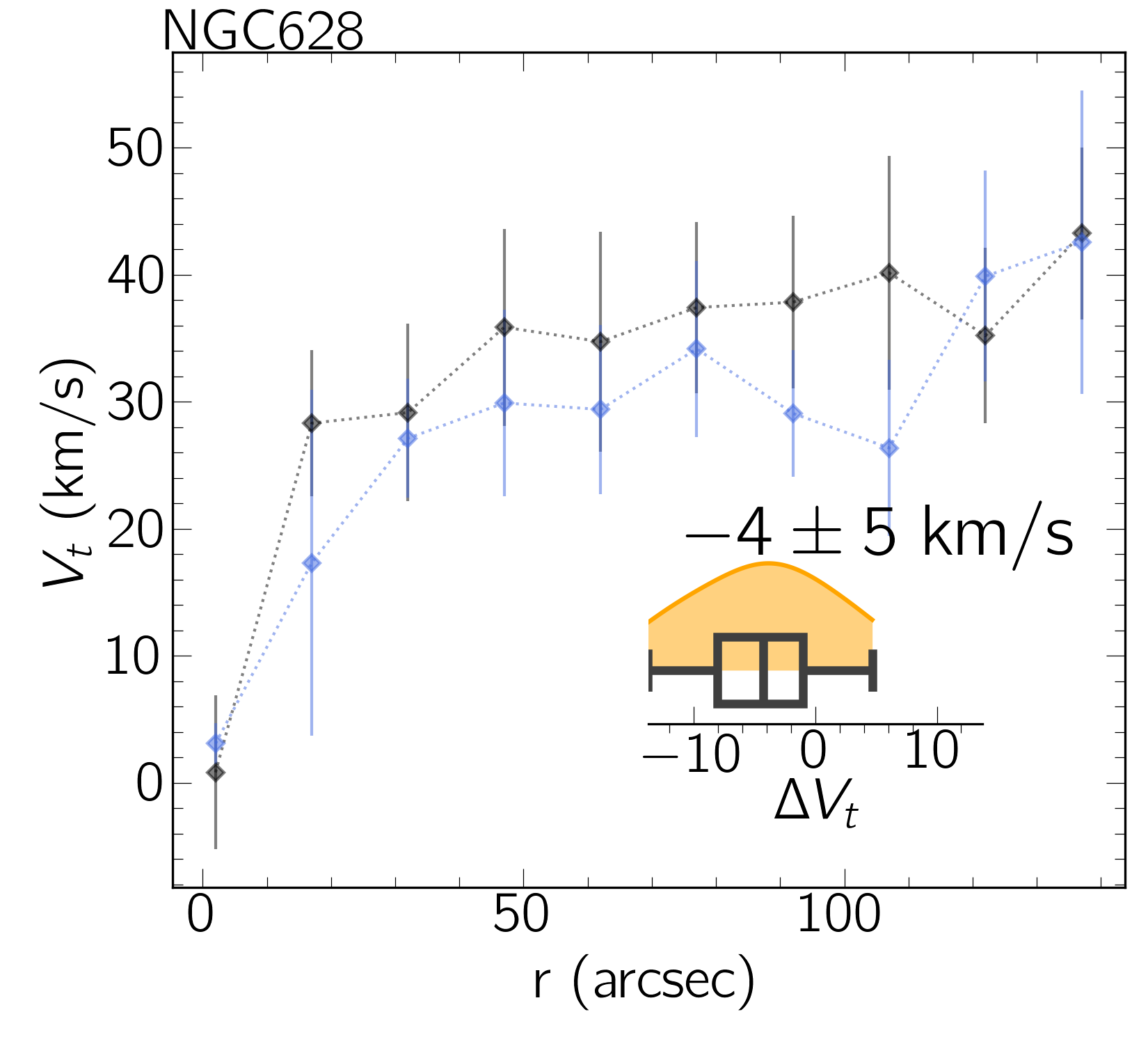}
	\includegraphics[width=0.24\textwidth,keepaspectratio]{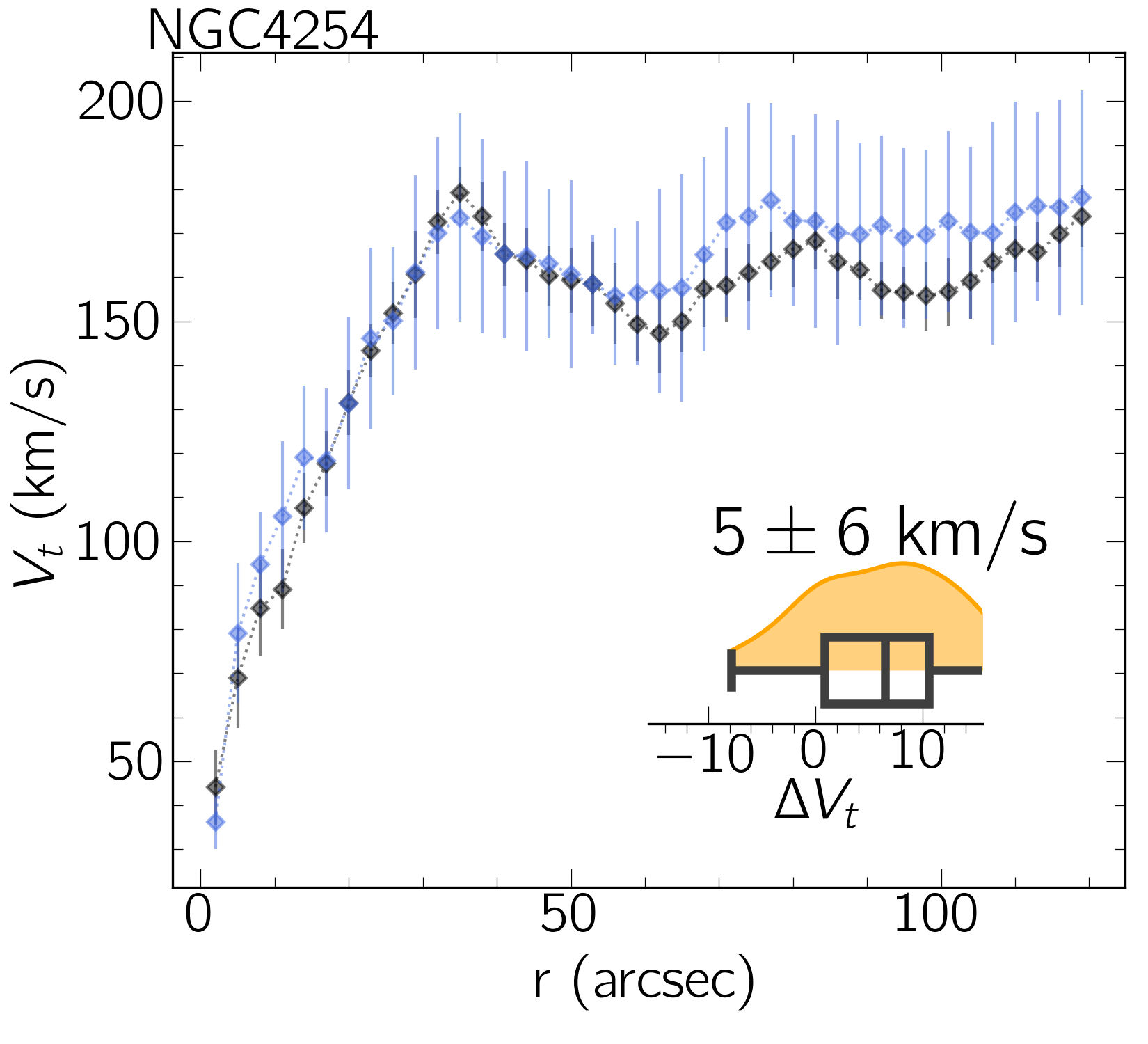}
	\includegraphics[width=0.24\textwidth,keepaspectratio]{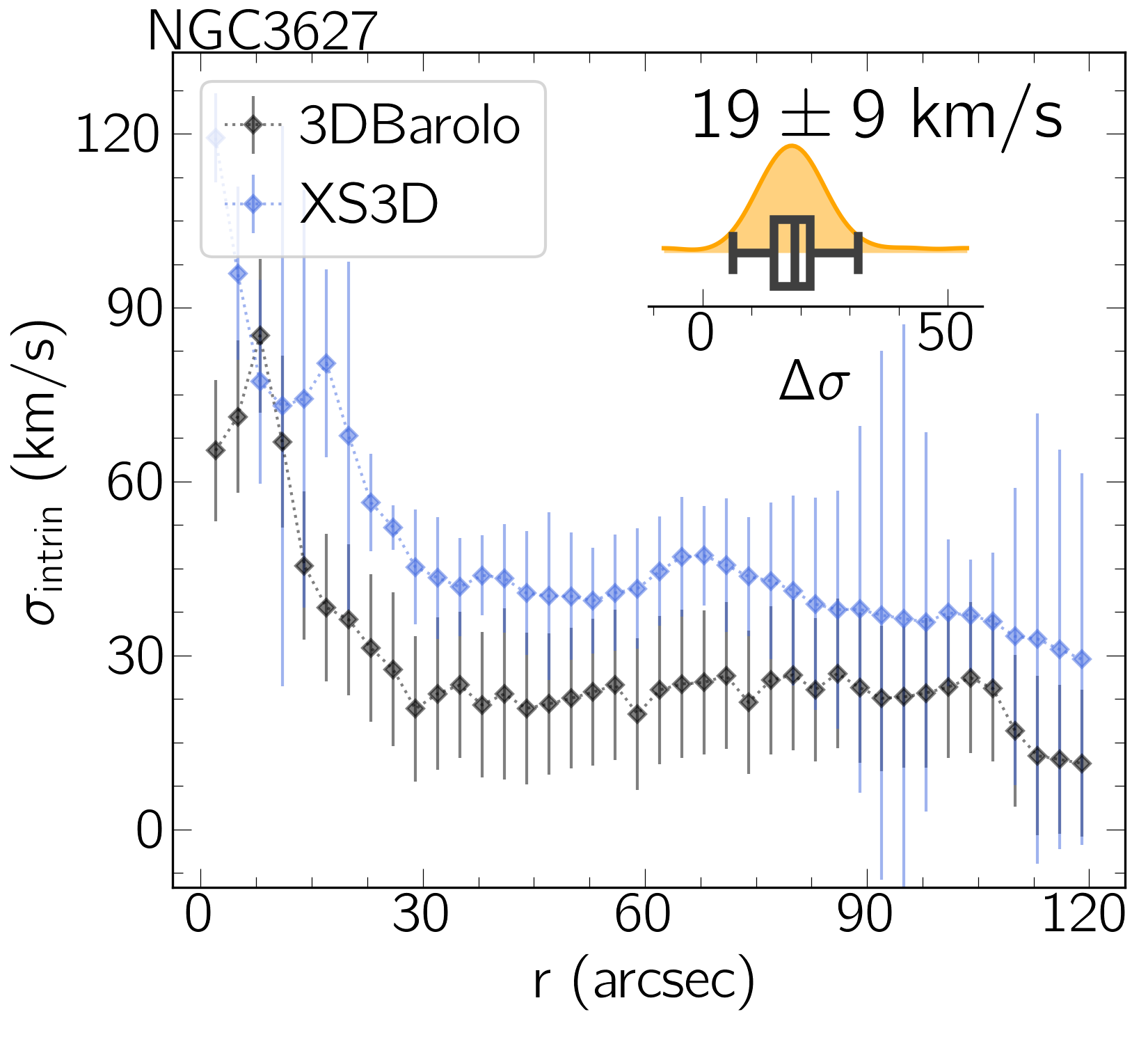}
	\includegraphics[width=0.24\textwidth,keepaspectratio]{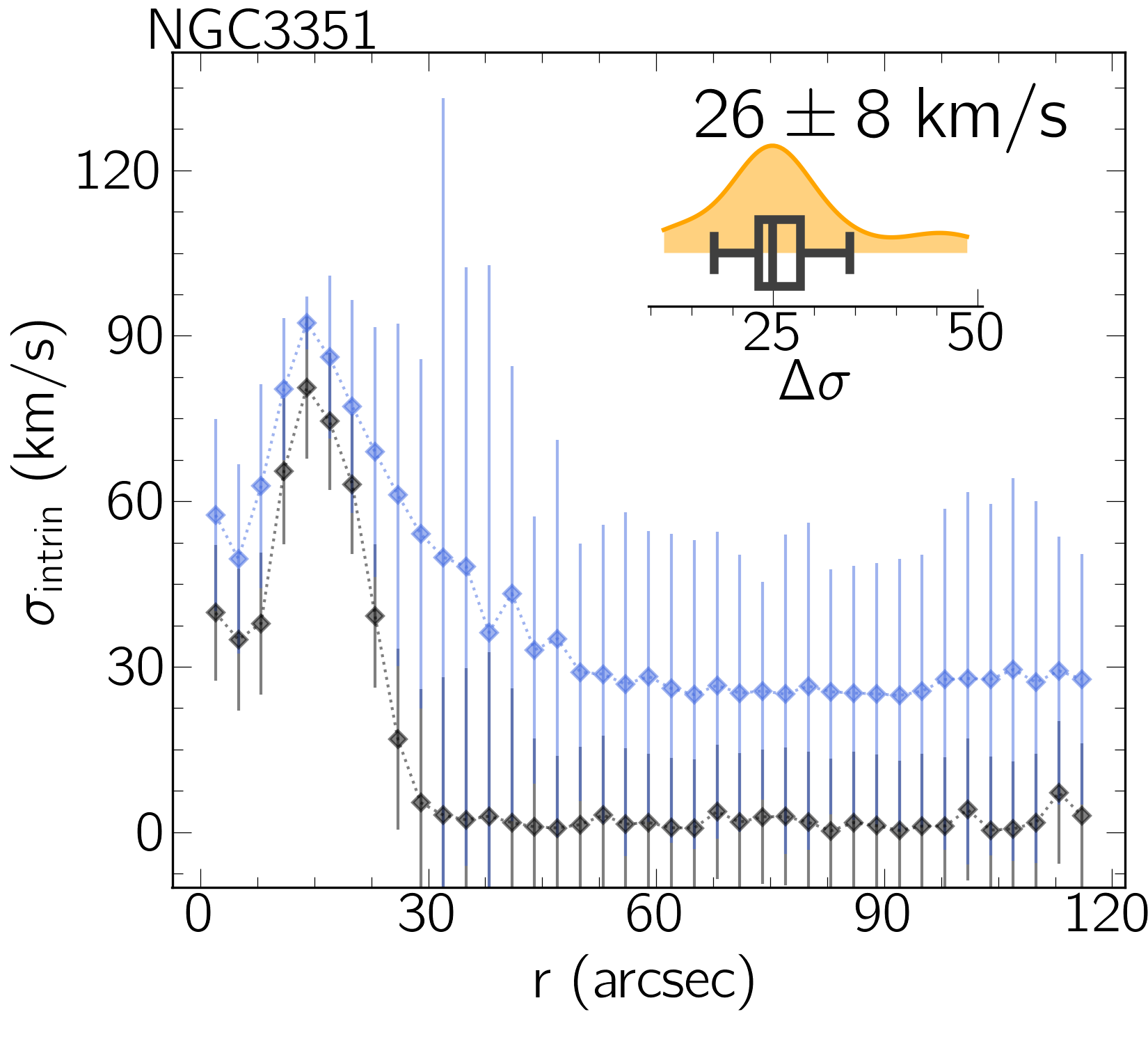}
	\includegraphics[width=0.24\textwidth,keepaspectratio]{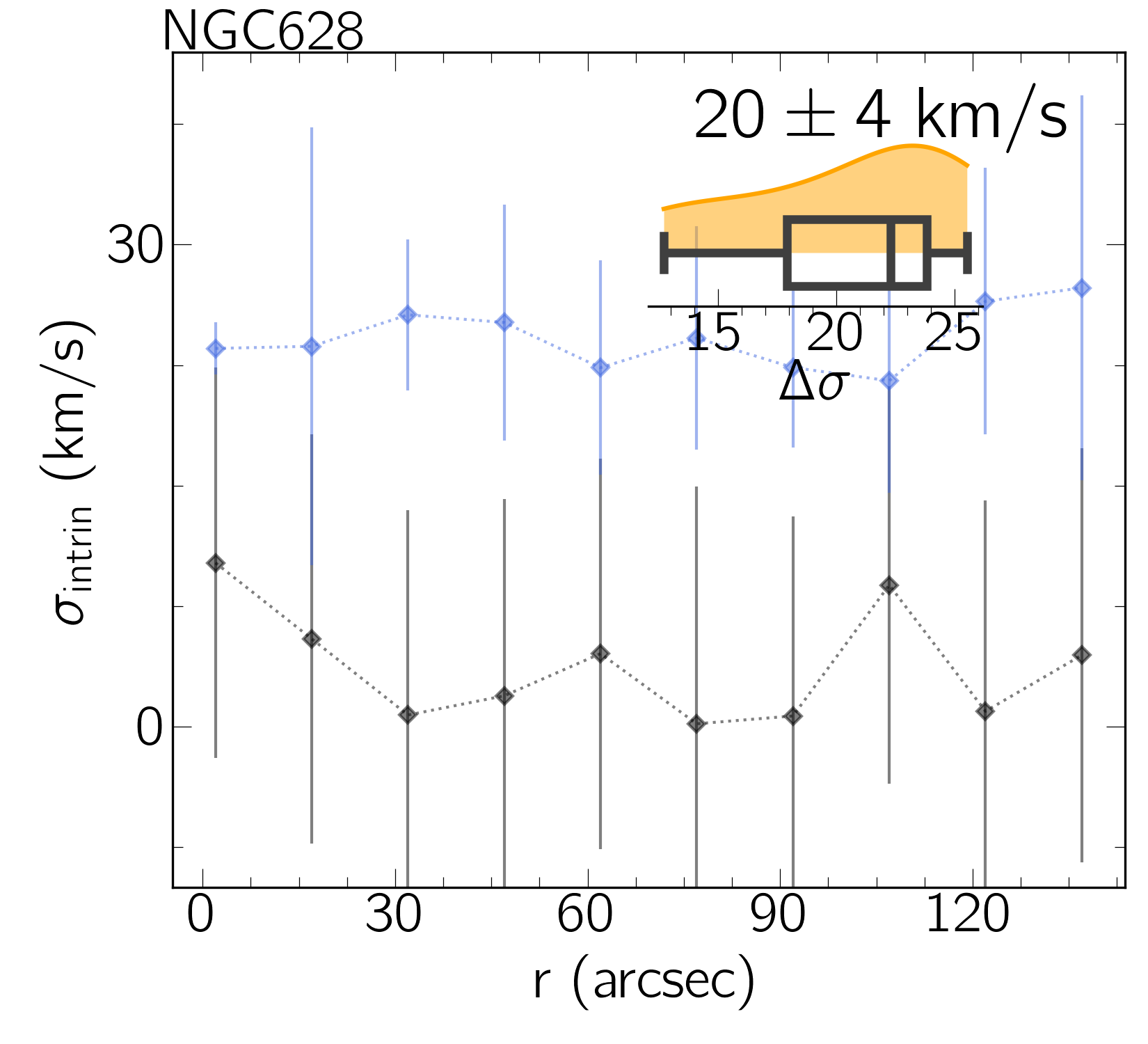}
	\includegraphics[width=0.24\textwidth,keepaspectratio]{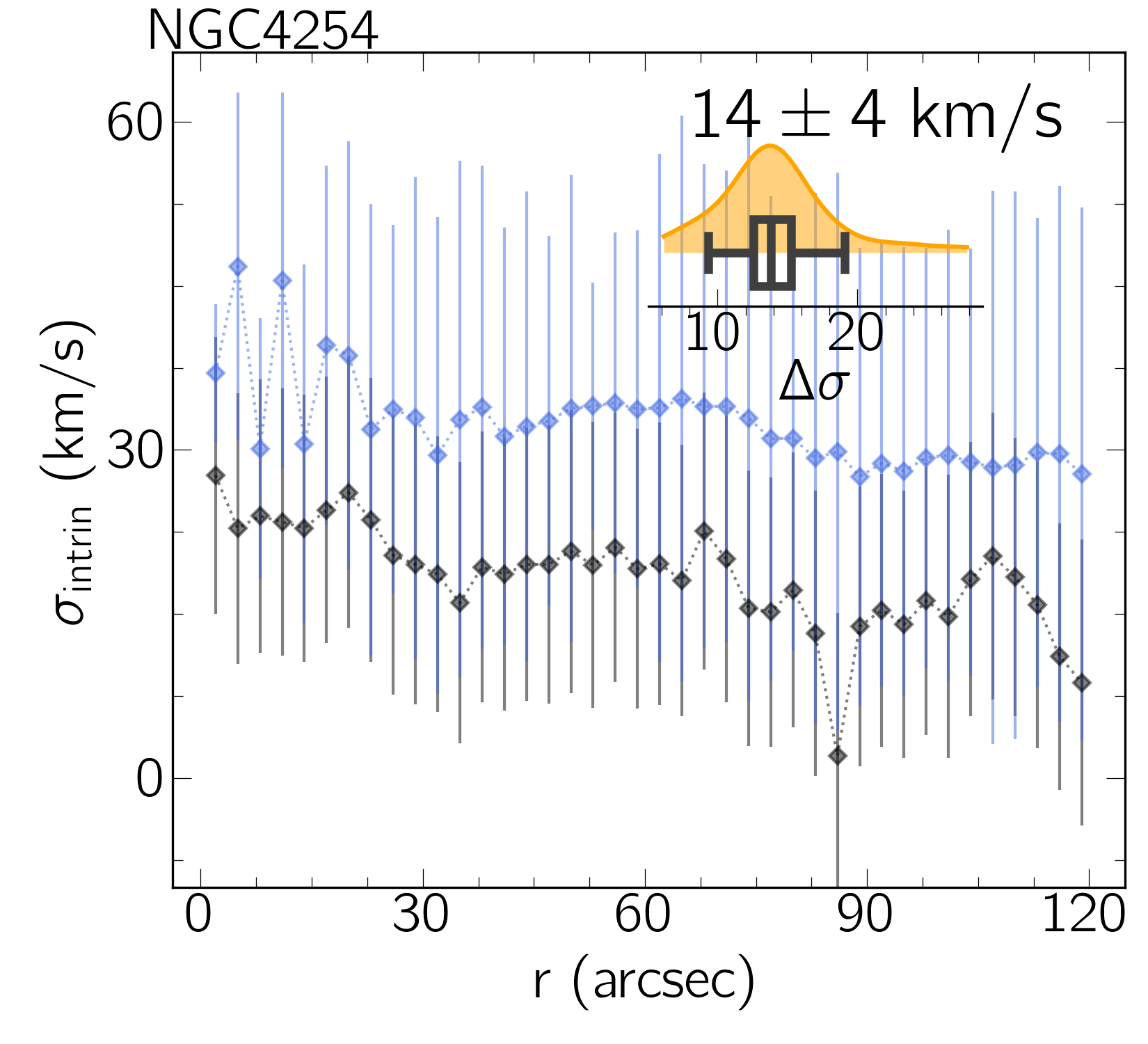}
	\caption{Comparison of $V_t(r)$ and $\sigma_\mathrm{intrin}(r)$ between {\tt XS3D} (blue dots) and {\tt 3DBarolo} (black dots) for a subsample of objects from PHANGS-MUSE. Top panels show the \ha~rotational curves while the bottom panels show the intrinsic velocity dispersion profiles. The boxplots represent the differences between {\tt 3DBarolo} and {\tt XS3D} (i.e., $\Delta=$ {\tt XS3D}-{\tt 3DBarolo}), with  the mean and standard deviation of the distribution indicated. }
	\label{fig:bbarolo_xs3d}
\end{figure*}

Rotational curves and intrinsic velocity dispersion profiles are shown for a set of objects in Figure~\ref{fig:bbarolo_xs3d}.
Results for the rotational curves with  {\tt XS3D} agree well  with those from {\tt 3DBarolo} within the quoted uncertainties, even in  complicated objects like NGC\,628 where reported  inclination angle range from as low as 5$^{\circ}$ to 35$^{\circ}$ \citep[see][and references therein]{Salak2019,Lang2020}. For this object {\tt XS3D} derives an inclination angle of $33\pm8^{\circ}$, while {\tt 3DBarolo} gives $37\pm1^{\circ}$ and S4G reports $31^{\circ}$.

\begin{figure*}[t!]
	\centering
	\includegraphics[width=0.42\textwidth,keepaspectratio]{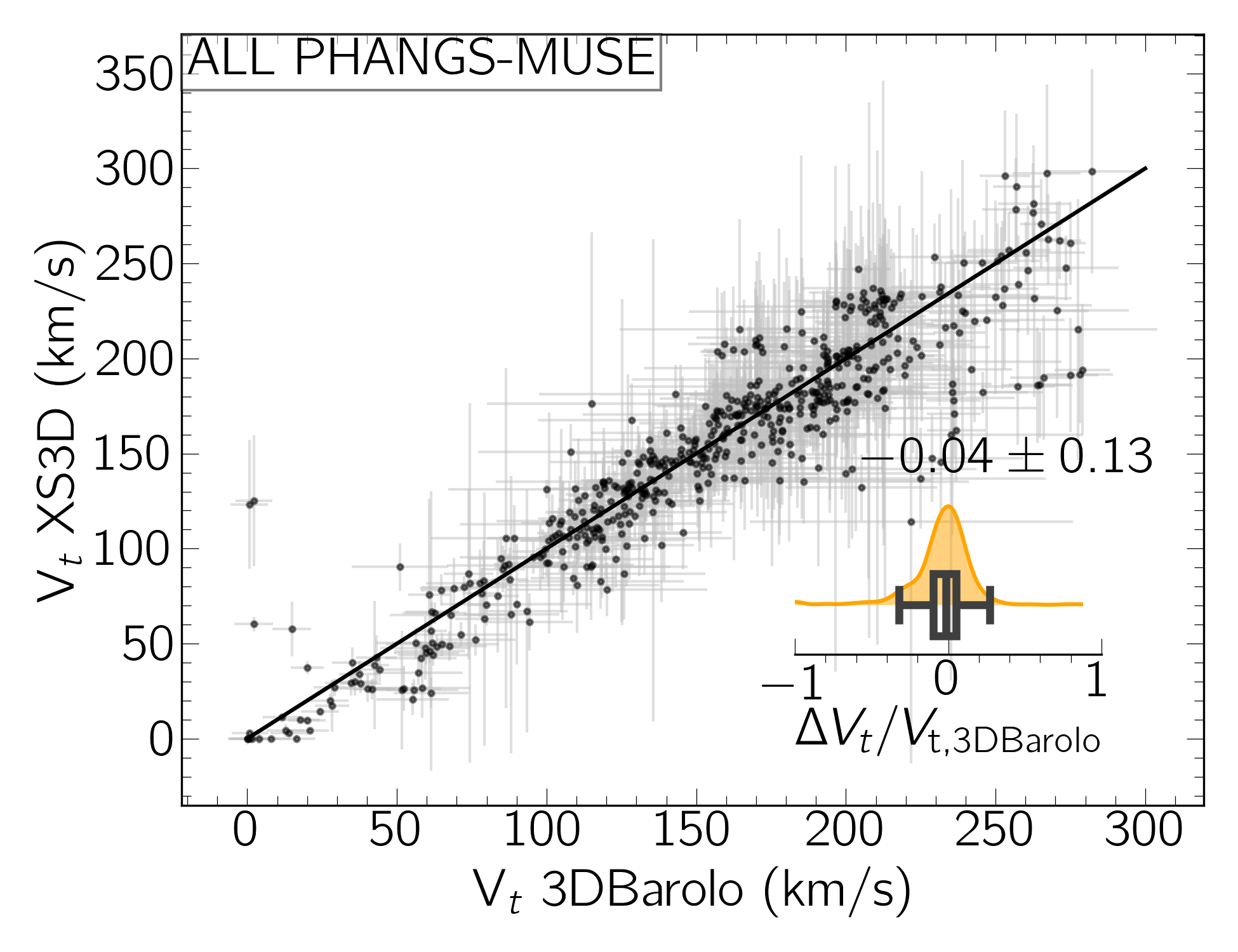}
	\includegraphics[width=0.42\textwidth,keepaspectratio]{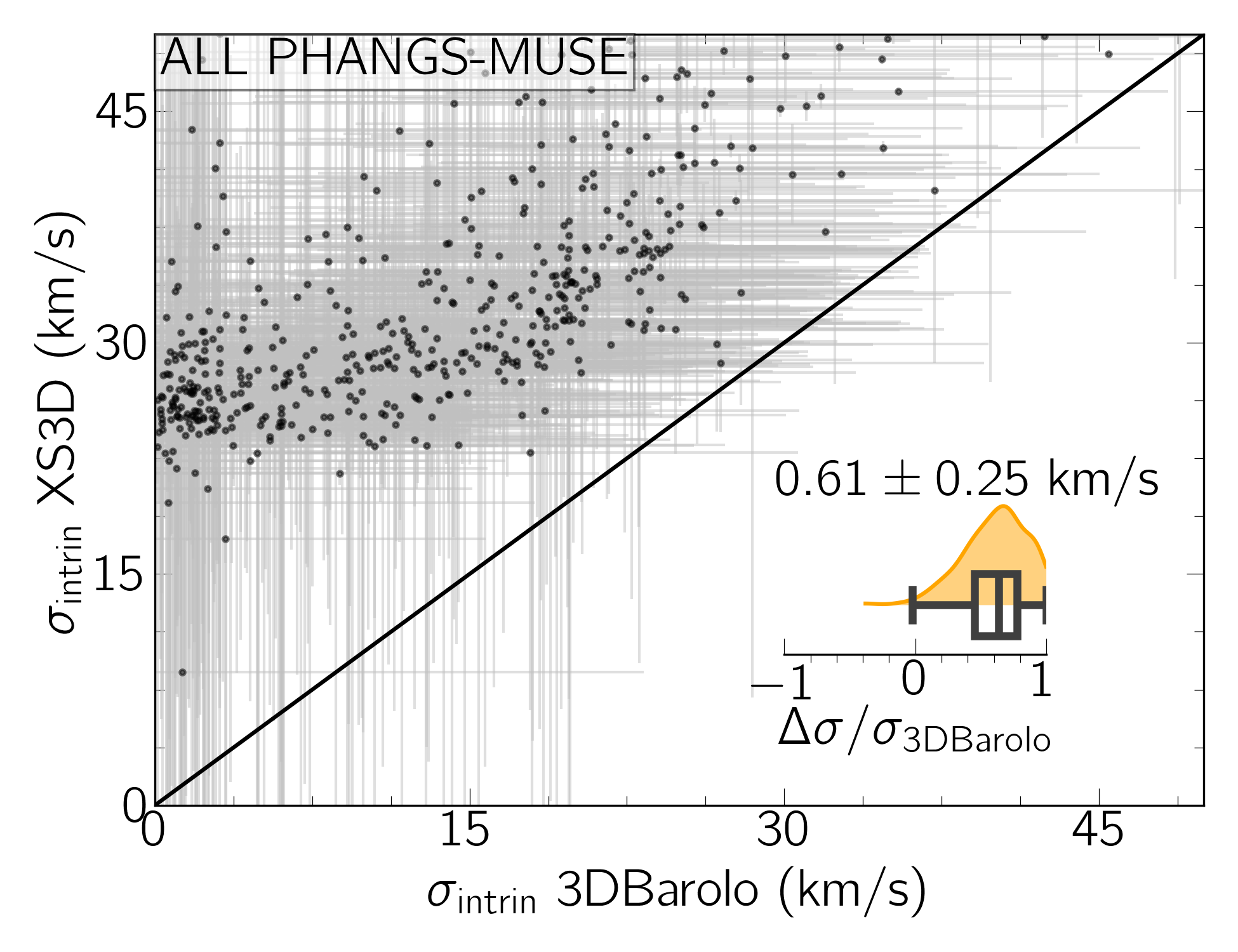}
	\caption{{ Global comparison of the 19 rotational curves and intrinsic dispersion profiles derived with {\tt XS3D} and {\tt 3DBarolo}}. {\it Left panel}: Rotational curves, with values derived using {\tt XS3D} on the y-axis and  {\tt 3DBarolo} on the x-axis. {\it Right panel:} Comparison of the average intrinsic velocity dispersion profiles. In both panels the solid line represents the 1-to-1 relation, while the boxplot indicates { the relative error with respect to {\tt 3DBarolo} (i.e., {\tt (XS3D-3DBarolo)/3DBarolo}).}}
	\label{fig:vt_sigma_all}
\end{figure*}

A global comparison of the 19 rotational curves  and dispersion profiles is illustrated in Figure~\ref{fig:vt_sigma_all}. The left side plot shows individual values of $V_t(r)$ for each object derived with both pipelines, while their relative differences with respect {\tt 3DBarolo}, namely $(V_\mathrm{t,XS3D}-V_\mathrm{t,3DBarolo})/V_\mathrm{t,3DBarolo}$, is shown in a box plot. From the latter plot we observe that {\tt XS3D} derives rotational curves with a $4\%$ error with respect to {\tt 3DBarolo}, which we attribute to the radial variations in $i(r)$ and $\phi(r)$ in {\tt 3DBarolo}. This result demonstrates that $V_t(r)$ derived with {\tt XS3D} produces consistent results as {\tt 3DBarolo}.

The largest discrepancy is observed in the intrinsic velocity dispersion $\sigma_\mathrm{intrin}$ of \ha. Our algorithm recovers intrinsic dispersions around $25\kms$ on average, while {\tt 3DBarolo} reports significantly lower values, although the radial profiles show similar behavior as observed in Figure~\ref{fig:bbarolo_xs3d}. The right panel of Fig~\ref{fig:vt_sigma_all} shows that {\tt XS3D} estimates intrinsic dispersions $62\%$ larger compared to {\tt 3DBarolo}.

To understand such discrepancy, we compared spaxel-by-spaxel the moment maps extracted from the original cubes and those obtained from the model gas cubes in both pipelines (i.e., Eq.~\ref{Eq:Gaussian_model} in the case of {\tt XS3D}). In this analysis we compared $\sim 2.5\times10^6$ spaxels corresponding to the 19 objects.
By comparing the observed intensity ($M_0$) and velocity ($M_1$) maps in both pipelines, we find relative differences with respect {\tt 3DBarolo} of $7\%$ and $\sim0\%$, respectively. Similar errors are found when comparing the $M_0$ and $M_1$ maps from the model cubes.

\begin{figure*}[t!]
	\centering
	\includegraphics[width=0.32\textwidth,keepaspectratio]{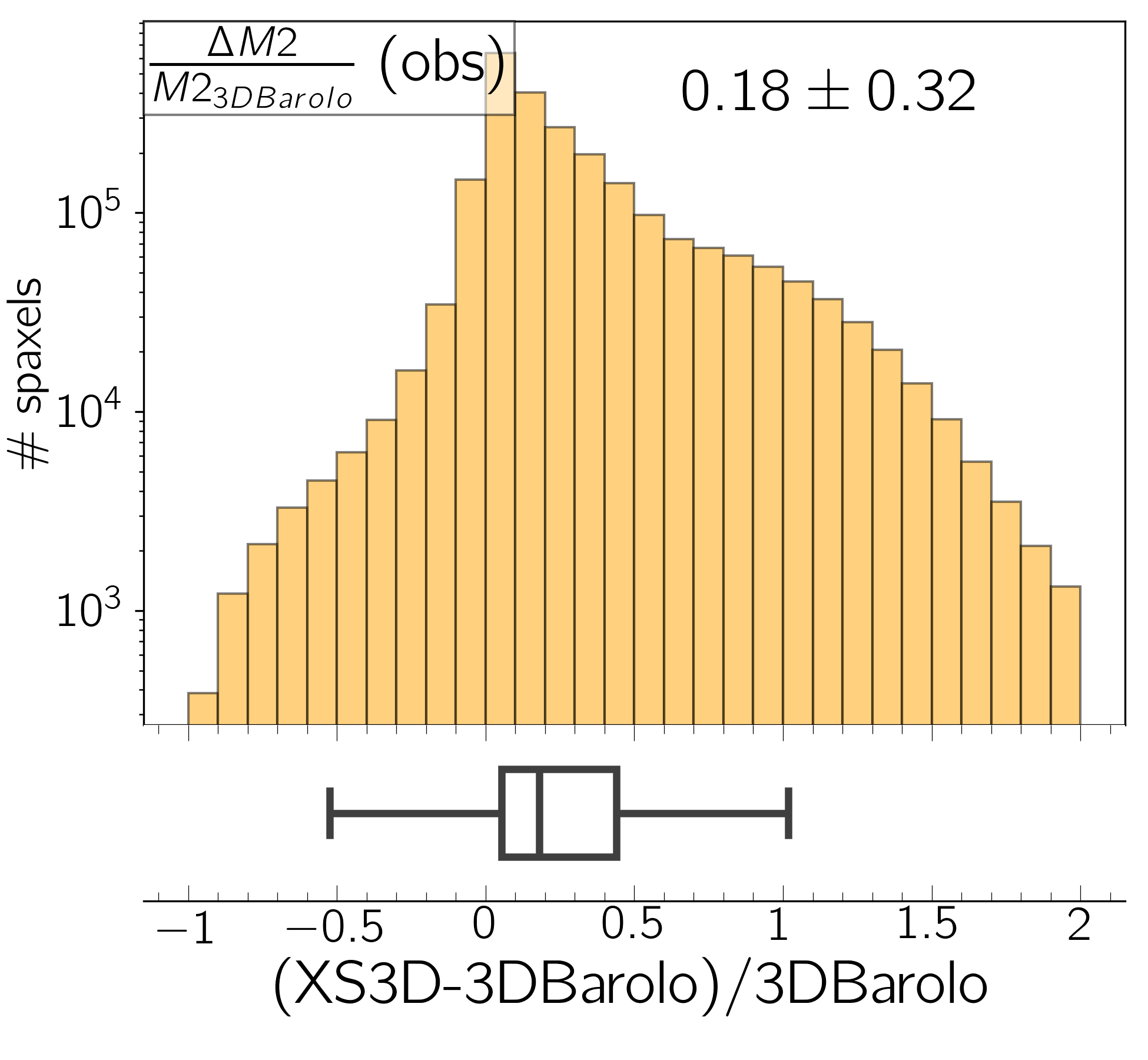}
	\includegraphics[width=0.32\textwidth,keepaspectratio]{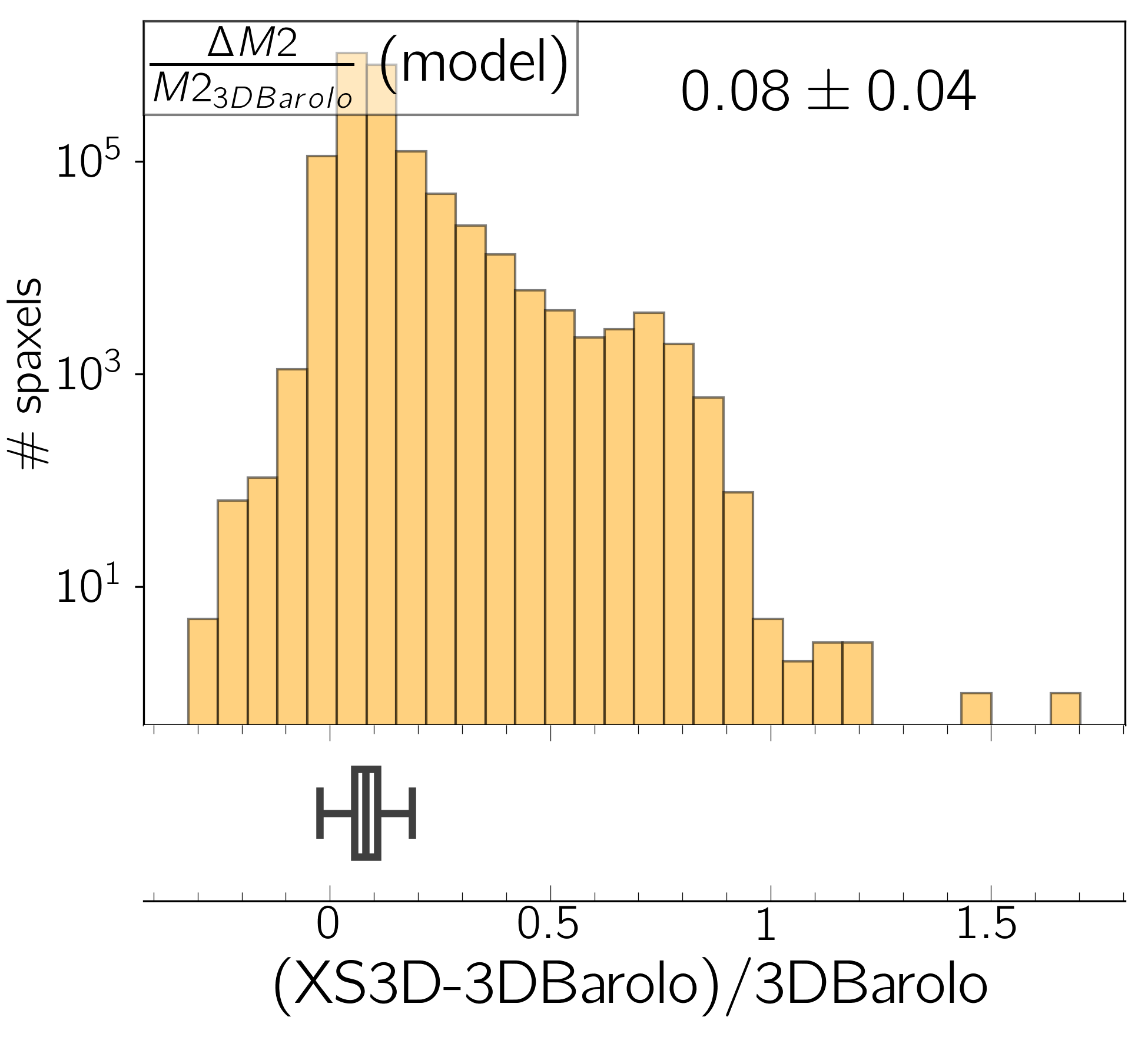}
	\includegraphics[width=0.32\textwidth,keepaspectratio]{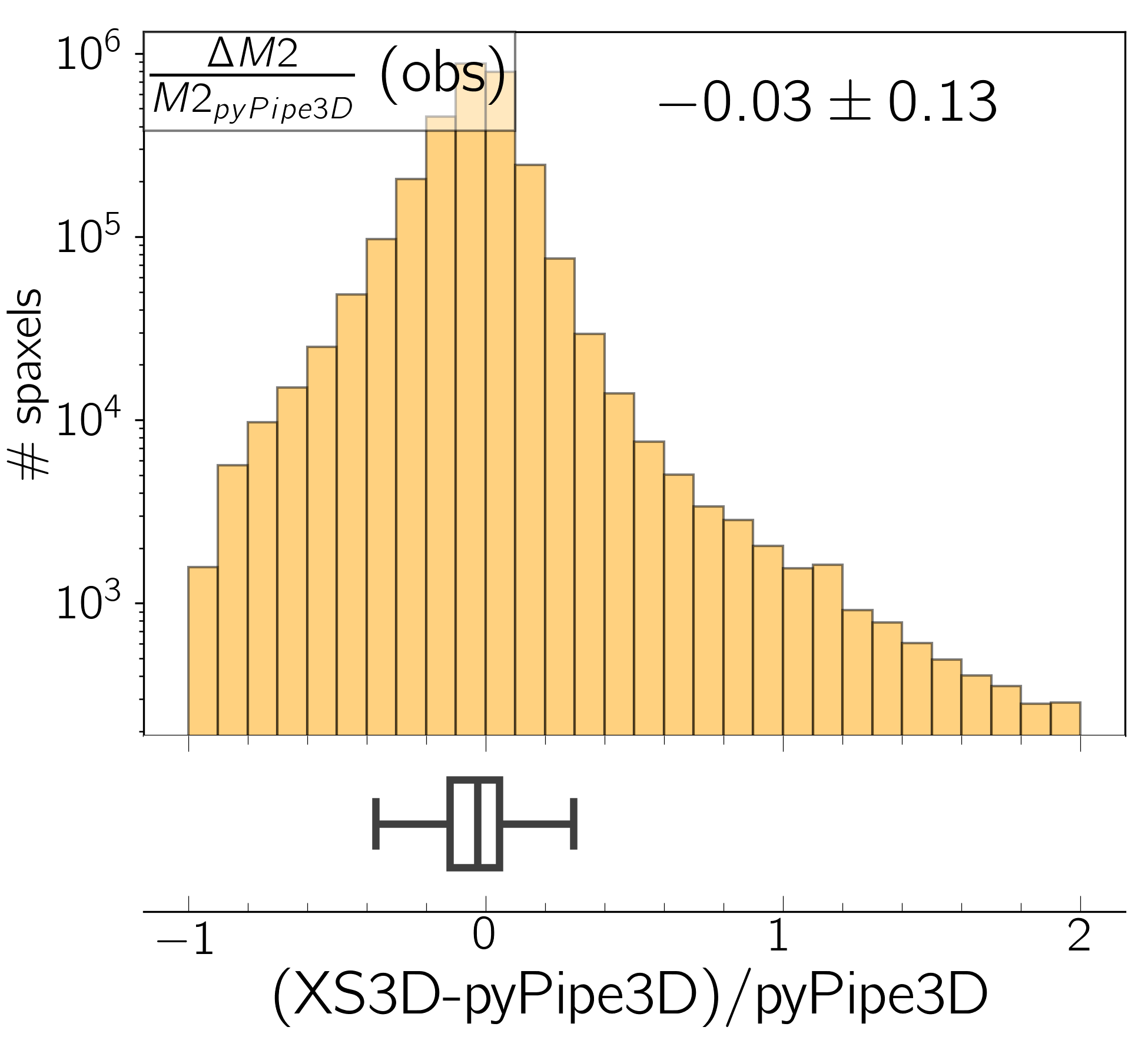}
	\caption{Spaxel-based comparison of the \Ha~velocity dispersion maps (2nd moments) from  {\tt 3DBarolo} and {\tt pyPipe3D} with those from {\tt XS3D}.
	Each histogram (in log scale) shows, from left to right, the relative error with respect to {\tt 3DBarolo} for the observed and model maps, while the third panel shows the relative error respect the {\tt pyPipe3D} dispersion maps.
	Box plots are shown below, with the median and standard deviation indicated in the upper right. Each histogram comprises approximately $2.5\times10^6$~spaxels.}
	\label{fig:moms_obs}
\end{figure*}

However, when comparing the velocity dispersion maps ($M_2$) we find larger errors, this can be observed in Figure~\ref{fig:moms_obs}. The comparison between the observed dispersion maps shows a relative error respect {\tt 3DBarolo} of $\sim 18\%$, while when comparing the models maps the error is $8\%$.

\begin{figure*}[t!]
	\centering
	\includegraphics[width=0.32\textwidth,keepaspectratio]{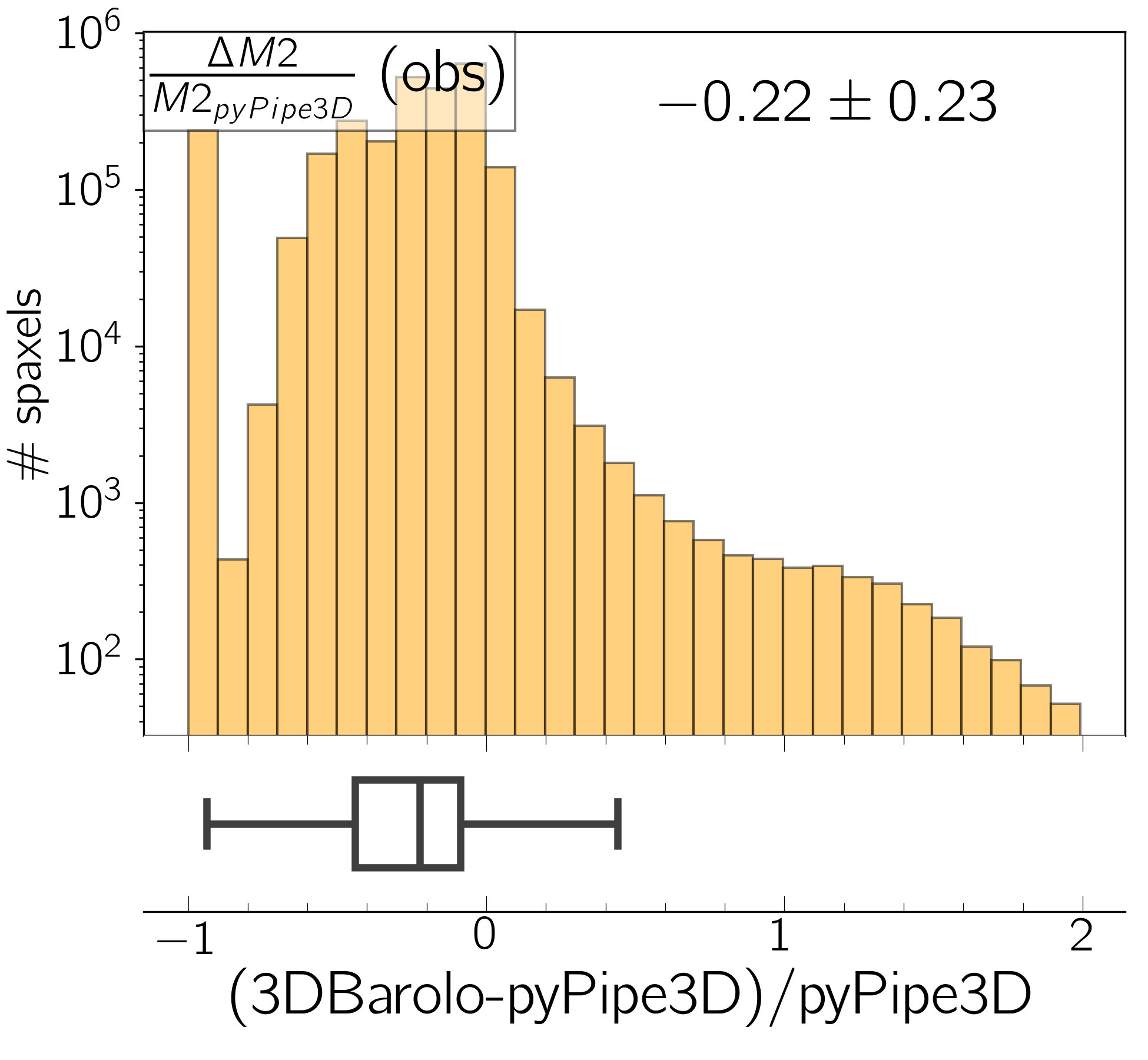}
	\caption{{ Similar to Figure~\ref{fig:moms_obs}, but comparing the observed dispersion maps from {\tt 3DBarolo} and {\tt pyPipe3D}. Relative errors are computed respect {\tt pyPipe3D}.}}
	\label{fig:moms_pipe3d}
\end{figure*}

To disentangle this discrepancy, we now compared  {\tt XS3D} with the dispersion maps from the {\tt pyPipe3D} pipeline, { which are derived from Gaussian fitting to the \ha~line}. The relative error in this case is of only $3\%$ as shown in the right panel of Figure~\ref{fig:moms_obs}. Figure~\ref{fig:moms_pipe3d} shows that indeed exist differences of up to $22\%$, in the velocity dispersion measurements between {\tt 3DBarolo} and {\tt pyPipe3D}.

The fact that the LOS dispersions from {\tt XS3D} are consistent with a third pipeline, {\tt pyPipe3D} in this case, suggests that the discrepancy  observed in  $\sigma_\mathrm{intrin}$ (i.e, Figure~\ref{fig:bbarolo_xs3d} and \ref{fig:vt_sigma_all}), must lie on how this parameter is defined in each pipeline. Note that in {\tt XS3D} the gas random motions are implicitly included in $\sigma_\mathrm{intrin}$, whereas {\tt 3DBarolo} treats these motions separately.

In terms of performance, we notice that {\tt XS3D} is always slower compared to {\tt 3DBarolo}. We find that on average {\tt 3DBarolo} is $\sim 2\times$ faster than  {\tt XS3D}. This is not surprising since {\tt XS3D}  is $100\%$ written in Python, which is an interpreter language. For reference, for the previously mentioned settings and without error computation, it takes $\sim 9$~min to analyze one of the PHANGS-MUSE \ha~subcubes in {\tt XS3D}, while the same analysis takes $\sim 5$~min in {\tt 3DBarolo}. In general, computational times depend on the number of independent velocities to estimate, and on the data cube dimensions.

Overall, we find that when comparing {\tt XS3D} and {\tt 3DBarolo} under the same conditions, all parameters yield equivalent results except for the intrinsic dispersion.

\bibliographystyle{aasjournal}
\bibliography{refs} 

\end{document}